Main Manuscript for

# Confinement determines transport of a reaction-diffusion active matter front

Nicolas Lobato-Dauzier[1*], Ananyo Maitra[1,2], André Estevez-Torres[1,3*], Jean-Christophe Galas[1*]

[1] Sorbonne Université, CNRS, Institut de Biologie Paris-Seine (IBPS), Laboratoire Jean Perrin (LJP), F-75005, Paris, France

[2] CY Cergy Paris Université, CNRS, Laboratoire de Physique Théorique et Modélisation, 95032 Cergy-Pontoise, France

[3] Université de Lille, CNRS, Laboratoire de spectroscopie pour les interactions, la réactivité et l'environnement (UMR 8516 - LASIRE), F-59655, Villeneuve d'Ascq, France

***Corresponding authors**: Nicolas Lobato-Dauzier, André Estevez-Torres, Jean-Christophe Galas
**Email:** {nicolas.lobato-dauzier ; andre.estevez-torres ; jean-christophe.galas}@sorbonne-universite.fr

**Author Contributions:** NL, AM, AET, JCG designed the research. NL performed the experiments and analyzed the data. NL, AM performed the analytical modelling. NL, AM, AET, JCG wrote the paper.

**Competing Interest Statement:** No competing interests.

**Classification:** Applied physical sciences (Biophysics and Computational Biology)

**Keywords:** Active turbulence, microtubule-kinesin, pattern formation, chemomechanics

**This PDF file includes:**
Main Text
Figures 1 to 6

**Abstract**
Couplings between biochemical and mechanical processes have a profound impact on embryonic development. However, in-vitro studies capable of quantifying these interactions have remained elusive. Here, we investigate a synthetic system where a DNA reaction-diffusion (RD) front is advected by a turbulent flow generated by active matter (AM) flows in a quasi-one-dimensional geometry. Whereas the dynamics of simple RD fronts solely depend on the reaction and diffusion rates, we show that RD-AM front propagation is also influenced by the confinement geometry. We first experimentally dissected the different components of the reaction-diffusion-advection process by knocking out reaction or advection and observed how RD-AM allows for faster transport over large distances, avoiding dilution. We then show how confinement impacts active matter flow: while changes in instantaneous flow velocities are small; correlation times are dramatically increased with decreasing confinement. As a result, RD-AM front speed increased up to 8-fold compared to a RD one, in quantitative agreement with a conveyor belt RDA theoretical model. The RD-AM experimental system described here provides a framework for the rational engineering of complex spatiotemporal processes observed in living systems. It will reinforce our understanding of how macro-scale patterns and structures emerge from microscopic components in non-equilibrium systems.

**Significance Statement**
Collections of molecules that react and diffuse create spatial patterns that may be used in living systems to provide positional information. Other collections of molecules, often made of filaments and molecular motors, generate mechanical forces that induce active flows in a closed container. These two processes are ubiquitously coupled in living systems. We designed a non-living experiment to study such coupling where the simplest reaction-diffusion pattern, a front, propagates in a solution of active flows. We found that the front propagates faster and blurs when the width of the container increases, which is not observed for classical reaction-diffusion fronts. This mechanism could be used in vivo for accelerating the transport of biochemical information.

## Introduction

Do cells and embryos harness confinement to optimize transport of chemical cues? Transport is singularly important during morphogenesis[1] where it can be achieved via diffusion, reaction-diffusion (RD) or advection by active flows[2]. However, the intrinsic complexity of living embryos[3] has led biophysicists and biochemists to develop synthetic systems where these transport mechanisms can be coupled in order to investigate general principles. Here we couple reaction-diffusion and active flows and show that confinement plays a major role in how transport occurs.

While transport by diffusion is fast at short length scales, it becomes slow at long distances as it scales as the square root of time. Two mechanisms allow going beyond this limitation. On the one hand, RD fronts allow for long-range transport at a constant velocity when an autocatalytic reaction sustains concentration levels[4,5]. RD fronts have been observed in a range of chemical systems from inorganic chemistry[6], to synthetic biochemical circuits[7–10]. DNA computing has proven to be a versatile tool to engineer RD patterns[11–13]. Adding enzymes[14,15] has enabled to program tunable RD fronts[16] and a variety of complex structures such as spiral waves[17], go-fetch fronts[18] and French flag patterns[19]. On the other hand, active flows provide a transport mechanism that is independent of both reaction and diffusion of a chemical species. In vivo, they may arise from a coupling between two types of cytoskeleton proteins: filaments and motors. In vitro systems combining such cytoskeletal proteins form active gels that share many similarities with living tissues[20], forming a variety of patterns[21–26] and generating flows, either chaotic[27] or coherent[28].

The coupling between advection and reaction-diffusion has primarily been studied for inertial flows[29,30]. In particular, work has focused on inertial turbulence[30–32] where hydrodynamic energy cascades down the spatial scales. The conclusion is that inertial turbulence aids the propagation of the reaction front. While such speed-up would undoubtedly be useful in multiple microbiological contexts, because microbiological materials have very small Reynolds numbers, they are Stokesian and inertial turbulence is impossible. However, molecular active matter produces a turbulence-like spatiotemporally chaotic state called active turbulence that creates large-scale flow structures driven by the small-scale injection of energy, potentially inducing a very different transport[33,34]. In principle, there is no reason why the couplings between active turbulence and reaction-diffusion should resemble those observed for inertial turbulence, though a recent numerical study of reaction-diffusion in a particular model of active, spatiotemporally chaotic fluids has suggested that there are certain similarities in the front dynamics[35]. Importantly, in both inertial turbulence-aided RD and active turbulence-aided RD, as exemplified by[32,35], the effect of the flow on the speed of the RD front can be captured by a renormalized effective diffusivity. Bate, *et al*[36] recently studied the coupling between diffusion and active flows, showing how fast transport of an inert small molecule using sustained turbulent active flows can be, but also how a chemical front cannot be sustained without reaction. Senoussi, *et al*[37] introduced reaction-diffusion active matter (RD-AM) fronts, but there the achieved active turbulence did not induce significant transport, because as our work will show, advection was acting at a time scale much smaller than reaction, which prevented real coupling.

In this experimental-theoretical work, we use a reconstituted in-vitro system to explore how reaction-diffusion kinetics are affected by spontaneous fluid flows due to active driving. We observe experimentally that the transport this induces is unexpected and strongly determined by the confinement-induced structure of the active flows. This contrasts with expectations based on reaction-diffusion systems in turbulent fluids and homogeneous, spatiotemporally chaotic active fluids. We show that the transport we observe cannot be fully understood in terms of a Fisher-Kolmogorov-Petrovskii-Piskunov[16,38] (FKPP) theory with a modified diffusivity. We develop an effective reaction-diffusion-advection model to rationalize these results and uncover two distinct regimes. At high confinement advection can be captured by a renormalized effective diffusivity. At low confinement, active flows have a novel spatial structure allowing for an 8-fold increase in transport speed. This property, which arises from active matter being ruled by geometry-sensitive instabilities[39–41] may be harnessed in vivo to optimize the transport of chemical cues.

## Results

### Experimental System

Our experiment combines a chemical sub-system incorporating reaction and diffusion and a mechanical sub-system that advects the chemical one via flow[37]. The former is a single-stranded DNA (ssDNA), denoted by α, coupled to a PEN DNA toolbox[14,15] autocatalytic reaction (Fig. 1A). The latter consists of microtubule filaments that form bundles in the presence of a depleting agent and are in continuous movement in the presence of clusters of kinesin molecular motors[27] (Fig. 1B).

The autocatalytic reaction leads to the exponential amplification of $\alpha$ through the reaction $\alpha \xrightarrow{f} 2\alpha$ with a reaction rate $f$. This reaction depends on a polymerase and a nicking enzyme that replicate the template ssDNA $T_{\alpha\alpha}$ in the presence of deoxyribonucleotides (dNTPs). Its rate is modeled with a Michaelis-Menten kinetics

$$f([\alpha]) = \frac{V[\alpha]}{K_M + [\alpha]}, \qquad (1)$$

where $[\alpha]$ is the concentration of α, $V$ the limiting rate and $K_M$ the Michaelis constant. We extracted the reaction rate in the limit of small concentrations from a control experiment in a fluorimeter and obtained a typical reaction time $\frac{1}{k}$ = 50 ± 15 min (see Fig. S1 & supplementary methods text), corresponding to a reaction rate constant $k = f'(0) = \frac{V}{K_m} = 0.02$ min$^{-1}$. When $\alpha$ is introduced on one side of a long rectangular channel, the chemical sub-system generates a RD front[16].

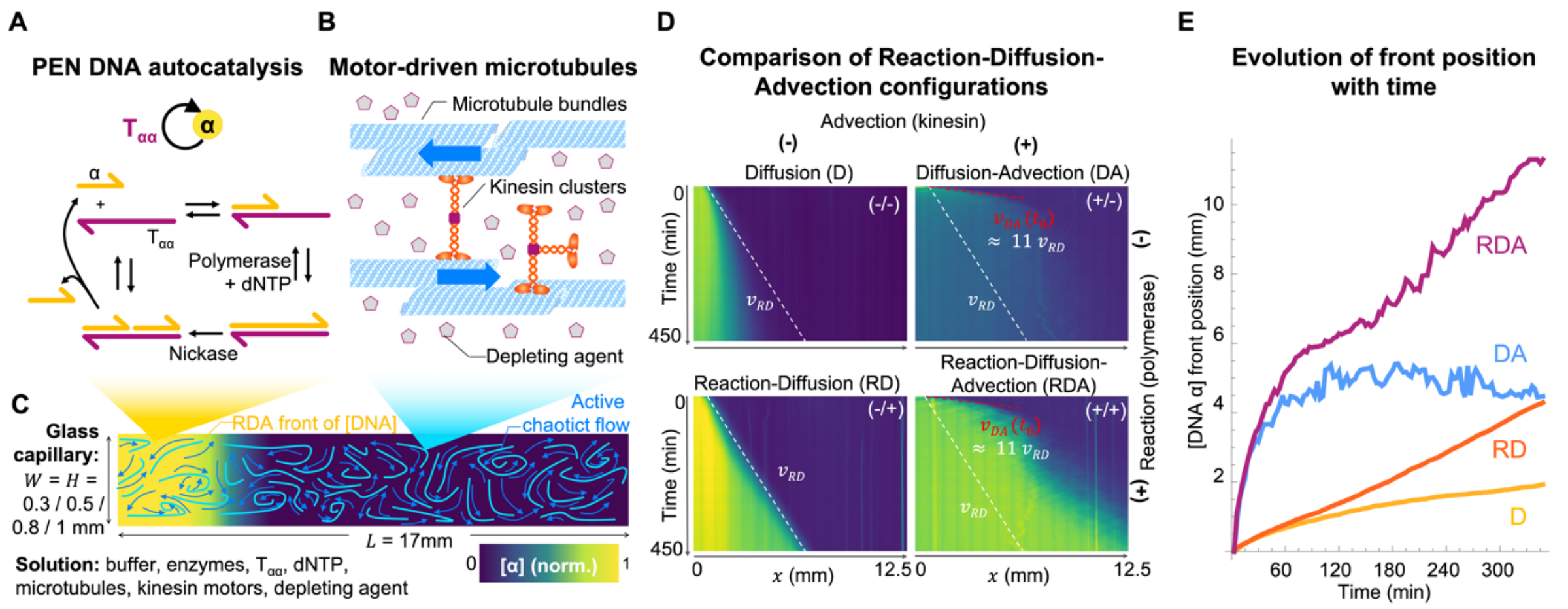

***Figure 1 – Experimental setup and its building sub-systems. A:*** *Sketch of the chemical sub-system where ssDNA $\alpha$ is amplified exponentially via the DNA template $T_{\alpha\alpha}$ and a polymerase and nickase enzymes.* ***B:*** *Sketch of the mechanical sub-system where microtubule bundles slide apart driven by clusters of kinesin motors.* ***C:*** *Experimental setup. Active chaotic flows interact with a front of $\alpha$ in a closed glass capillary.* ***D****: Kymographs of $[\alpha]$ for the RD-AM system in the four different reaction-diffusion-advection regimes (channel width W=0.8 mm). (+) and (-) indicate respectively the presence or the absence of kinesin or polymerase. The white and red dotted lines correspond to the velocity $v_{RD}$ of the front in the reaction-diffusion (RD) regime and the initial velocity $v_{DA}(t_0)$ in the diffusion-advection (DA) regime, respectively.* ***E:*** *Front position over time for the four regimes from a single experiment (threshold chosen at $[\alpha]$ = 0.2. to measure a front even without reaction).*

In the mechanical sub-system, kinesins use ATP as a fuel to generate microtubule flows. In this work we use caged-ATP to trigger the activity of motors on demand using ultraviolet (UV) light[36] (Fig. S2). Both subsystems are mixed in an aqueous solution in the absence of α and injected into glass capillaries of 17 mm length and varying square sections of 0.3-1 mm widths, plunged in oil to prevent evaporation and observed with a fluorescence microscope (Fig. 1C). A small quantity of α is added on the left side of a channel to trigger the front. Microtubules are fluorescently-labelled and the DNA concentration is monitored thanks to a fluorescent DNA intercalator.

**Experimental dissection of the reaction-diffusion-advection model**

The fluid flow generated by the mechanical subsystem advects the chemical subsystem. Taking this into account, the concentration of $\alpha$ satisfies the equation:

$$\frac{\partial[\alpha]}{\partial t} = f([\alpha]) + \left(\vec{u} \cdot \vec{\nabla}\right)[\alpha] + D\Delta[\alpha], \quad (2)$$

with $[\alpha]$ the DNA concentration, $f([\alpha])$ the reaction rate, $\vec{u}$ the velocity of the fluid, and $D$ the diffusion coefficient. In order to experimentally study the different contributions of each term in the RDA equation, we successively knock out reaction and/or advection by suppressing DNA polymerase and/or kinesin motors in the experiments and investigated the dynamics of the front of α in a channel of width W = 0.8 mm. Figures 1D,E respectively show the kymographs and front position over time. The threshold to determine the front position is chosen at the low value $[\alpha]$ = 0.2 in order to be measurable even in absence of reaction.

When both kinesin and DNA polymerase are absent, α diffuses freely (Figs. 1D top left and S3). Transport slows down quickly and the concentration front is nearly stopped after ~2 mm. From this experiment we extracted the bare diffusion coefficient of α $D = 1900 \pm 200$ $\mu m^2$/min, in agreement with previous work[16]. Adding polymerase turns on the DNA autocatalytic reaction and generates a reaction-diffusion (RD) front propagating with constant speed $v_{RD} = 12.6 \pm 0.6$ μm/min and width $\delta_{RD} = 0.3 \pm 0.1$ mm (Figs. 1D bottom left and S4), as previously reported[16]. The width $\delta_{RD}$ corresponds to the characteristic length of a sigmoidal profile $1/(1 + e^{x/\delta_{RD}})$ and is measured using the difference in position of two levels of concentration of α (Fig S4D & supplementary methods). The Fisher-Kolmogorov-Petrovskii-Piskunov[16] (FKPP) model predicts the speed of such RD fronts to be

$$v_{FKPP} = 2\sqrt{kD}, \quad (3)$$

where *k* is the reaction rate constant and *D* the diffusion coefficient. The associated width is given by

$$\delta_{FKPP} = \sqrt{\frac{D}{k}}. \quad (4)$$

Using the independently measured value of $k$ and $D$ we find $v_{FKPP} = 13 \pm 3$ μm/min and $\delta_{FKPP} = 0.3 \pm 0.05$ mm, in excellent agreement with the experimental values $v_{RD}$ and $\delta_{RD}$. Eq. (4) means that for a given chemistry (i.e. $k$ fixed), any gain in speed via a change in $D$ involves also an increase in the width of the front. We will see in the following that our active advection flows decouple $v$ and $\delta$. This decoupling is also true for advection due to chaotic flows where the speed of the front is nevertheless described by Eq. (3),with a renormalized $D$[35].

A diffusion-advection (DA) system can be obtained in the presence of kinesin and in the absence of polymerase (Figs. 1D top right and S5). Advection allows for fast transport over large scales. At short times, the speed is 11-fold higher than in the pure RD system. But this induces a dilution, which attenuates the signal making its use in further downstream computation impossible. We observe that the DA front is stopped half-way through the channel. This sharp stoppage looks like a flow barrier that blocks substantial concentrations of DNA to be advected further.

When both polymerase and kinesin are present (Figs. 1D bottom right and S6), the RD front propagates within an active convective medium. This reaction-diffusion-advection (RDA) front resembles the DA front at short times ($t$ < 30 min) with similar propagation (Fig 1E). However, in contrast with DA, the reaction term allows, on the one hand, the DNA concentration to recover behind the front and, on the other hand, to reach a high steady state at long times. While speed of the RD front is constant during its propagation, the RDA is not: the speed of the front decreases when it reaches the middle of the channel like in the DA case, as the front positions over time show (Fig 1E). However, unlike the DA case, propagation is not stopped at the middle of the channel and continues to propagate beyond that with a seemingly larger width (Fig. S6).

In summary, at short times (t < 30 min) advection is the most discriminating factor, while for t > 1h the difference in transport comes from reaction (Fig 1E). Before studying further the RD-AM dynamics we will investigate in detail the effect of confinement in the active sub-system alone.

## Effect of confinement on active turbulent flows

Confinement is known to dramatically change active-matter systems[39–43], in contrast to reaction-diffusion ones. In particular, channel width has been reported to induce a transition from locally chaotic to coherent flows in microtubule-kinesin active matter[28]. In Wu *et al*, coherent flows in the XY plane were possible because of the closed loop geometry used. This result is not directly transposable to our 17 mm long straight capillaries.

To study the impact of confinement in the active advection flows, we used 1 µm-diameter fluorescent beads as velocimetry tracers (Figs. 2, S7-8, Movies S2-3) for the mechanical sub-system alone in square channels of different widths $W$. Over a 30 min time-scale, bead trajectories are very different in the highly confined channel ($W$ = 0.3 mm) compared to the wider channel ($W$ = 1 mm, Fig. 2A). In the former, bead trajectories are eddies of a typical length of 0.5 mm. In the latter, trajectories are much more rectilinear, the flow velocity along *x* not changing sign for up to 4 mm. This result is not trivial as one could expect the boundaries to funnel trajectories in the longitudinal direction or confinement to allow less directional changes. We observe the opposite. While complex, the variation in the flow field when varying confinement can be described with two scalars: the maximal speed $u_{max}$ and the correlation time, $\tau_{cor}$. $u_{max}$ is the maximum velocity along the $x$ axis, extracted using complete trajectories of a single bead. $\tau_{cor}$ corresponds to the characteristic time between a direction switch along the x axis and is extracted from the changes in sign of velocity correlation function (see supplementary methods). Fig. 2B & Fig. S10 show that the velocity correlation function for each bead greatly changes as W is increased. In contrast, we observe that $u_{max}$ increases linearly with W with a significant but modest 1.5-fold increase from 0.3 to 1.0 mm (Fig. 2C). When we plot the distribution of bead velocity correlation times, $\tau_{cor}$, vs. *W* we observe a sharp non-linear increase in the range $W$ = 0.5-0.8 mm up to 10-fold (Fig. 2D). As $\tau_{cor}$increases, the flow trajectories become more rectilinear as observed in Fig. 2A. The typical vortex size is $u_{max} \cdot \tau_{cor}$, which also increases sharply in the range *W* = 0.5-0.8 mm from 0.2 to 2 mm (Fig. 2E). Importantly, the typical vortex size is smaller than $W$ in narrow channels ($W \leq 0.5$) but larger than $W$ in broader channels ($W \geq 0.8$). To quantify the relative strength of the active advection compared to diffusion we calculate the Péclet number, $Pe = {u_{max}}^2\tau_{cor}/D$, taking $u_{max}$ as the characteristic speed and the vortex size $u_{max}\tau_{cor}$ for the characteristic length (Fig S11D). For $W \leq 0.5$, we obtain $Pe$ = 1-10, while for $W \geq 0.8$, $Pe$ = 20-200. This means that advection is never negligible compared to diffusion, but more importantly there is a factor 20 between the effect of advection in narrow and large channels.

Representative trajectories of beads sorted by position along $x$ for the 0.3 mm (Fig. 2F, left) and 1 mm wide channel (Fig. 2F, right) confirm this difference in vortex sizes for these two extreme channel widths. Moreover, they suggest that while there is no large-scale spatial organization in the 0.3 mm channel, with directions switches happening evenly along $x$, the flow in the 1 mm channel is mainly constituted of two hydrodynamic cells, resembling convection cells, with beads switching directions on each side and in the middle. Figure 2G shows the average orientation of the velocity vector in the narrowest and broadest channels. Red areas correspond to zones where beads move along the *x* axis, while blue ones correspond to a motion in the transverse direction. In the 0.3 mm channel such zones are numerous, small and not very sharply defined whereas in the 1 mm channel we easily observe two red zones on each side and a blue zone in the middle. Considering the channel as a 1D system, bead transport is only achieved in the red zones. Blue zones act as hard to cross barriers that force most beads to switch directions (Fig. S8 Movies S2-3).

In summary, the results in Fig. 2 show a dramatic change in the structure of the flow as the width increases from 0.5 to 0.8 mm.

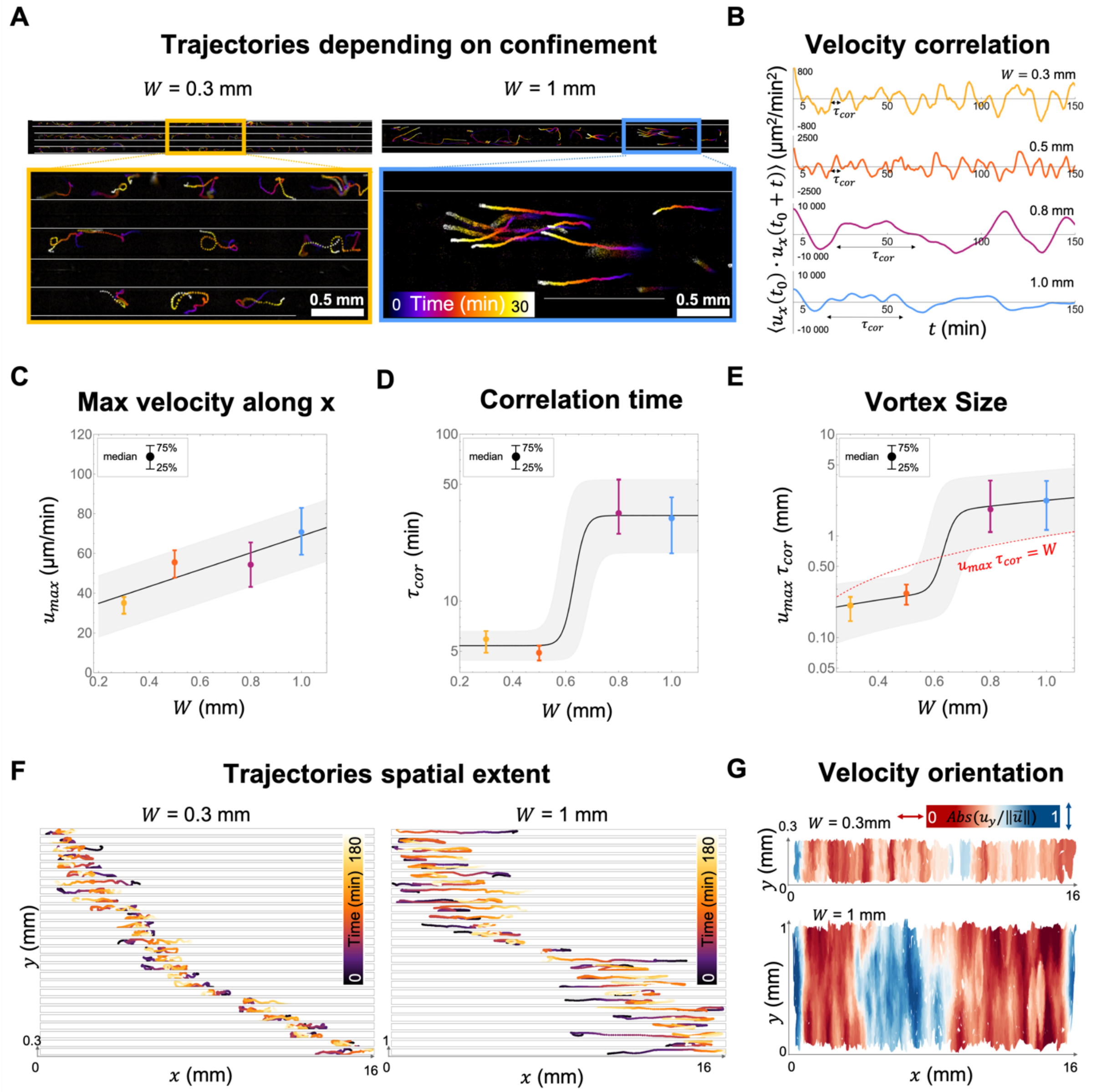

***Figure 2 – Characterization of chaotic active flows in closed channels of different widths. A:*** *Trajectories followed with fiducial beads in square channels of 0.3 mm (left) and 1 mm width (right). Time is color coded from purple to white.* ***B:*** *Bead velocity correlation along the x axis vs. time for a representative bead for each channel. The correlation time is indicated as* $\tau_{cor}$. ***CDE****: Distributions of bead maximal velocity along the x axis* ***(C)****, bead correlation time along the x axis* ***(D)*** *and vortex sizes* ***(E)*** *for different channel widths. Central dots in panels C, D, E correspond to the median value, and bars to the 25% and 75% quantiles.* ***F:*** *Representative bead trajectories sorted by position along x for the 0.3 mm (left) and 1 mm wide channel (right).* ***G****: Average bead velocity orientation map for the 0.3 mm (top) and 1 mm wide channel (bottom). Red and blue indicate bead velocity respectively parallel and orthogonal to the x.*

Such spatially heterogeneous flows along the length of the channel are not captured within simple analytical treatments of active gels We attempted to model this using a well-established model of active fluids[40,44–47] (See Theoretical supplement). Such traditional treatments of active matter would suggest that an aligned state of microtubules is destroyed by activity[39,48–50] (Fig. S12), to be replaced by a well-mixed (turbulent) gel with a characteristic vortex size taken to be the wavelength of the fastest growing mode of the instability $\lambda_f$. This strongly depends on the confinement geometry. In a long channel with a square cross-section of size $W$, it scales as

$$\lambda_f^2 = \frac{2\,W^2}{\left(\frac{W}{l_a}\right) - 1} \sim W l_a, \tag{5}$$

where we have defined the active length scale

$$l_a = \pi \sqrt{\frac{3K[4\Gamma\eta + (\lambda - 1)^2\,] + 4\zeta_2(\lambda - 1)}{3\zeta(\lambda - 1)}}, \tag{6}$$

that is a complicated function of the active stresses $\zeta$ and $\zeta_2$, passive elasticity $K$, viscosity $\eta$, rotational viscosity $\Gamma^{-1}$ and a flow-alignment coefficient accounting for the tendency of microtubules to align with the fluid, $\lambda$. While $\lambda_f$does increase with $W$ as observed in the experiment, it doesn't account for the flow structure seen in the experiment which is much more intricate. Crucially, the flow we observe is heterogeneous along the channel with clearly defined hydrodynamic cells or vortices as seen in Figure 2. The scaling of the vortex lengths is far from $\sqrt{W}$ scaling predicted by this active model. It is also far from $W$ scaling obtained by assuming that the flow structures assume the largest possible size allowed by the lateral confinement.

While this linear theory is insufficient for the explanation of the detailed flow structure, it may well be recapitulated by a numerical simulation of the active fluid model with the experimentally relevant boundary condition. However, our primary interest here is not to understand the flow structure but its effect on the propagation of the reaction front. We will show that the front speed in our system does not fit the naive estimate obtained from the FKPP theory obtained by replacing the molecular diffusivity with a renormalized one –(as would be expected if the advecting flow was the standard active turbulent one[35]) Instead displaying a more intricate behaviour. We will construct a heuristic model to rationalize this.

**Front speed increases with channel width**

We now investigate the effect of the active flow on the RD process. Figure 3 shows the propagation of RD-AM fronts for the two limit cases of low and high transport studied in Fig. 2 ($W$ = 0.3 and 0.8 mm), while Figs. 4 and S13 provide data for all channel widths. Corresponding microtubules images are presented in Fig. S14. To show that the difference is solely due to the active chaotic flows, we first let the experiment run with caged-ATP until $t_0$ = 77 min, when it is uncaged with UV light. Note that, while for RD fronts the front shape is stationary with a speed independent of the chosen concentration level threshold, for RD-AM fronts it is difficult to define the actual position of the front due to its inhomogeneities and instabilities. Choosing different concentration thresholds leads to different (but coherent) results (Fig 4 & Fig. S16). In the following we have chosen to measure the front position using a concentration threshold of $[\alpha]$ = 0.5 corresponding to half its maximum value (See fig S16), as described in the supplementary methods text. For each width, several front speeds are measured sequentially: $v_0$ is the speed before ATP was uncaged while $v_{1-3}$ correspond to the constant speeds measured at different transport phases (Fig. S15).

Before UV exposure, the front speed is identical in all channels, irrespective of *W*, with $v_0$= 9±1 µm/min. This value is 30% lower than the $v_{RD}$ measured in Fig. 1D, and we attribute the difference to batch-to-batch variability of enzymes. After UV exposure, the front dynamics is only slightly changed for the 0.3 mm channel, while it is significantly disrupted for the 0.8 mm one (Fig 3A-D). In the thinner 0.3 mm channel, the front speed after UV exposure, $v_1$ is identical to $v_0$ while the front width increases linearly over time up to twofold at 200 min (Fig. 3E). Both observations are in line with the small extent trajectories observed in Fig. 2A and advection only acting as mixing. In contrast, in the 0.8 mm channel, right after uncaging, the front speed increases 9-fold to $v_1$ = 85 µm/min (Fig. 3F). During this first phase, that lasts 40-50 min, strong inhomogeneities appear in the transverse region: low fluorescence spots behind the front show backward dilution while bright zones ahead of the front indicate forward transport of DNA (Fig 3D). This corresponds to transport by advection in a single hydrodynamic cell, with concentration levels sustained by the reaction.

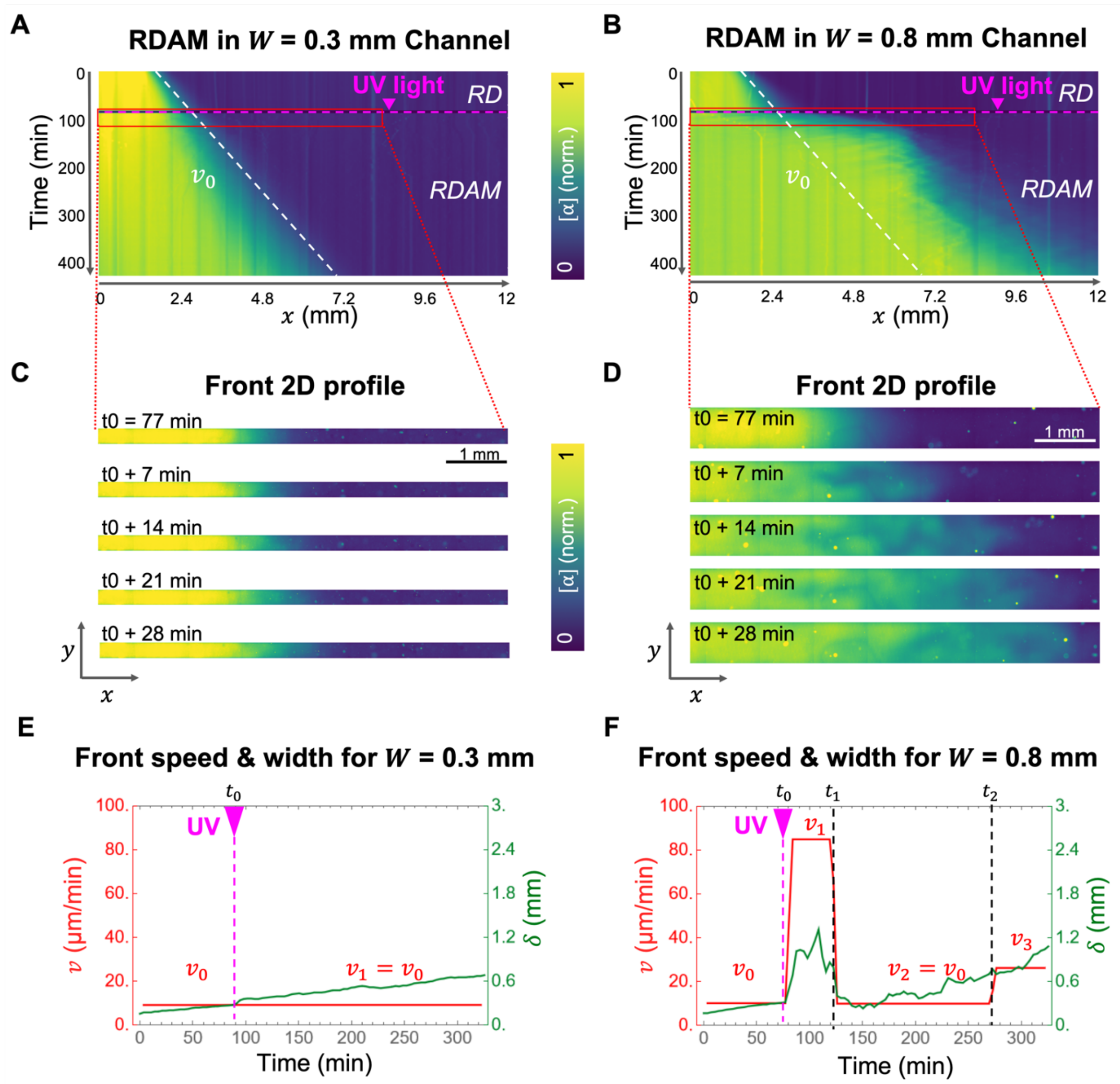

***Figure 3 - Geometry-dependent Reaction-Diffusion-Advection. A:*** *Kymograph of the front propagation in the 0.3 mm width channel.* ***B:*** *Kymograph of the front propagation in the 0.8 mm width (See Fig. S13 for the 0.5 mm and 1 mm channels).* ***C-D:*** *Time-lapse fluorescence images showing front 2D profile in the $xy$ plane.* ***E-F:*** *Evolution of front speed (left axis, red) and width (right axis, green) for the 0.3 mm (E) and 0.8 mm (F) channel. See Fig. S17 for data of all channel widths.*

A second propagation phase is observed for the fronts in the channels with widths $W \geq 0.5$ mm (Fig. 4A). This happens when the fronts reach the middle of the channel ($t_1$ = 110-120 min, for $W$ = 0.8-1.0 mm and $t_1$ = 190 min for $W$ = 0.5 mm in Figure 4A), where it is severely slowed down back to $v_2 = v_0$ = 9 µm/min, while the front width decreases abruptly. We interpret this second phase as the front reaching the limit of a convection cell: advection becomes negligible, and the front behaves as an RD one. As it happens for the first phase in $W$ = 0.3 mm, $v_2$ remains constant during this second phase and $\delta$ increases steadily (Fig 3E-F).

For channels $W$ = 0.8-1.0 mm a third phase is observed at $t_2$ = 280 min, with speed $v_3$ about two-fold the RD one and $\delta$ still increasing over time (Figs. 3F and 4A). As the experimental configuration is similar to that observed in the first hydrodynamic cell when ATP is uncaged, one could expect $v_3$ to be equal to $v_1$. However, while $v_1$ corresponds to the advection of a previously well-established RD front, this is not the case for $v_3$: advection in the second cell washes away any DNA that diffuses through the hydrodynamic cell barrier, preventing concentration build-up and proper propagation of a stationary front. To test if the difference between $v_1$ and $v_3$ was related to the spatial organization of the hydrodynamic flow or to temporal variations of the active matter, we changed the position of the initial condition of the front in $W$=0.8 mm channels. We observed that propagation of a front starting at the middle of the channel, and thus sitting at the border between two hydrodynamic cells, is not modified by active matter and its speed remained equal to $v_{FKPP}$ (Fig. S18). Instead, a front starting on the right side of the channel, and thus sitting on the second hydrodynamic cell, propagated similarly to a front starting on the left side of the channel, and thus sitting on the first hydrodynamic cell, both travelling at $v_1$ (Fig S18). We thus conclude that the different speeds for each phase are related to the spatial organization of the hydrodynamic flow. At longer times (t > 250 min), we observe in the 1 mm channel an increase in DNA concentration happening almost simultaneously everywhere. We interpret it as a kinematic wave[51] of DNA amplification of very dilute concentrations advected across the whole length of the channel.

Figure 4B sums up the different speeds obtained in each channel, extracted from a linear piecewise decomposition of Figure 4A (see also Figure S15).

In summary, phases 0 and 2 for all channels and phase 1 for $W$ = 0.3 mm correspond to a reaction-diffusion front of speed $v_{RD}$ although, in contrast to RD fronts $\delta_1$ and $\delta_2$ in these cases increases linearly with time, probably due to mixing by active matter with a low correlation length.

Phase 1 for $W \geq 0.5$ mm corresponds to fronts disrupted by active matter that reach a stationary phase with $v_1$ increasing with $W$ until reaching a plateau at $W$ = 0.8-1.0 mm.

Phase 3 is visible for W ≥ 0.8 mm and corresponds to the advection of fronts in a second hydrodynamic cell that have not reached a stationary level. This initial condition in the second hydrodynamic cell originates from small amounts of $\alpha$ diffusing from the first cell. It thus differs significantly from phase 1 initial condition which is a well-established FKPP front. Indeed, the front speed for a threshold of $[\alpha]$ = 0.15, i.e. a low concentration, is fast (up to 5 times $v_0$ in the 1 mm channel) showcasing that advection is still strong (upper bar in Figure 4B). In contrast, the speed for a threshold of $[\alpha]$ = 0.85, i.e. a high concentration, is lower, close to the RD speed. To sum up, $v_3$ corresponds to the propagation of diffusion-limited fronts in the second hydrodynamic cell.

In the following we will focus on explaining the advection of a well-established FKPP front as a function of confinement, occurring during phase 1.

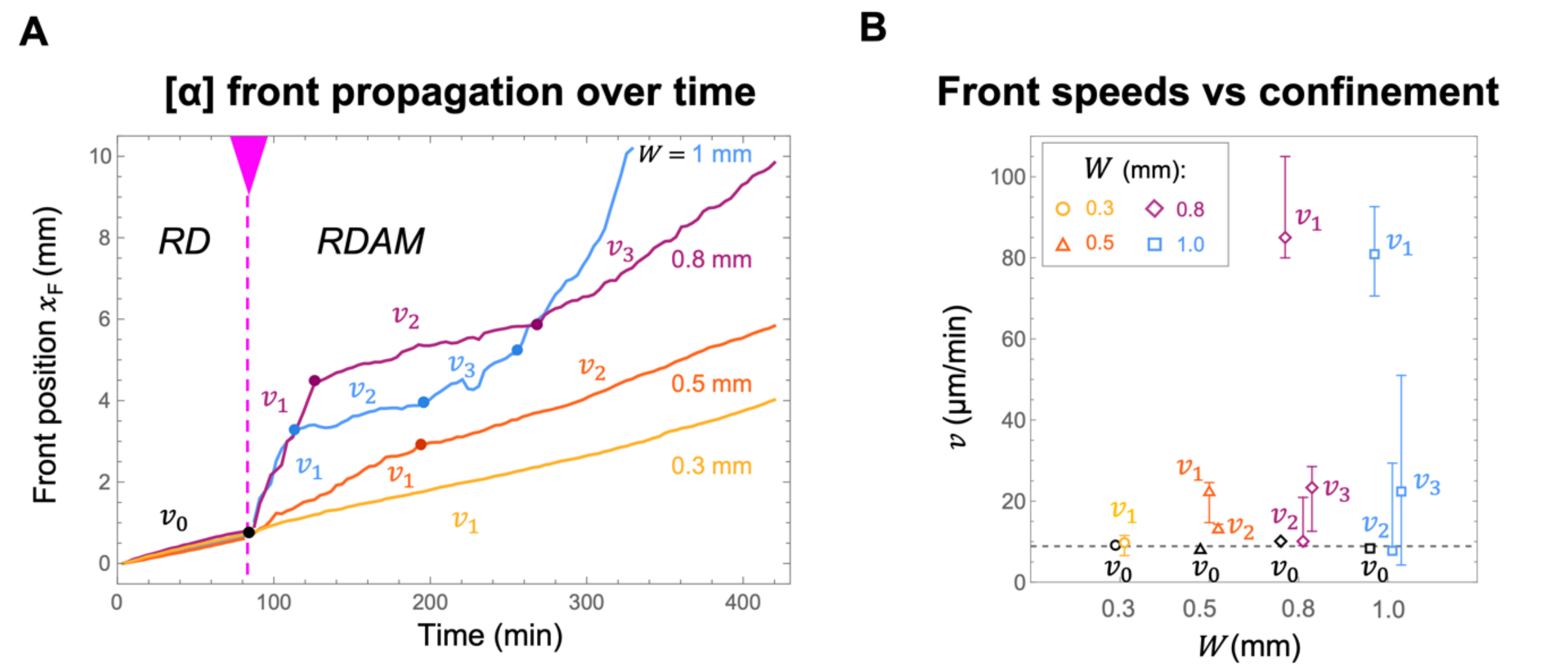

**Figure 4 -** ***Front speeds vs. channel width***. ***A:*** *Evolution of front position for different channel widths (Threshold $[\alpha]$ = 0.5). The speeds of each different transport phase are marked for each channel.* ***B:*** *Front speeds vs. channel width. For each width, several front speeds are measured sequentially. $v_0$ is the speed before ATP was uncaged while $v_{1-3}$ correspond to the constant speeds measured at different transport phases. The dashed line corresponds to the average of all $v_0$. The central symbol corresponds*

*to the speed of the front for a threshold taken at 50% of the maximum concentration, and bars respectively to 15% (upper bar) and 85% (bottom bar).*

**Towards a heuristic rationalization of the RDAM front behavior: a conveyor-belt RDA model**

In order to rationalize these results, we need an effective model of the flow that recapitulates features seen in in Fig. 1 and couples to the RD front. We therefore propose a conveyor-belt RDA model which is not based on a fundamental understanding of the flow structure starting from an active hydrodynamic theory. Instead, it heuristically accounts for the flow behaviour described in Fig. 2. We restrict our analysis to the transport within a single hydrodynamic cell, which corresponds to phase 1, with front velocity $v_1$.

The flow within the cell is modeled as a conveyor belt in the $xy$ plane, with the upper stream having velocity $+u_0\vec{x}$ and the lower stream $-u_0\vec{x}$. Species α can switch from one stream to the opposite stream through the hydrodynamic perturbations (smaller scale vortices observed experimentally, with a characteristic time $\tau$). The corresponding equations are derived from Eq. (2) by considering a uni-dimensional problem where α switches between two populations, $\alpha_+$ and $\alpha_-$ through a fictional reaction of characteristic time $\tau$ (Fig. 5A). $\alpha_+$ is advected with velocity $+u_0$, while $\alpha_-$ is advected with velocity $-u_0$, both diffusing with coefficient $D$. This can be modeled as

$$\begin{cases} \dfrac{\partial[\alpha]_+}{\partial t} + u_0 \dfrac{\partial[\alpha]_+}{\partial x} = k[\alpha]_+ - \dfrac{1}{\tau}([\alpha]_+ - [\alpha]_-) + D\Delta[\alpha]_+ & (7) \\ \dfrac{\partial[\alpha]_-}{\partial t} - u_0 \dfrac{\partial[\alpha]_-}{\partial x} = k[\alpha]_- + \dfrac{1}{\tau}([\alpha]_+ - [\alpha]_-) + D\Delta[\alpha]_- & (8) \end{cases}$$

Defining $[\alpha] = [\alpha]_+ + [\alpha]_-$ and $\widetilde{[\alpha]} = [\alpha]_+ - [\alpha]_-$ , (7) and (8) become

$$\begin{cases} \dfrac{\partial[\alpha]}{\partial t} + u_0 \dfrac{\partial\widetilde{[\alpha]}}{\partial x} = k[\alpha] + D\Delta[\alpha] & (9) \\ \dfrac{\partial\widetilde{[\alpha]}}{\partial t} + u_0 \dfrac{\partial[\alpha]}{\partial x} = k\widetilde{[\alpha]} - \dfrac{2}{\tau}\widetilde{[\alpha]} + D\Delta\widetilde{[\alpha]} & (10) \end{cases}$$

Two asymptotic solutions exist for Eq. (9-10) (See Theoretical supplement).

For $k\tau \ll 1$, the autocatalytic reaction is slow compared to the switching time between the two advection streams. In this case $\widetilde{[\alpha]}$ relaxes fast and we consider that it reaches a stationary state that is injected in Eq. (9) yielding, to 2nd order in gradients,

$$\frac{\partial[\alpha]}{\partial t} = k[\alpha] + \left(D - \frac{{u_0}^2}{\gamma} - \frac{{u_0}^2 k}{\gamma^2}\right)\frac{\partial^2[\alpha]}{\partial x^2} \tag{11}$$

where $\gamma = k - \frac{2}{\tau} < 0$. Eq. (11) is an FKPP equation with Deff $D_{eff} = D - \frac{{u_0}^2}{\gamma} - \frac{{u_0}^2 k}{\gamma^2}$ and thus the speed of the front is

$$v = 2\sqrt{kD_{eff}}. \tag{12}$$

When $k\tau \gg 1$, we neglect the transfer between the + and – streamlines. In this regime, we expect the front to display an S-shape in the x-y plane. In the vicinity of the front (Fig. 5B), $[\alpha]_- \ll [\alpha]_+$, thus $[\alpha] \sim [\alpha]_+$ and only Eq. (1) remains. In the co-moving frame of the front moving at speed $v$ and defining $\beta = x - vt$

$$-(v - u_0)\frac{\partial[\alpha]}{\partial\beta} = \left(k - \frac{1}{\tau}\right)[\alpha] + D\frac{\partial^2[\alpha]}{\partial\beta^2} \tag{13}$$

Which is again an FKPP equation with $k_{eff} = k - \frac{1}{\tau}$, which solution is a moving front of speed

$$v = u_0 + 2\sqrt{k_{eff}D}. \tag{14}$$

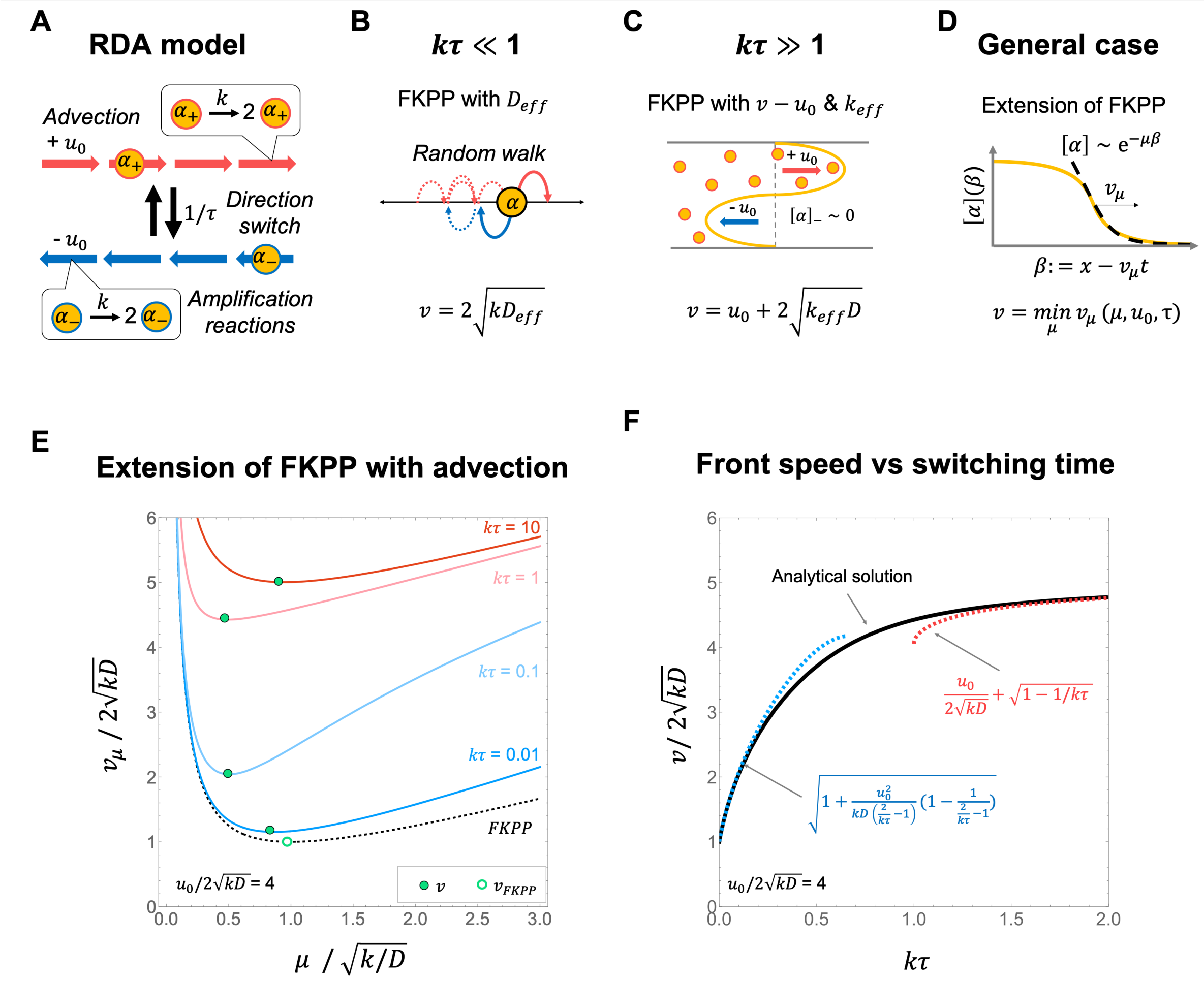

***Figure 5 - Theoretical model proposed to describe the propagation of a reaction-diffusion front advected by an active matter flow. A:** Scheme of the conveyor belt RDA model: species $\alpha$ reacts autocatalytically at rate constant $k$, and diffuses with diffusivity $D$. In addition, active matter flows advect $\alpha$ either to the right (at velocity $+u_0$) or to the left (at velocity $-u_0$). $\alpha$ switches direction - from $+u_0$ to $-u_0$ streamlines - with a characteristic time $\tau$. Note that when an $\alpha$ is produced, it remains on the same streamline and is thus advected in the same way as its parent, positively or negatively. Moreover, when considering a cross section of the channel perpendicular to the streamlines, the net advective flow is zero. **B:** For $k\tau$ small, the model is equivalent to a random walk and leads to FKPP fronts with an effective diffusion coefficient $D_{eff}$, Eq. (12). **C:** For $k\tau$ large, the contribution of switches between opposite streamlines is modelled as an effective reaction constant $k_{eff}$ The solution of the model leads to a propagation equal to the sum of advection and effective reaction-diffusion, Eq. (14). **D:** The general case is a front, approximated as an exponential decay shape travelling at a speed $v_\mu$ that is a function of $u_0$, $\tau$ and $\mu$. **E:** Extension of FKPP with advection. The front speed $v_\mu$ in Eq. (15), is plotted as a function of $\mu$ for different $k\tau$. By analogy with FKPP[38], for a given $k\tau$, the selected speed of the front $v$ is the minimum. **F:** Front speed prediction: asymptotic solutions, respectively Eq. (12) and Eq. (14), predict $v$ for both small (blue) and large (red) values of $k\tau$. The asymptotic solutions agree with the general analytical solution, Eq (16) (See Theoretical supplement).*

In the general case, to solve Eq. (9 -10) we assume that the front has an exponential decay shape and we look for solutions in the form of $A\,e^{-\mu\beta}$ fronts travelling at speed $v_\mu$ (Fig. 5D). In the Supporting Materials we demonstrate that

$$v_\mu(\mu) = \frac{k}{\mu} + D\mu - \frac{1}{\tau\mu}\left(1 - \sqrt{1 + (\tau u_0 \mu)^2}\right) \quad (15)$$

represented Fig. 5E for five cases: reaction diffusion only (black dotted line), and reaction diffusion advection for a normalized speed $u_0/\sqrt{2kD}$ = 4 and values of the normalized switching time $k\tau$ from 0.01 to 10 (colored lines). This corresponds to $u_0$ = 50 µm/min, the typical speed $u_{max}$ in Fig 2C.
By analogy with FKPP[38] we suppose that the speed of the front $v$ is unique and equal to

$$v = \min_{\mu} \left[ \frac{k}{\mu} + D\mu - \frac{1}{\tau\mu}\left(1 - \sqrt{1 + (\tau u_0 \mu)^2}\right) \right]. \tag{16}$$

Green dots mark these velocities on Fig. 5E, showing, in particular, that an RD-AM front in a channel with a small $k\tau$ (corresponding to narrow channels) and an RD front propagate at very similar velocities. The analytical solution (16) appears as a black line in Fig. 5F, in excellent agreement with the asymptotic solutions (12) and (14) described above. Taking $\tau$ = 1 min leads to $k\tau$ = 0.02, and for $u_0$ = 4 $v_{FKPP}$, it only gives a $v$ superior to $v_{FKPP}$ by 13%. A normalized $k\tau$ of 0.6, (corresponding to $\tau$ = 30 min) is enough to reach $v \sim u_0$. Further increasing $k\tau$ allows us to approach the maximal $v = u_0 + v_{FKPP}$.

### The conveyor-belt model is in good agreement with the front velocity in phase 1

We repeated the experiments in Fig. 4 and measured $v_1$ and $\delta_1$ for a total of 21 channels (Fig S19, S20), including two intermediate widths, 0.4 and 0.6 mm (Figure 6). $v_1$ increases with $W$ in a non-linear fashion (Figure 6A). At large $W$, $v_1$ is up to 8-fold higher than $v_0$, the speed without advection. At $W \leq$ 0.4 mm $v_1$ is similar within experimental error to $v_0$. An important increase in $v_1$ occurs in the interval $W$= 0.5-0.8 mm. Our theoretical model describes well this behavior, predicting a sigmoid-like increase of $v_1$ as a function of $W$. This sigmoid-like function is the product of two contributions, a linear increase of $v_1/v_0$ with *W* essentially coming from the linear increase of $u_{max}$ (Fig. 2C) and a step-wise increase $v_1/v_0$ with *W* around *W* = 0.6 mm, which we attribute to the step-wise increase of $\tau_{cor}$ with *W* (Fig. 2D).

Note that the RDA model has various free parameters whose values must be inferred from the experiments. These have a wide uncertainty as velocities and correlation times plugged in the model actually correspond to distributions and choosing the median value as the most relevant value from the distribution is somehow arbitrary. To obtain the black line $v_1(W)$ in Fig. 6A we calculated the solutions to Eq. (16) taking for $u_0$ and $\tau$ the median values of respectively $u_{max}$ and $\tau_{cor}$ measured for each $W$ in Figure 2. The shaded area around the line estimates the error in the theoretical solution coming from the measures of $k$, $D$, $u_{max}$ and $\tau_{cor}$. Alternatively, the green dashed line is obtained for values of $u_{max}$ and $\tau_{cor}$ that still fall inside the measured intervals but closer to the extrema (Fig S11). The green dashed line in Fig. 6A allows for a closer fit with the experimental data, suggesting that the physical reality is better captured by extrema values in the distributions of $u_{max}$ and $\tau_{cor}$. This is in agreement with a vision of an autocatalytic front which dynamics is controlled by the head of the front, i.e. the particles of $\alpha$ that travel the furthest ahead of the front that are subsequently exponentially amplified by the chemical reaction.

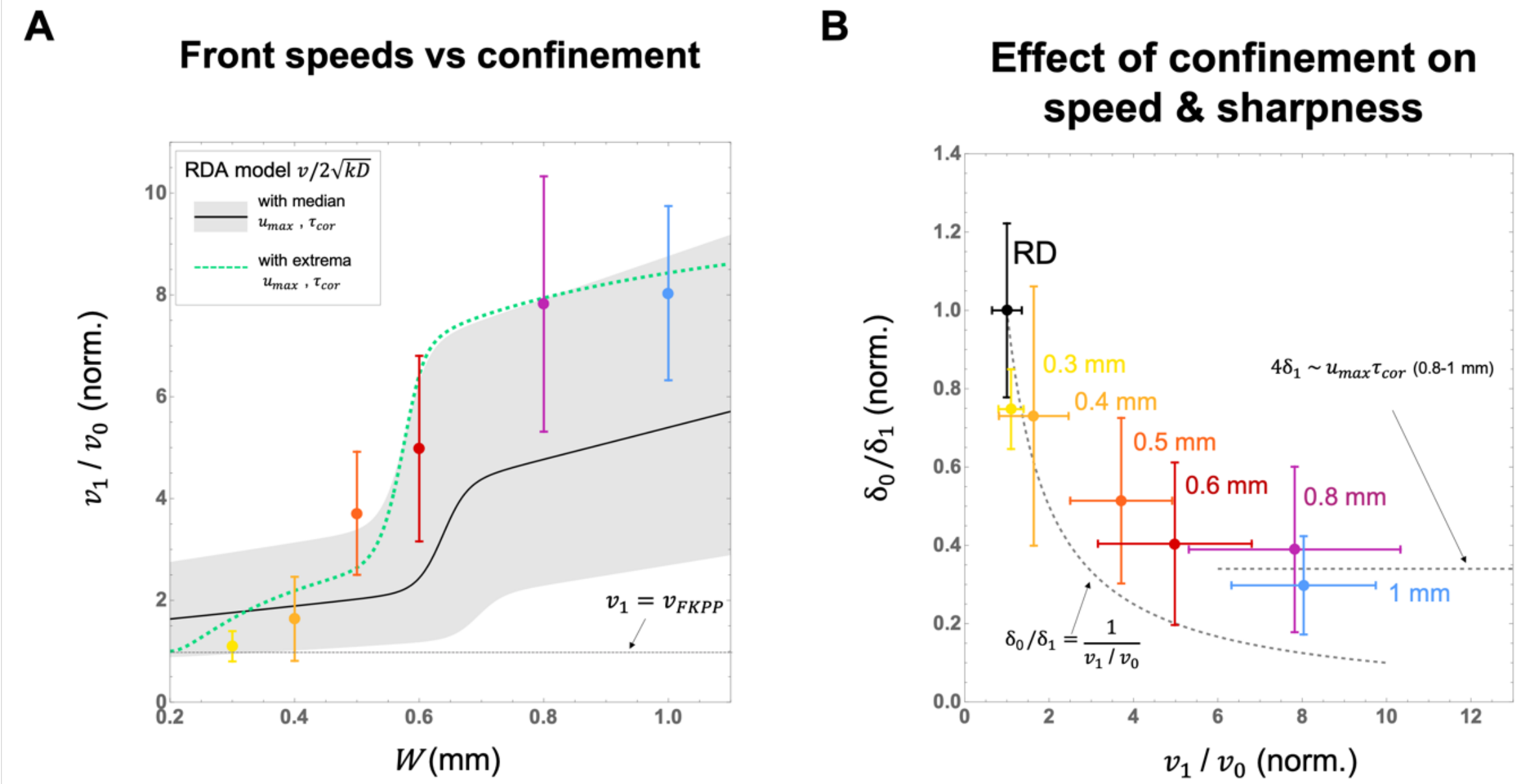

***Figure 6 – Geometry-dependent front propagation: experiments vs RDA model. A:** Front speeds during the first phase ($v_1$, hydrodynamic cell) vs. channel width for n = 21 experiments. The dots and the bars correspond respectively to the mean and one standard deviation. The black line is calculated from Eq. (16) for the median values of $u_{max}$ and $\tau_{cor}$ measured for each $W$ in Fig. 2, with the shaded area corresponding to the 25% and 75% quantiles. The green dashed line corresponds to taking values of $u_{max}$ and $\tau_{cor}$ close to the extrema as shown in Fig. S11. Experimental values are normalized by $v_0$, the RD speed, while models are normalized by $v_{FKPP}$, shown as a gray dashed line. **B:** [DNA] front sharpness vs speed for varying channel widths. The dots and the bars correspond respectively to the mean and one standard deviation. The dashed asymptotes correspond respectively to the sharpness of a FKPP front with an effective diffusivity (left), and the sharpness of a front that expands to the size of the largest vortexes (right).*

For $W$ = 0.8-1 mm, the prediction of the median model (black line) is below the data which is in line with an undervaluation of $\tau$. $\tau$ corresponds to switches in direction inside a hypothetically infinite hydrodynamic cell while our measured $\tau_{cor}$ is made smaller by the finite size of our hydrodynamic cells that force direction switches at their border. For $W$ = 0.3 mm, the model is above the data, this is in line with the real $\tau$ being smaller than our measured $\tau_{cor}$ due to the low sampling frequency of experimental trajectories (every 20 s) and their smoothing to remove measurement noise. The green line that takes minimal values of $\tau_{cor}$ for small $W$ corrects this bias partially. The aim of a fully developed active fluid model of the flow would be to remove such uncertainties and connect the heuristic model discussed here to standard theories of active matter.

A simple phenomenological model based linearly on the Péclet allows, from one empirical measurement of $v_1$ at a given channel width $W_0$, to predict speeds for other channel widths:

$$v_1(W) = a \cdot Pe(W) + v_{RD}, \tag{17}$$

with $a = (v_1(W_0) - v_{RD})/Pe(W_0)$ a prefactor obtained empirically. This model captures quite well the data, except for $W$ = 0.5 mm, showing our phenomenon is mainly ruled by the balance between advection and diffusion (Fig. S21).

Figure 6B shows the trade-off in front speed and sharpness when varying the channel width. Sharpness of the front decreases with increasing speed. For narrow channels ($W$ = 0.3-0.4 mm), the sharpness is roughly inversely proportional to front speed as shown by the hyperbolic trend $\delta_0/\delta_1 = 1/(v_0/v_1)$. This agrees with advection acting as an effective diffusivity at $k\tau \ll 1$ (and thus low, Fig. 5B). This relationship is no longer true for $W \geq 0.5$ mm, as experimental points move away from the hyperbolic trend and seem to plateau around $\delta_0/\delta_1$ = 1/3. We posit that while advection tends to spread the front and thus diminish its sharpness, such spread is also limited by the extension of large vortexes that tend to prevent molecules from escaping too far. Indeed, considering a typical sigmoidal profile for the front $[\alpha] = 1/(1 + e^{-x/\delta_1})$, the full switch between the high and low concentration levels happens over a typical length of roughly $4\delta_1$ (see Fig S4D). For our plateauing sharpness of 1/3, $4\delta_1 = 12\,\delta_0$ = 3.6 mm which corresponds well to the upper typical vortex length $u_{max}\,\tau_{cor}$ for $W$ = 0.8 -1 mm. Such a limitation in spreading means that RD-AM fronts in large channels can propagate up to eight times faster than pure reaction diffusion while only diminishing their sharpness by a factor of three.

To conclude, the transport of RD-AM fronts is complex and associated with an increase of both speed and width when the activity increases as confinement is reduced. The piecewise front speed observed indicates that the process is inhomogeneous in space due to the hydrodynamic cells observed in Fig. 2. However, for the phase 1 where the front moves within a single hydrodynamic cell, we have shown that the results can be rationalized by an extension of the FKPP model where species alpha switches between two advection regimes associated to speeds $\pm u_0$ .

## Discussion and conclusion

Within cells and embryos, biochemical cues are transported through mechanisms that may combine reaction, diffusion and advection. At this scale, if the advective flow is turbulent it is due to activity and not inertia. While the effect of advection by inertial turbulence on reaction-diffusion dynamics is well understood, little is known about the effect of activity-induced flows on it. We have shown that reaction-diffusion dynamics advected by active flows are strongly geometry-dependent and more complex than RD dynamics advected by turbulent flows. This is due to the nature of active fluid flows which, depending on the confining scale and geometry can go from being homogeneous (but chaotic) spatiotemporally to spatially structured.

We characterized chaotic active flows in closed channels of different widths and observed very different and complex flow fields. As channel width increases, the speed of RD-AM fronts increases up to 8-fold in respect to simple RD fronts while only decreasing sharpness by ~60%. With a pure RD mechanism, such an increase in speed would require elevating the diffusion coefficient or the reaction rate constant ~60-fold. Increasing molecular diffusion $D$ while maintaining chemical activity is almost impossible while reducing diffusion is possible but hard[16]. Increasing $k$ in vivo is possible by mutation. The RDA mechanism that we have elucidated is a simple alternative to accelerate transport without chemically modifying the species at play. In particular, this mechanism could be particularly significant during development where the size increases dramatically and thus, potentially, active flows and their influence in RDA dynamics.

The influence of geometry on active flows has been highlighted previously[28,39–41,43,46,47,52–58]. Within circular channels, or race-tracks, microtubule-kinesin active matter is known to transition from turbulent to coherent flow when the cross section of the channel is sufficiently symmetric, with the velocity of the coherent flow increasing as the height of the channel increases[28]. Within rectangular cuboid channels, the activity level and the velocity correlation length have been shown to increase with channel height[42]. However, in contrast with those experiments happening in pseudo-2D and 3D channels, our pseudo-1D experiments showed both non-null flow velocities and long-time advection transport with coherent eddies or "rivers" sustained for tens of minutes.

The coupling between active matter flows with diffusion or reaction-diffusion has very recently been investigated. With diffusion only, when convective transport dominates mixing is fast but dilution precludes long-range transport[36]. When both reaction and diffusion were present, experiments were limited to conditions where convective transport remained limited, similar to our observations in 0.3 mm channels[37]. Our results are both coherent and extend this previous work.

The crucial role of confinement is best appreciated in the context of the hierarchical architecture observed in adult organisms. For instance, vessel diameters in our vascular system span from cm to µm while in the respiratory system they are in the cm-0.1 mm range. Morphogenesis is responsible for this architecture, which raises a number of questions. Firstly, how does this multi-scale morphogenesis interact with reaction-diffusion processes? Could active transport dramatically transform the landscape of chemical cues and by doing so sharpen the formed shapes? Secondly, as a wide range of signal transport velocities can be achieved solely through confinement, can we ultimately consider it as a biochemical reaction network simplifier? Our results demonstrate that a 3-fold change in the confinement of RD-AM processes, ubiquitous in vivo, have a strong impact on the transport of a chemical cue. They reinforce our understanding of how macro-scale patterns and structures emerge from microscopic components in non-equilibrium systems and may help us engineer living materials with unprecedented properties.

## Materials and Methods

**Kinesin and Microtubule preparation.** We used K401, a truncated kinesin-1 from Drosophila melanogaster[59], which bears a biotin function allowing it to form clusters in the presence of streptavidin[21]. Kinesins were expressed in E. coli and purified as previously described[23]. Tubulin and TRITC-labeled tubulin (Cytoskeleton) were dissolved at 10 mg/mL in 1× PEM buffer (80 mM Pipes, pH 6.9, 1 mM EGTA, 1 mM MgSO4 ) supplemented with 1 mM GTP, flash frozen, and stored at −80 ◦C. Microtubules were polymerized in 1× PEM, 1 mM GTP, 10% (wt/vol) glycerol, 5 mg/mL tubulin (including 2.5% TRITC-labeled tubulin), 0.5 mg/mL GMPCPP (Jena Bioscience), first at 37°C for 30 min, then at 25°C for 5h, flash frozen, and stored at −80 °C.

**Active mix preparation.**The active mix consisted of (see Table S1) 0.6X PEM buffer (pH = 6.9 KOH adjusted), 0.2X PEM buffer (pH = 7.2 NaOH adjusted), 8 mM MgSO4, 1.5% (wt/vol) Pluronic F-127, 5 µg/mL creatine kinase, 10 mM creatine phosphate, 2 mM NPE-Caged ATP, 0.5 mg/mL BSA, 1 mM trolox, 20 mM D-glucose, 3 mM DTT, 150 µg/mL glucose oxidase, 25 µg/mL catalase, 10 µM dATP, 25 µM dGTP, 500 µM dCTP, 500 µM dTTP, 0.4X SYBR GreenI, 100 nM DNA template Tαα, 0 or 1 µM DNA α, 0.5 % (vol/vol) BST Large Fragment polymerase (NEB), 1.5 % (vol/vol) Nt. BstNBI nickase (NEB), 75 nM Kinesin motors, 75 nM Streptavidin, 1 mg/mL Microtubules. DNA sequences (Biomers) are presented in Table S2. For the beads tracking experiments, Tαα , DNA α, polymerase and nickase were replaced with 1 µm diameter beads (Estapor Fluorescent Functionalized Microspheres F1-XC 100, Merck).
All elements were pipetted in this order, on ice, except for the addition of microtubules step that was performed at room temperature. Solution was homogenized by a 10s vortexing step before adding the enzymes. Final homogenization was carefully done using the pipette to avoid damaging the microtubules.

**Experiments in capillaries.** Glass capillaries (Vitrocom) of 50 mm length were cut into 17 mm pieces and coated with an acrylamide brush[60]. They were filled at 95% with the active mix without DNA α and the last 5% with DNA α. Capillaries were immersed in a (70 × 20 × 5 mm) glass pool filled with mineral oil to prevent evaporation and keep a homogeneous temperature. The glass pool was placed on a transparent Tokai Hit Thermo Plate set at 24 °C. To ensure thermal equilibrium, this was done 30 min before immersing the capillaries. For ATP uncaging, channels were exposed to UV light with a UV LED (15335337AA350, 365nm, Wurth Elektronik), using a current of 300 mA at ≈ 5 mm distance for 3 min.

**Imaging.** Epifluorescence images were taken on a Zeiss Observer 7 automated microscope equipped with a Andor iXon Ultra camera, a 10X objective (or 4X for beads tracking). The stage is motorized and controlled with MicroManager 1.4. Images were recorded automatically every 3-3.5 min (or 20-30s for beads tracking) using an excitation at 470 nm (for the DNA intercalator SYBR Green I or the beads) and 550 nm for microtubules with a CoolLED pE2 combined to a four bands filter.

**Image Analysis.** Single images were stitched together and raw fluorescence was normalized to account for different channel thicknesses (See Fig. S22). Time-lapse images were analyzed using Fiji/ImageJ and Mathematica. Beads tracking in Fig. 2 was performed using Trackmate plugin[61]. Definitions of the front width and position are given in Fig S4.

**Acknowledgments**

We thank Anis Senoussi, Maxime Deforet, Chloe Dal Degan, Romain Leroux, Samuel Bell and Raphaël Voituriez for insightful discussions. This work has been funded by the European Research Council (ERC) under the European's Union Horizon 2020 programme (grant No 770940, A.E.-T.) and ANR LiliMat (ANR-22-CE06-0004, J.-C. G.). A.M. was supported by a TALENT fellowship from CY Cergy Paris Université.

## References

(1) Collinet, C.; Lecuit, T. Programmed and Self-Organized Flow of Information during Morphogenesis. *Nature Reviews Molecular Cell Biology* **2021**, *22* (4), 245–265.
(2) Burkart, T.; Wigbers, M. C.; Würthner, L.; Frey, E. Control of Protein-Based Pattern Formation via Guiding Cues. *Nature Reviews Physics* **2022**, *4* (8), 511–527.
(3) Liu, Y.; Xue, X.; Sun, S.; Kobayashi, N.; Kim, Y. S.; Fu, J. Morphogenesis beyond in Vivo. *Nature Reviews Physics* **2023**, 1–17.
(4) Kondo, S.; Miura, T. Reaction-Diffusion Model as a Framework for Understanding Biological Pattern Formation. *science* **2010**, *329* (5999), 1616–1620.
(5) Murray, J. D.; Murray, J. D. *Mathematical Biology: II: Spatial Models and Biomedical Applications*; Springer, 2003; Vol. 3.
(6) Hanna, A.; Saul, A.; Showalter, K. Detailed Studies of Propagating Fronts in the Iodate Oxidation of Arsenous Acid. *Journal of the American Chemical Society* **1982**, *104* (14), 3838–3844.
(7) Semenov, S. N.; Markvoort, A. J.; de Greef, T. F.; Huck, W. T. Threshold Sensing through a Synthetic Enzymatic Reaction–Diffusion Network. *Angewandte Chemie International Edition* **2014**, *53* (31), 8066–8069.
(8) Tayar, A. M.; Karzbrun, E.; Noireaux, V.; Bar-Ziv, R. H. Propagating Gene Expression Fronts in a One-Dimensional Coupled System of Artificial Cells. *Nature Physics* **2015**, *11* (12), 1037–1041.
(9) Loose, M.; Fischer-Friedrich, E.; Ries, J.; Kruse, K.; Schwille, P. Spatial Regulators for Bacterial Cell Division Self-Organize into Surface Waves in Vitro. *Science* **2008**, *320* (5877), 789–792.
(10) Dupin, A.; Simmel, F. C. Signalling and Differentiation in Emulsion-Based Multi-Compartmentalized in Vitro Gene Circuits. *Nature chemistry* **2019**, *11* (1), 32–39.
(11) Chirieleison, S. M.; Allen, P. B.; Simpson, Z. B.; Ellington, A. D.; Chen, X. Pattern Transformation with DNA Circuits. *Nature chemistry* **2013**, *5* (12), 1000–1005.
(12) Zenk, J.; Scalise, D.; Wang, K.; Dorsey, P.; Fern, J.; Cruz, A.; Schulman, R. Stable DNA-Based Reaction–Diffusion Patterns. *RSC advances* **2017**, *7* (29), 18032–18040.
(13) Abe, K.; Murata, S.; Kawamata, I. Cascaded Pattern Formation in Hydrogel Medium Using the Polymerisation Approach. *Soft Matter* **2021**, *17* (25), 6160–6167.
(14) Montagne, K.; Plasson, R.; Sakai, Y.; Fujii, T.; Rondelez, Y. Programming an in Vitro DNA Oscillator Using a Molecular Networking Strategy. *Molecular systems biology* **2011**, *7* (1), 466.
(15) Baccouche, A.; Montagne, K.; Padirac, A.; Fujii, T.; Rondelez, Y. Dynamic DNA-Toolbox Reaction Circuits: A Walkthrough. *Methods* **2014**, *67* (2), 234–249.
(16) Zadorin, A. S.; Rondelez, Y.; Galas, J.-C.; Estevez-Torres, A. Synthesis of Programmable Reaction-Diffusion Fronts Using DNA Catalyzers. *Physical review letters* **2015**, *114* (6), 068301.
(17) Padirac, A.; Fujii, T.; Estévez-Torres, A.; Rondelez, Y. Spatial Waves in Synthetic Biochemical Networks. *Journal of the American Chemical Society* **2013**, *135* (39), 14586–14592.
(18) Gines, G.; Zadorin, A.; Galas, J.-C.; Fujii, T.; Estevez-Torres, A.; Rondelez, Y. Microscopic Agents Programmed by DNA Circuits. *Nature nanotechnology* **2017**, *12* (4),

351–359.
(19) Zadorin, A. S.; Rondelez, Y.; Gines, G.; Dilhas, V.; Urtel, G.; Zambrano, A.; Galas, J.-C.; Estevez-Torres, A. Synthesis and Materialization of a Reaction–Diffusion French Flag Pattern. *Nature chemistry* **2017**, *9* (10), 990–996.
(20) Koenderink, G. H.; Dogic, Z.; Nakamura, F.; Bendix, P. M.; MacKintosh, F. C.; Hartwig, J. H.; Stossel, T. P.; Weitz, D. A. An Active Biopolymer Network Controlled by Molecular Motors. *Proceedings of the National Academy of Sciences* **2009**, *106* (36), 15192–15197.
(21) Nédélec, F.; Surrey, T.; Maggs, A. C.; Leibler, S. Self-Organization of Microtubules and Motors. *Nature* **1997**, *389* (6648), 305–308.
(22) Ideses, Y.; Erukhimovitch, V.; Brand, R.; Jourdain, D.; Hernandez, J. S.; Gabinet, U.; Safran, S. A.; Kruse, K.; Bernheim-Groswasser, A. Spontaneous Buckling of Contractile Poroelastic Actomyosin Sheets. *Nature communications* **2018**, *9* (1), 2461.
(23) Senoussi, A.; Kashida, S.; Voituriez, R.; Galas, J.-C.; Maitra, A.; Estevez-Torres, A. Tunable Corrugated Patterns in an Active Nematic Sheet. *Proceedings of the National Academy of Sciences* **2019**, *116* (45), 22464–22470.
(24) Sarfati, G.; Maitra, A.; Voituriez, R.; Galas, J.-C.; Estevez-Torres, A. Crosslinking and Depletion Determine Spatial Instabilities in Cytoskeletal Active Matter. *Soft Matter* **2022**, *18* (19), 3793–3800.
(25) Najma, B.; Varghese, M.; Tsidilkovski, L.; Lemma, L.; Baskaran, A.; Duclos, G. Competing Instabilities Reveal How to Rationally Design and Control Active Crosslinked Gels. *Nature communications* **2022**, *13* (1), 6465.
(26) Keya, J. J.; Suzuki, R.; Kabir, A. M. R.; Inoue, D.; Asanuma, H.; Sada, K.; Hess, H.; Kuzuya, A.; Kakugo, A. DNA-Assisted Swarm Control in a Biomolecular Motor System. *Nature communications* **2018**, *9* (1), 453.
(27) Sanchez, T.; Chen, D. T.; DeCamp, S. J.; Heymann, M.; Dogic, Z. Spontaneous Motion in Hierarchically Assembled Active Matter. *Nature* **2012**, *491* (7424), 431–434.
(28) Wu, K.-T.; Hishamunda, J. B.; Chen, D. T.; DeCamp, S. J.; Chang, Y.-W.; Fernández-Nieves, A.; Fraden, S.; Dogic, Z. Transition from Turbulent to Coherent Flows in Confined Three-Dimensional Active Fluids. *Science* **2017**, *355* (6331), eaal1979.
(29) Wu, W.; Wang, L.; Calzavarini, E.; Schmitt, F. G. Reactive Scalars in Incompressible Turbulence with Strongly out of Equilibrium Chemistry. *Journal of Fluid Mechanics* **2022**, *938*, A19.
(30) Colombi, R.; Schlüter, M.; Kameke, A. von. Three Dimensional Flows beneath a Thin Layer of 2D Turbulence Induced by Faraday Waves. *Experiments in Fluids* **2021**, *62*, 1–13.
(31) von Kameke, A.; Huhn, F.; Fernández-García, G.; Muñuzuri, A. P.; Pérez-Muñuzuri, V. Propagation of a Chemical Wave Front in a Quasi-Two-Dimensional Superdiffusive Flow. *Physical Review E* **2010**, *81* (6), 066211.
(32) von Kameke, A.; Huhn, F.; Muñuzuri, A. P.; Pérez-Muñuzuri, V. Measurement of Large Spiral and Target Waves in Chemical Reaction-Diffusion-Advection Systems: Turbulent Diffusion Enhances Pattern Formation. *Physical review letters* **2013**, *110* (8), 088302.
(33) Alert, R.; Casademunt, J.; Joanny, J.-F. Active Turbulence. *Annual Review of Condensed Matter Physics* **2022**, *13*, 143–170.
(34) Martínez-Prat, B.; Ignés-Mullol, J.; Casademunt, J.; Sagués, F. Selection Mechanism at the Onset of Active Turbulence. *Nature physics* **2019**, *15* (4), 362–366.
(35) Chatterjee, R.; Joshi, A. A.; Perlekar, P. Front Structure and Dynamics in Dense Colonies of Motile Bacteria: Role of Active Turbulence. *Physical Review E* **2016**, *94* (2), 022406.
(36) Bate, T. E.; Varney, M. E.; Taylor, E. H.; Dickie, J. H.; Chueh, C.-C.; Norton, M. M.;

Wu, K.-T. Self-Mixing in Microtubule-Kinesin Active Fluid from Nonuniform to Uniform Distribution of Activity. *Nature communications* **2022**, *13* (1), 6573.
(37) Senoussi, A.; Galas, J.-C.; Estevez-Torres, A. Programmed Mechano-Chemical Coupling in Reaction-Diffusion Active Matter. *Science Advances* **2021**, *7* (51), eabi9865.
(38) Van Saarloos, W. Front Propagation into Unstable States. *Physics reports* **2003**, *386* (2–6), 29–222.
(39) Voituriez, R.; Joanny, J.-F.; Prost, J. Spontaneous Flow Transition in Active Polar Gels. *Europhysics Letters* **2005**, *70* (3), 404.
(40) Chandrakar, P.; Varghese, M.; Aghvami, S. A.; Baskaran, A.; Dogic, Z.; Duclos, G. Confinement Controls the Bend Instability of Three-Dimensional Active Liquid Crystals. *Physical review letters* **2020**, *125* (25), 257801.
(41) Hardoüin, J.; Hughes, R.; Doostmohammadi, A.; Laurent, J.; Lopez-Leon, T.; Yeomans, J. M.; Ignés-Mullol, J.; Sagués, F. Reconfigurable Flows and Defect Landscape of Confined Active Nematics. *Communications Physics* **2019**, *2* (1), 121.
(42) Fan, Y.; Wu, K.-T.; Aghvami, S. A.; Fraden, S.; Breuer, K. S. Effects of Confinement on the Dynamics and Correlation Scales in Kinesin-Microtubule Active Fluids. *Physical Review E* **2021**, *104* (3), 034601.
(43) Opathalage, A.; Norton, M. M.; Juniper, M. P.; Langeslay, B.; Aghvami, S. A.; Fraden, S.; Dogic, Z. Self-Organized Dynamics and the Transition to Turbulence of Confined Active Nematics. *Proceedings of the National Academy of Sciences* **2019**, *116* (11), 4788–4797.
(44) Lemma, L. M.; Varghese, M.; Ross, T. D.; Thomson, M.; Baskaran, A.; Dogic, Z. Spatio-Temporal Patterning of Extensile Active Stresses in Microtubule-Based Active Fluids. *PNAS nexus* **2023**, *2* (5), pgad130.
(45) Varghese, M.; Baskaran, A.; Hagan, M. F.; Baskaran, A. Confinement-Induced Self-Pumping in 3D Active Fluids. *Physical review letters* **2020**, *125* (26), 268003.
(46) Genkin, M. M.; Sokolov, A.; Lavrentovich, O. D.; Aranson, I. S. Topological Defects in a Living Nematic Ensnare Swimming Bacteria. *Physical Review X* **2017**, *7* (1), 011029.
(47) Maitra, A.; Srivastava, P.; Marchetti, M. C.; Lintuvuori, J. S.; Ramaswamy, S.; Lenz, M. A Nonequilibrium Force Can Stabilize 2D Active Nematics. *Proceedings of the National Academy of Sciences* **2018**, *115* (27), 6934–6939.
(48) Simha, R. A.; Ramaswamy, S. Hydrodynamic Fluctuations and Instabilities in Ordered Suspensions of Self-Propelled Particles. *Physical review letters* **2002**, *89* (5), 058101.
(49) Chepizhko, O.; Saintillan, D.; Peruani, F. Revisiting the Emergence of Order in Active Matter. *Soft Matter* **2021**, *17* (11), 3113–3120.
(50) Maitra, A. Two-Dimensional Long-Range Uniaxial Order in Three-Dimensional Active Fluids. *Nature Physics* **2023**, 1–8.
(51) Epstein, I. R.; Pojman, J. A. *An Introduction to Nonlinear Chemical Dynamics: Oscillations, Waves, Patterns, and Chaos*; Oxford university press, 1998.
(52) Theillard, M.; Saintillan, D. Computational Mean-Field Modeling of Confined Active Fluids. *Journal of Computational Physics* **2019**, *397*, 108841.
(53) Giomi, L. Geometry and Topology of Turbulence in Active Nematics. *Physical Review X* **2015**, *5* (3), 031003.
(54) Fürthauer, S.; Neef, M.; Grill, S. W.; Kruse, K.; Jülicher, F. The Taylor–Couette Motor: Spontaneous Flows of Active Polar Fluids between Two Coaxial Cylinders. *New Journal of Physics* **2012**, *14* (2), 023001.
(55) Wioland, H.; Woodhouse, F. G.; Dunkel, J.; Kessler, J. O.; Goldstein, R. E. Confinement Stabilizes a Bacterial Suspension into a Spiral Vortex. *Physical review letters* **2013**, *110* (26), 268102.
(56) Woodhouse, F. G.; Goldstein, R. E. Spontaneous Circulation of Confined Active

Suspensions. *Physical review letters* **2012**, *109* (16), 168105.
(57) Suzuki, K.; Miyazaki, M.; Takagi, J.; Itabashi, T.; Ishiwata, S. Spatial Confinement of Active Microtubule Networks Induces Large-Scale Rotational Cytoplasmic Flow. *Proceedings of the National Academy of Sciences* **2017**, *114* (11), 2922–2927.
(58) Singh, A.; Vagne, Q.; Jülicher, F.; Sbalzarini, I. F. Spontaneous Flow Instabilities of Active Polar Fluids in Three Dimensions. *Physical Review Research* **2023**, *5* (2), L022061.
(59) Subramanian, R.; Gelles, J. Two Distinct Modes of Processive Kinesin Movement in Mixtures of ATP and AMP-PNP. *The Journal of general physiology* **2007**, *130* (5), 445–455.
(60) Sanchez, T.; Dogic, Z. Engineering Oscillating Microtubule Bundles. In *Methods in enzymology*; Elsevier, 2013; Vol. 524, pp 205–224.
(61) Ershov, D.; Phan, M.-S.; Pylvänäinen, J. W.; Rigaud, S. U.; Le Blanc, L.; Charles-Orszag, A.; Conway, J. R.; Laine, R. F.; Roy, N. H.; Bonazzi, D. TrackMate 7: Integrating State-of-the-Art Segmentation Algorithms into Tracking Pipelines. *Nature Methods* **2022**, *19* (7), 829–832.

Supporting Materials for:

# Confinement determines transport of a reaction-diffusion active matter front

Nicolas Lobato-Dauzier[1*], Ananyo Maitra[1,2], André Estevez-Torres[1*], Jean-Christophe Galas[1*]

[1] Sorbonne Université, CNRS, Institut de Biologie Paris-Seine (IBPS), Laboratoire Jean Perrin (LJP), F-75005, Paris, France

[2] CY Cergy Paris Université, CNRS, Laboratoire de Physique Théorique et Modélisation, 95032 Cergy-Pontoise, France

[3] Université de Lille, CNRS, Laboratoire de spectroscopie pour les interactions, la réactivité et l'environnement (UMR 8516 - LASIRE), F-59655, Villeneuve d'Ascq, France

***Corresponding authors**: Nicolas Lobato-Dauzier, André Estevez-Torres, Jean-Christophe Galas
**Email:** {nicolas.lobato-dauzier ; andre.estevez-torres ; jean-christophe.galas} @sorbonne-universite.fr

## 1. Supplementary Figures

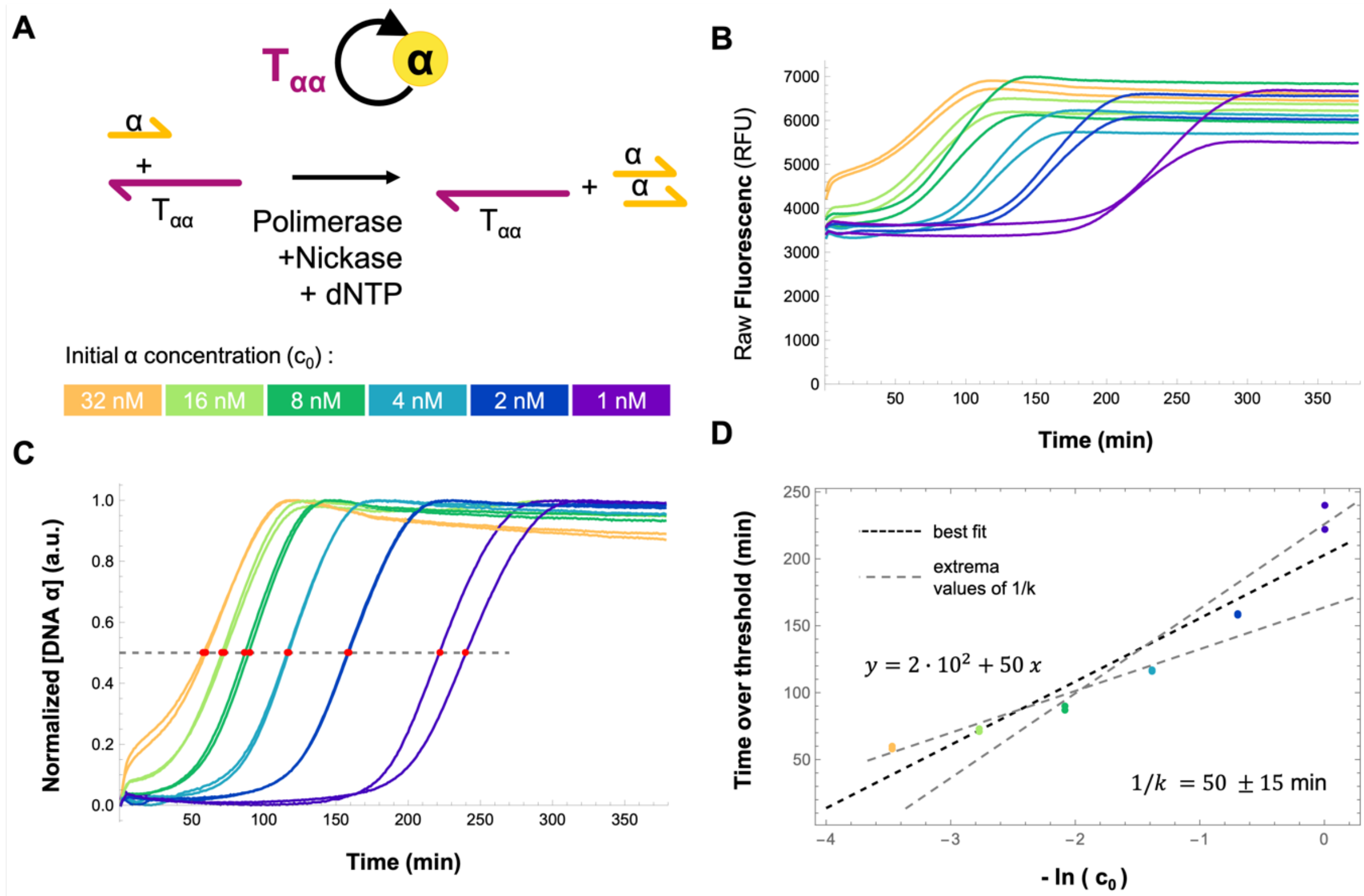

***Figure S1 - Extraction of reaction rate. A:*** *PEN DNA Toolbox reaction.* ***B****: Raw Fluorescence amplification curves for varying initial α concentration from 1 nM to 32 nM. Acquisition was performed with a CFX96 (Biorad), on 10 µL of active mix (without microtubules or kinesin).* ***C****: Normalized [DNA α] curves. The dashed line corresponds to the thresholding at 50%.* ***D****: Amplification times for each initial α concentration. The slope gives 1/k, the inverse of the reaction rate (see supplementary methods). The gray dashed lines correspond to the extrema values of the slope 1/k.*

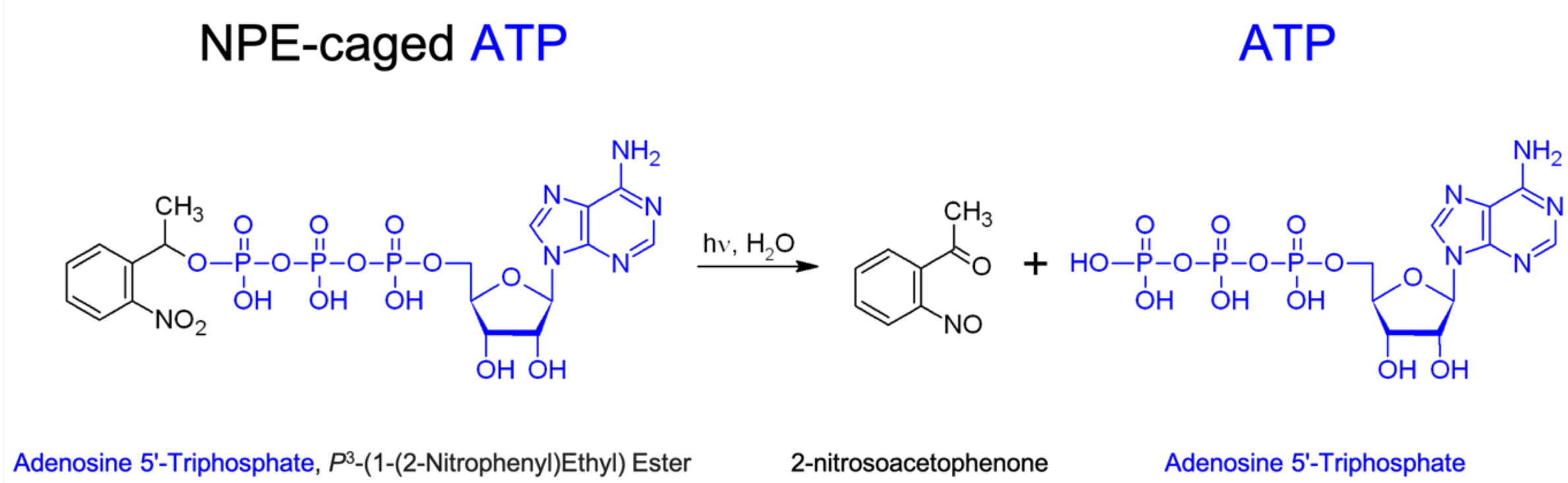

***Figure S2- Photolytic reaction of NPE caged-ATP.*** In the presence of UV Light NPE-caged ATP is dissociated and ATP is released in a usable form. (Adapted from: Wikimedia Commons, Georgwille)

# Diffusion

Polymerase (-) / Kinesin (-)

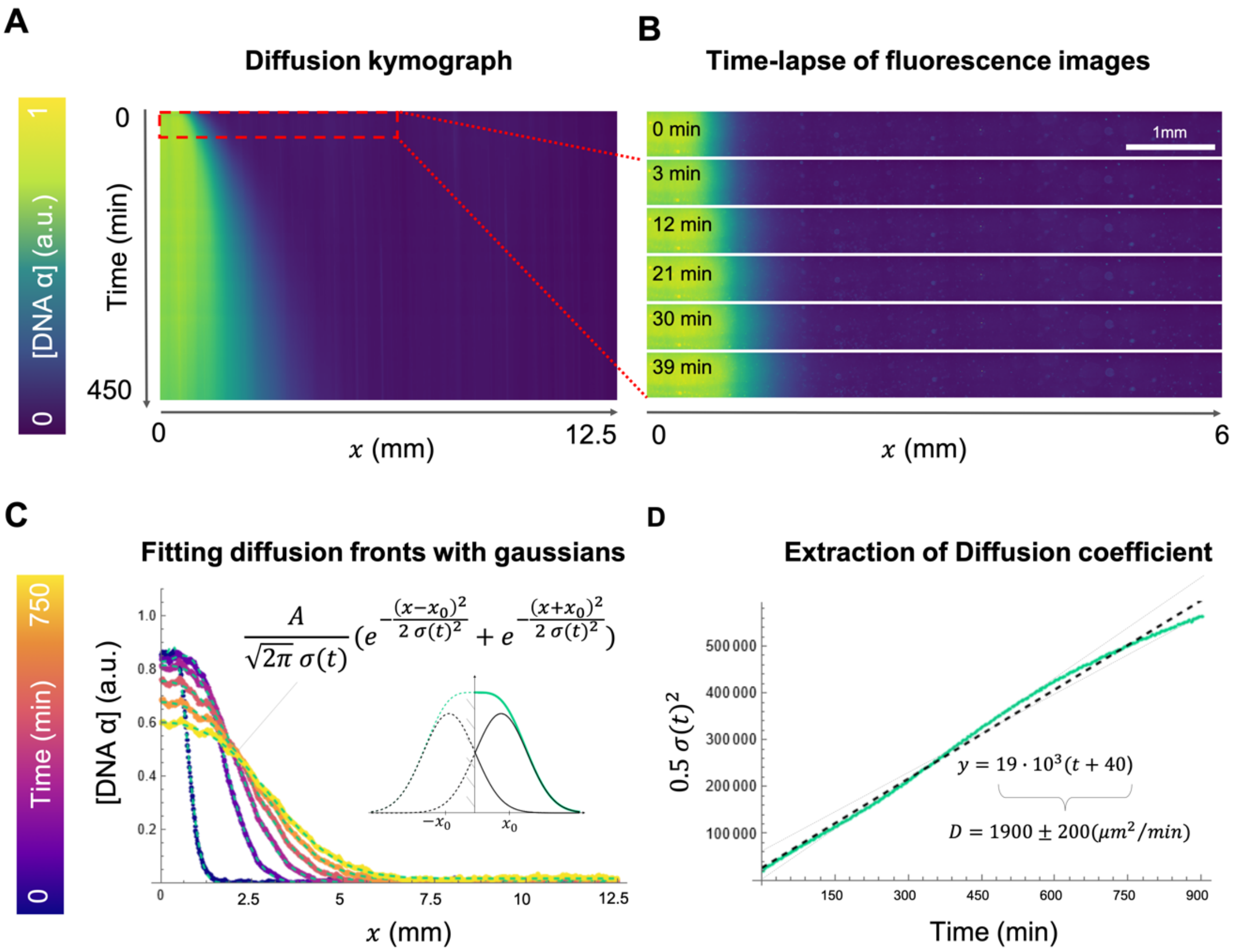

***Figure S3 – Diffusion. A:*** *Kymograph of DNA concentration diffusion in space over time.* ***B****: Fluorescence images showing slow transport of DNA by diffusion.* ***C****: Fitting concentration profiles to obtain the diffusion coefficient (See supplementary methods).* ***D****: Extraction of the diffusion coefficient by linear regression. (See supplementary methods).*

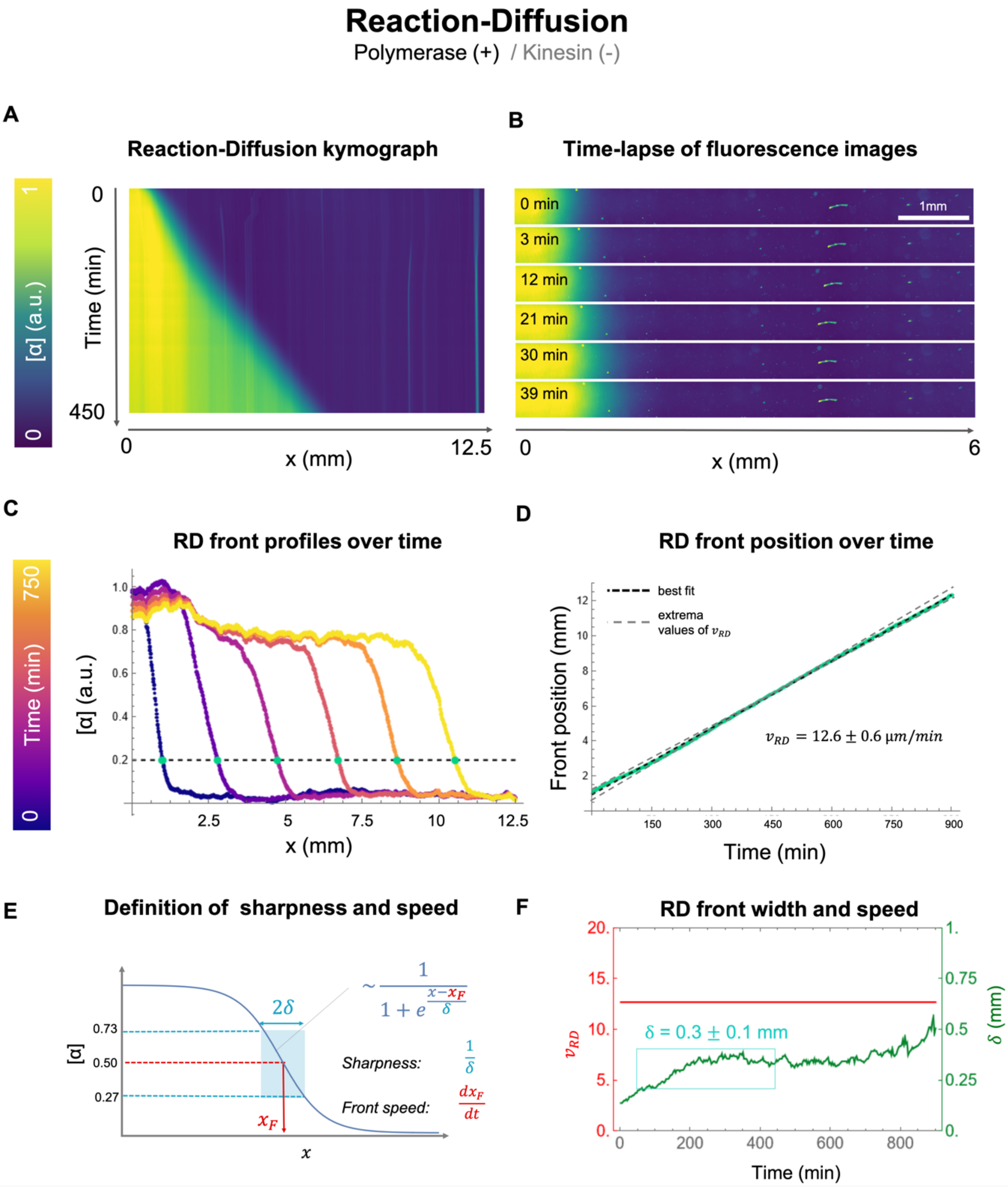

***Figure S4 – Reaction-Diffusion.*** **A**: *Kymograph of RD DNA concentration front propagation in space over time.* ***B:*** *Fluorescence images showing slow but steady transport of DNA by RD.* ***C:*** *Measuring first passage time over $[\alpha] = 0.2$ a.u. threshold to measure front position over time* ***D:*** *Extraction of the front propagation speed. Gray dashed lines correspond to the extrema values for the slope.* ***E:*** *Sketch of the DNA front profile with definitions of the sharpness ($1/\delta$), the front position ($x_f$), and the speed. Taking 0.27 and 0.73 as thresholds allows to easily extract $\delta$ (see supplementary methods).* ***F:*** *Evolution of front speed and width over time. The cyan box corresponds to the variation of $\delta_{RD}$ during the experiment.*

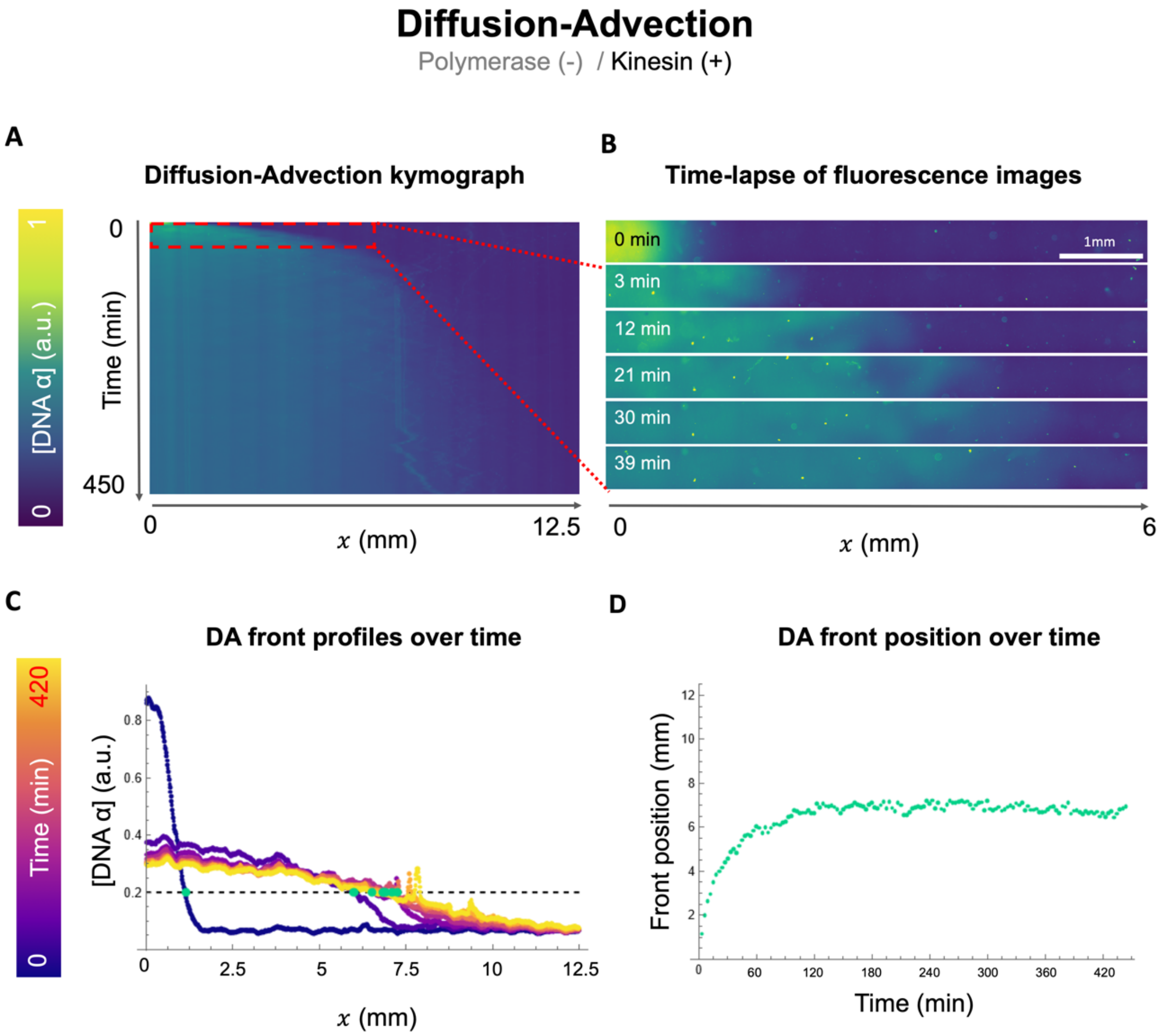

***Figure S5 – Diffusion-Advection. A:*** *Kymograph of DA DNA concentration front propagation in space over time.* ***B:*** *Fluorescence images showing fast short-range transport of DNA by DA.* ***C:*** *Measuring first passage time over 0.2 a.u threshold to measure front position over time* ***D:*** *Front goes from high velocity on the left to complete stop at the center of the channel.*

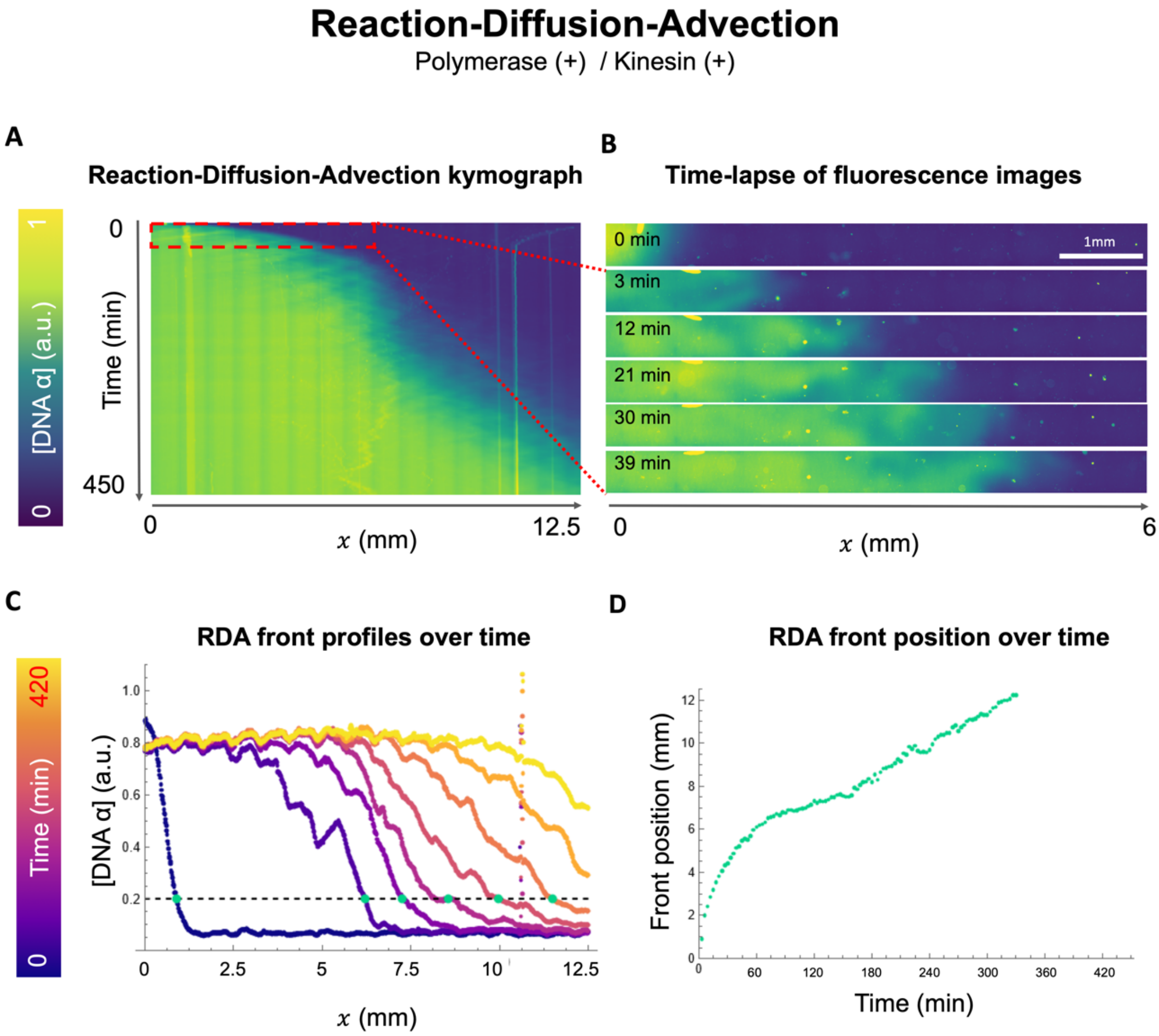

***Figure S6 – Reaction-Diffusion-Advection A:*** *Kymograph of RDA DNA concentration front propagation in space over time.* ***B:*** *Fluorescence images showing fast long-range transport of DNA by RDA.* ***C:*** *Measuring first passage time over 0.2 a.u threshold to measure front position over time* ***D:*** *Front keeps a high velocity for several hours.*

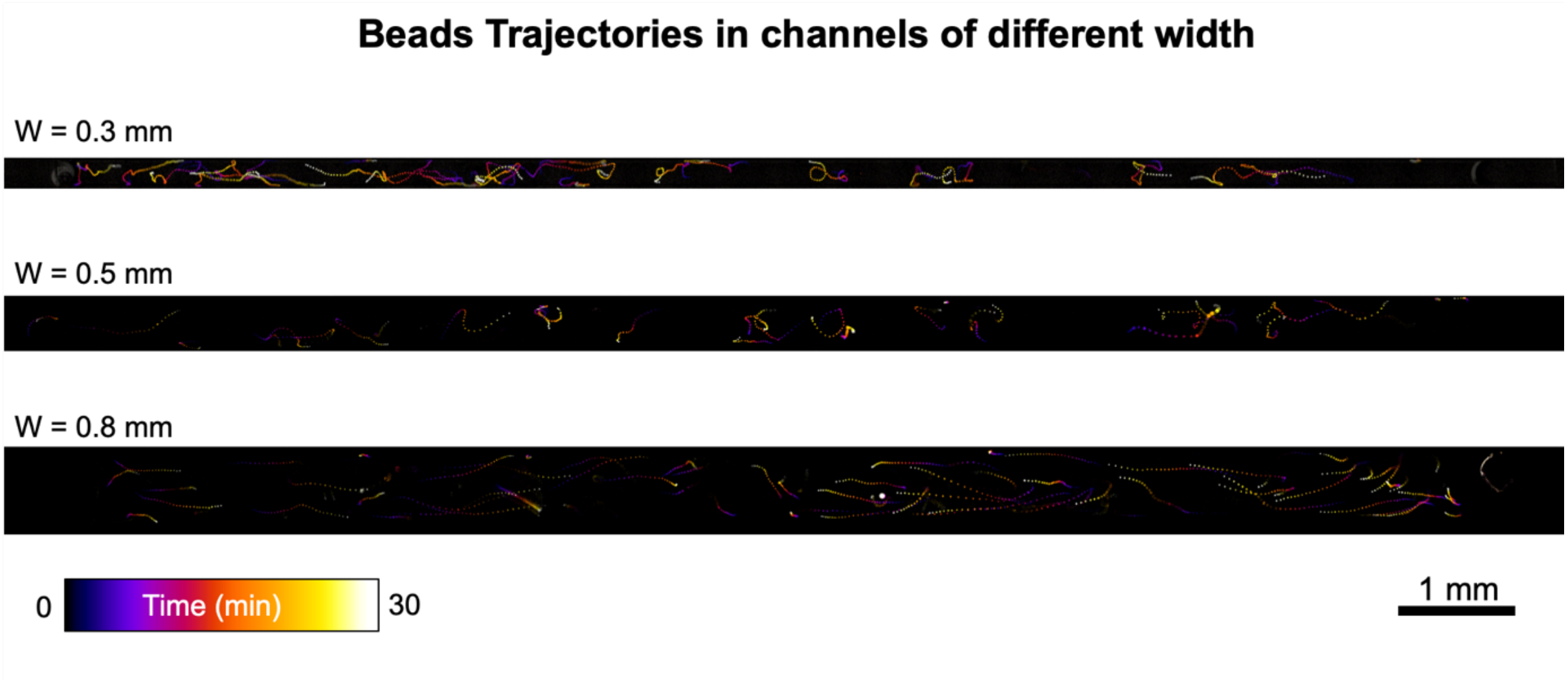

***Fig S7 – Beads trajectories over 30 min in square capillaries of width W = 0.3 / 0.5 / 0.8 mm.*** *See Movie S2 for full trajectories analyzed in Figure 2*

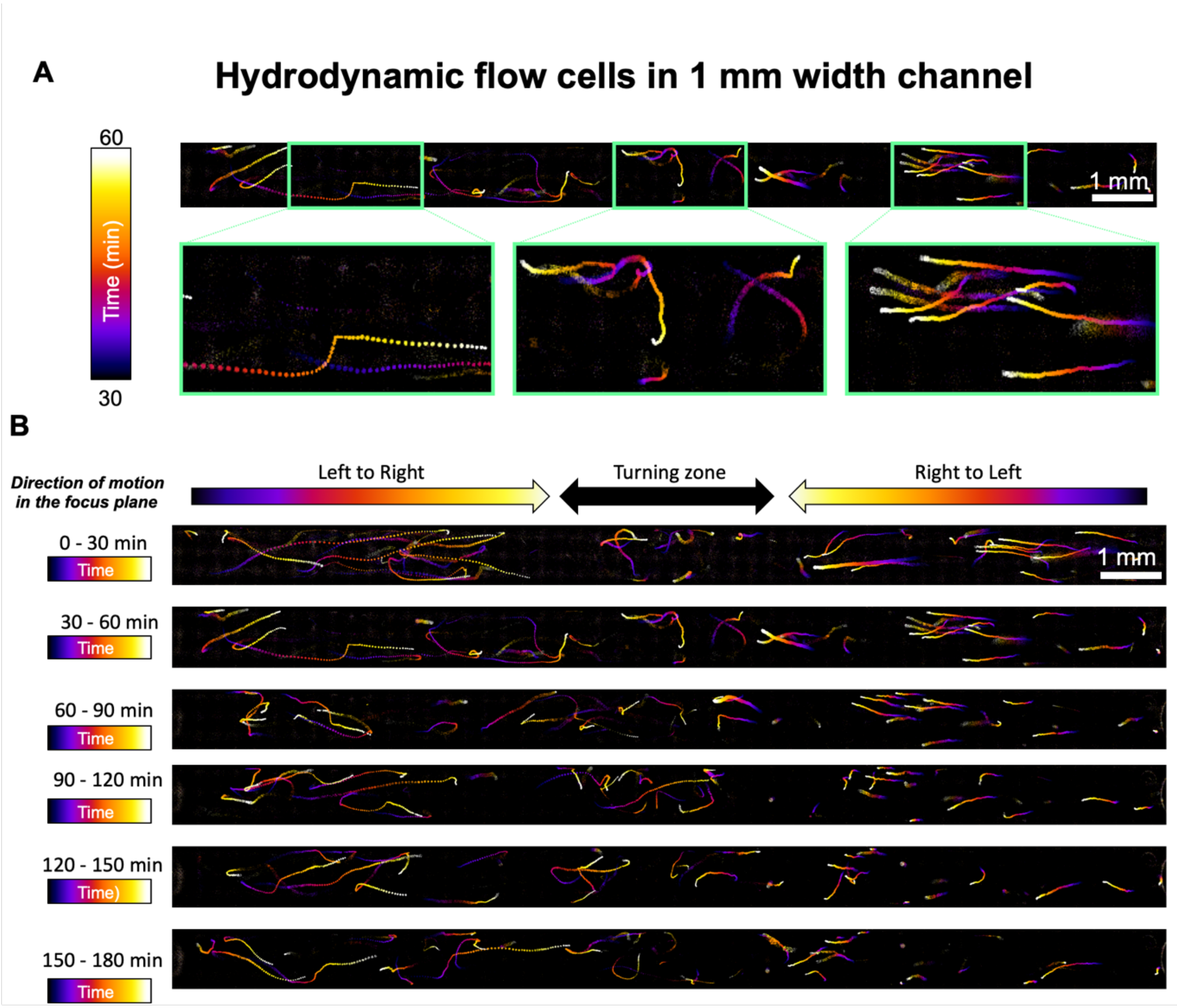

***Figure S8 – Characterization and stability of hydrodynamic flow cells in W = 1 mm channel. A:*** *Beads trajectories between t = 30 and t = 60 min, color coded by time. We observe two zones (left and right) with motion along the x axis and one zone (center) where beads change direction and move along the y axis.* ***B:*** *Stability of the three zones over 3h.*

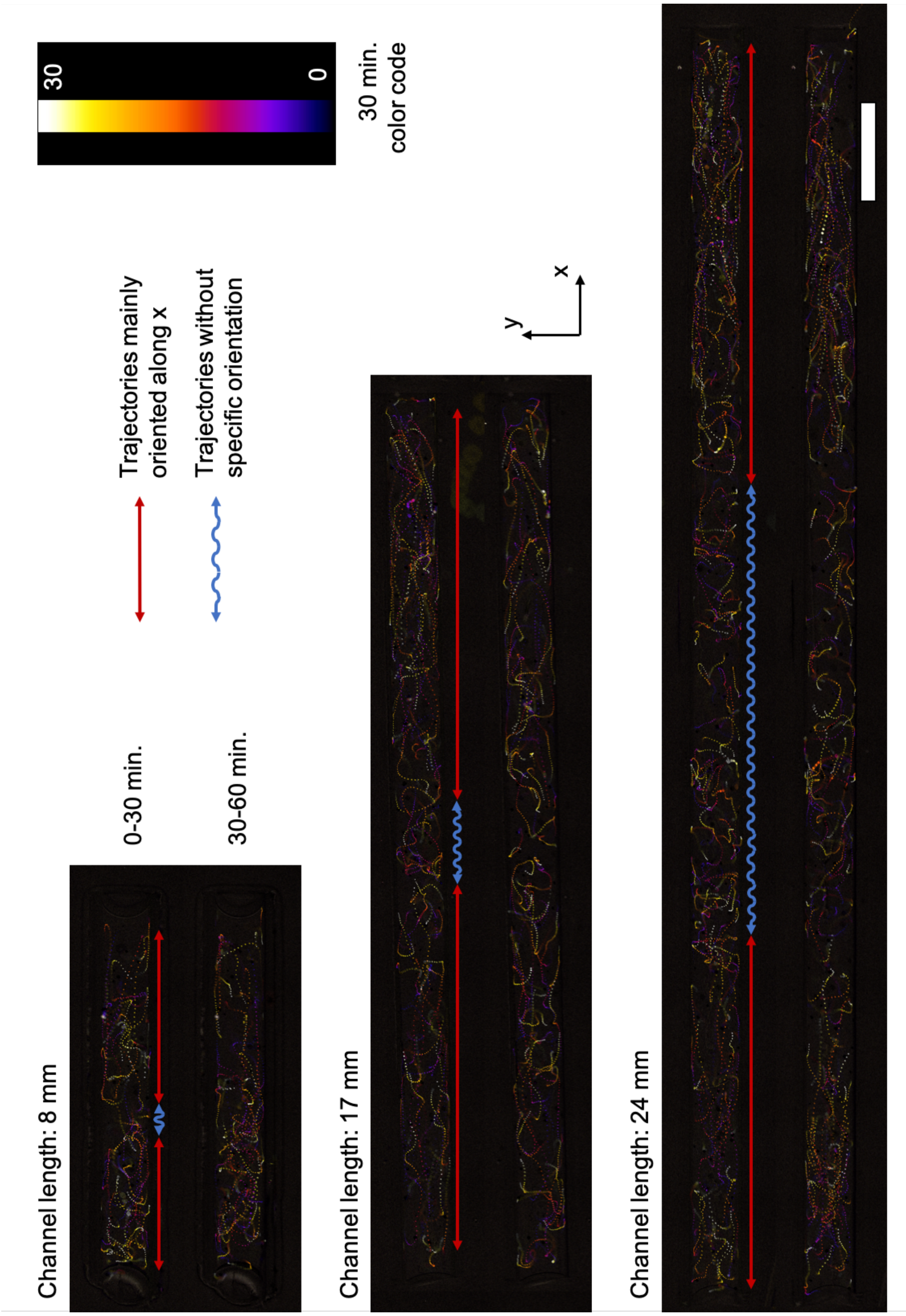

***Figure S9 – Characterization of chaotic active flows in closed channels of different lengths.*** *Trajectories followed with fiducial beads in square channels of 1 mm width (square section) and 8, 17 and 24 mm length. The 60-minute experiments are divided into 2 stages, each lasting half an hour, for time color-coding. The red arrows correspond to channel zones where trajectories, in the first 30 minutes, are predominantly oriented along the x axis. The blue arrows correspond to zones where trajectories have no specific orientations.*

## Velocity correlation depending on confinement

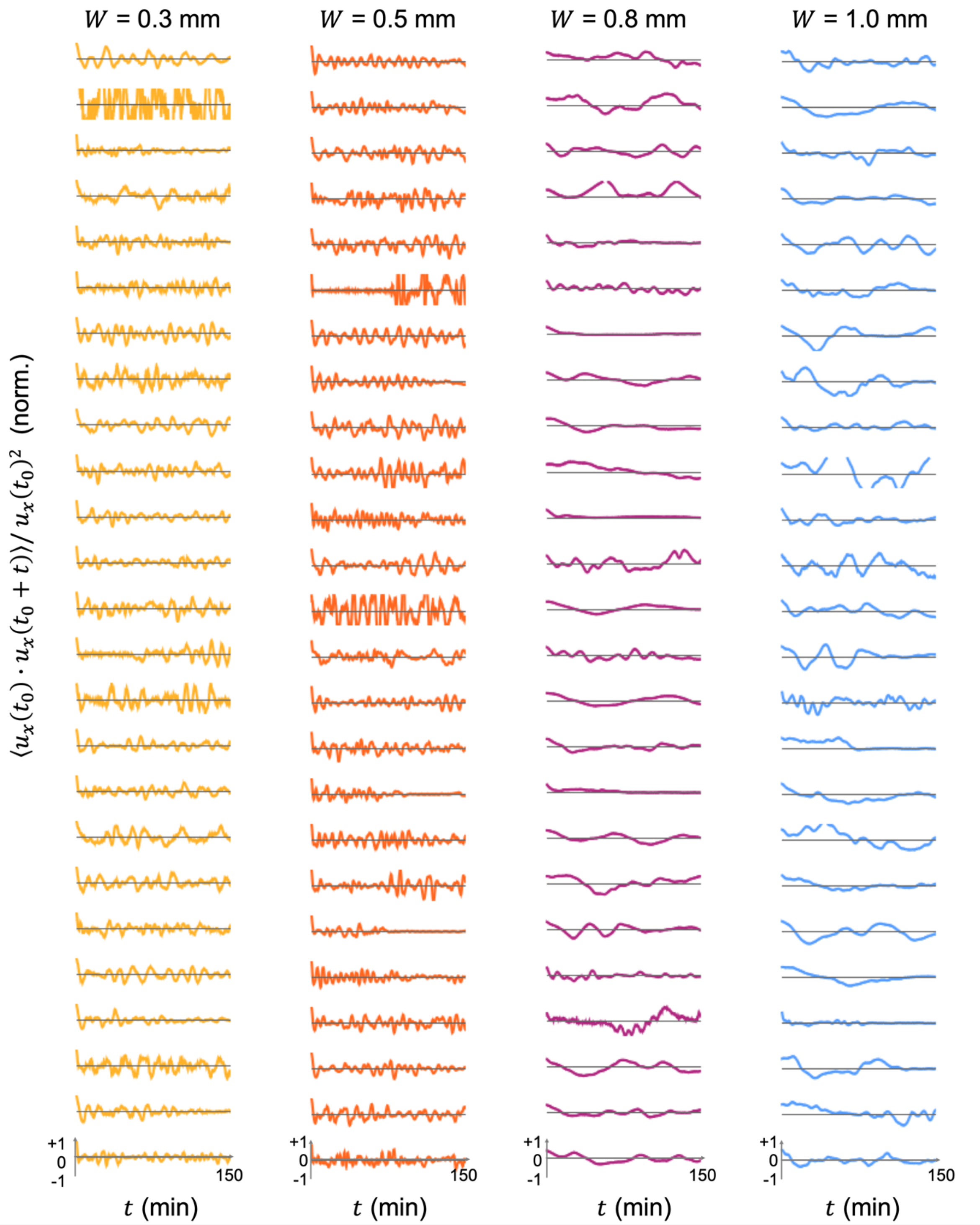

***Figure S10 – Velocity correlation functions for different channel widths.*** *Each plot corresponds to one bead. Changes in sign of the function correspond to switching in direction along the x axis (see supplementary methods).*

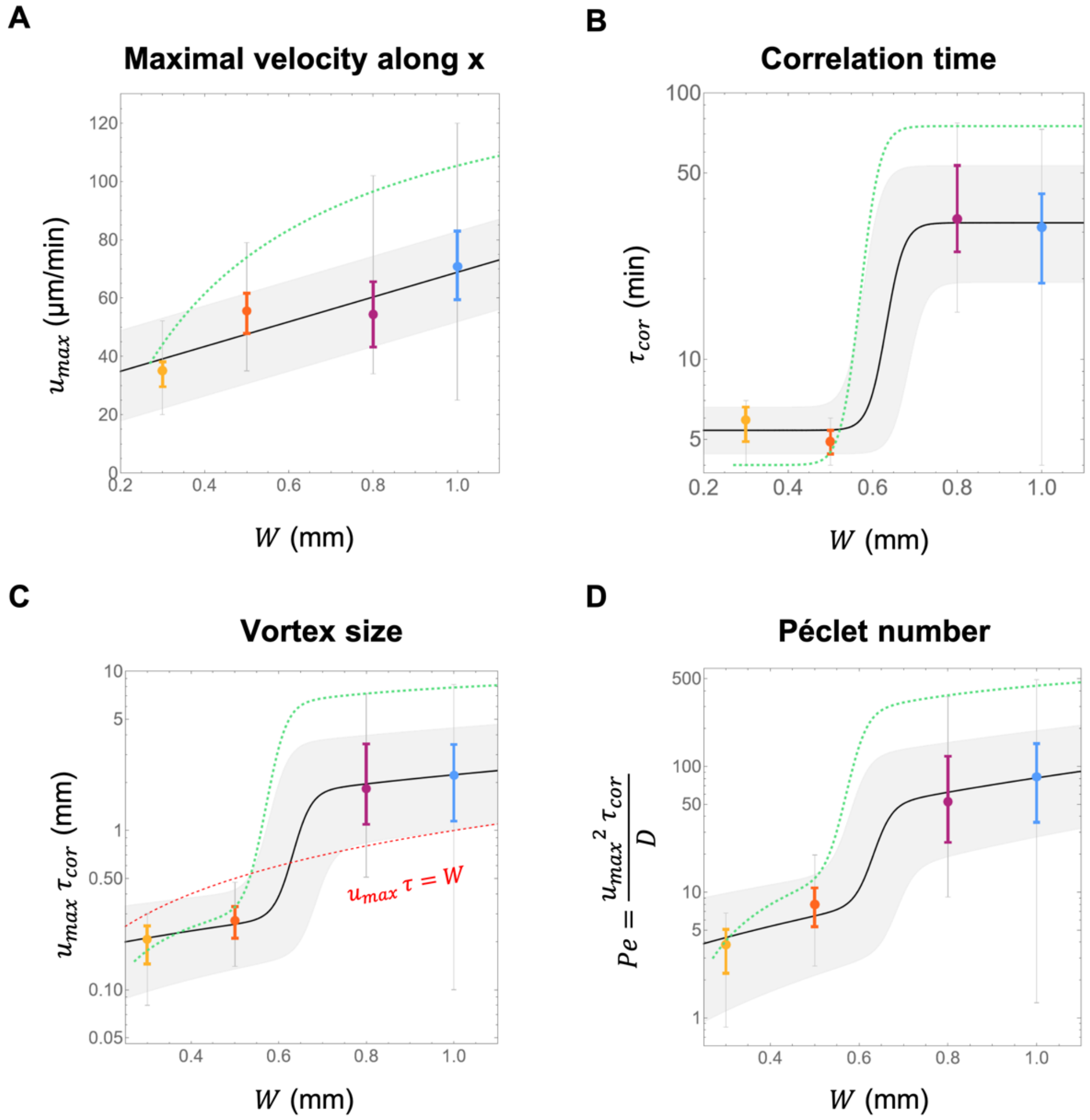

***Figure S11 - Active flows' parameters.*** *Distributions of bead maximal velocity along the x axis **(A)**, bead correlation time along the x axis **(B)**, vortex sizes **(C)** and the Péclet number **(D)** for different channel widths. Central dots in panels correspond to the median value, and colored bars to the 25% and 75% quantiles. Grey bars correspond to the maximal values. The black line corresponds to respectively linear or sigmoidal extrapolations of the median values of* $u_{max}$ *and* $\tau_{cor}$*, with the shaded area corresponding to the 25% and 75% quantiles. The green dashed line corresponds to taking values of* $u_{max}$ *and* $\tau_{cor}$ *close to the extrema. (see supplementary methods)*

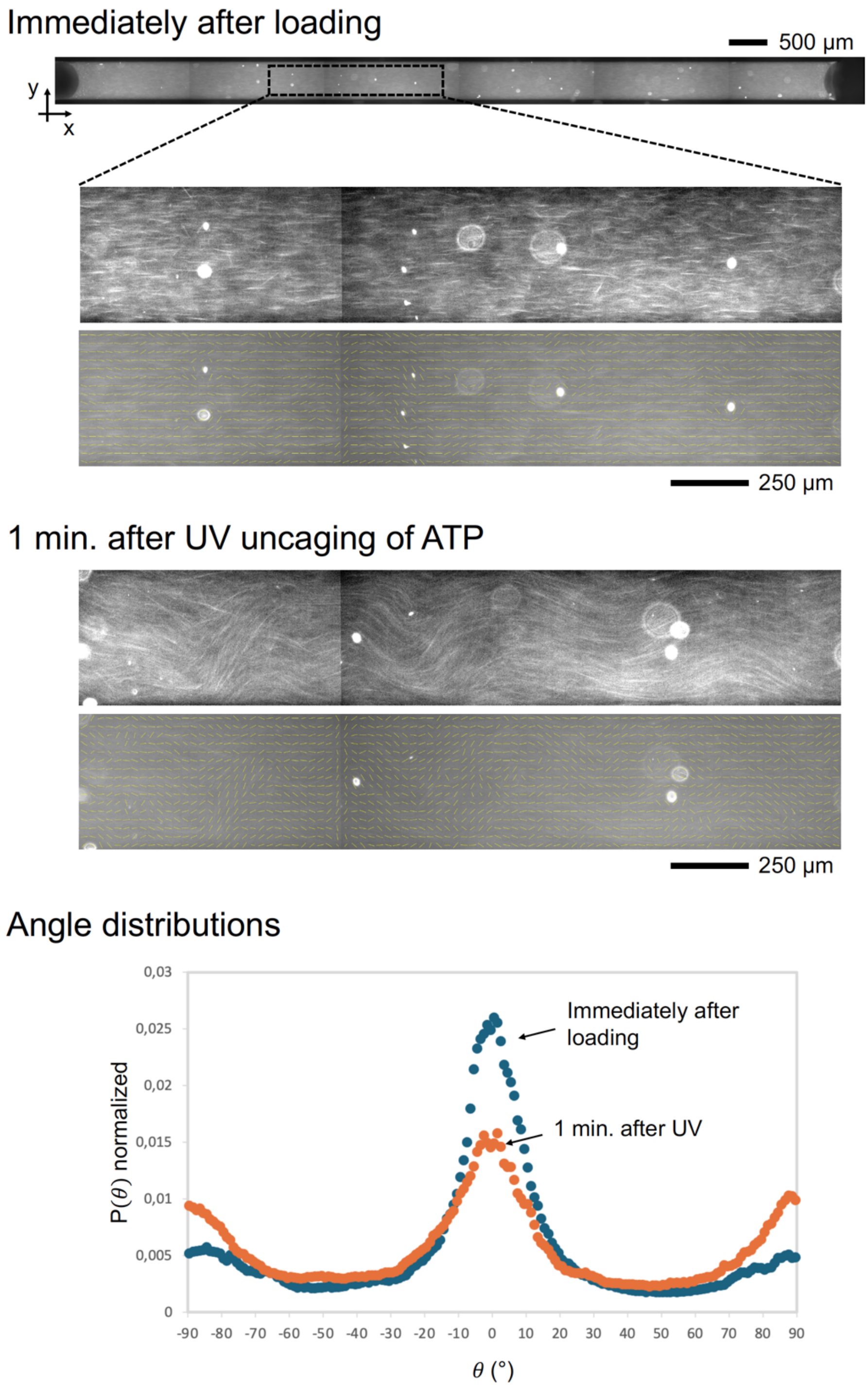

***Figure S12 - Nematic orientation of the microtubules in the xy plane*** *immediately after loading the solution in the 0.5 mm channel, and 1 minute after ATP was uncaged with UV light to trigger activity. Orientation vectors and angle distributions were obtained using. OrientationJ plugin* for ImageJ. $\theta$ is the angle between the filaments and the x axis. It should be noted that the addition of the initial condition that triggers the reaction-diffusion front locally breaks this nematic order.*

* Rezakhaniha, R., Agianniotis, A., Schrauwen, J. T. C., Gria, A., Sage, D., Bouten, C. v., Van de Vosse, F., Unser, M., and Stergiopulos, N. (2012) Experimental investigation of collagen waviness and orientation in the arterial adventitia using confocal laser scanning microscopy. *Biomechanics and modeling in mechanobiology* 11, 461-473.

## Reaction-Diffusion-Advection in channels of varying width

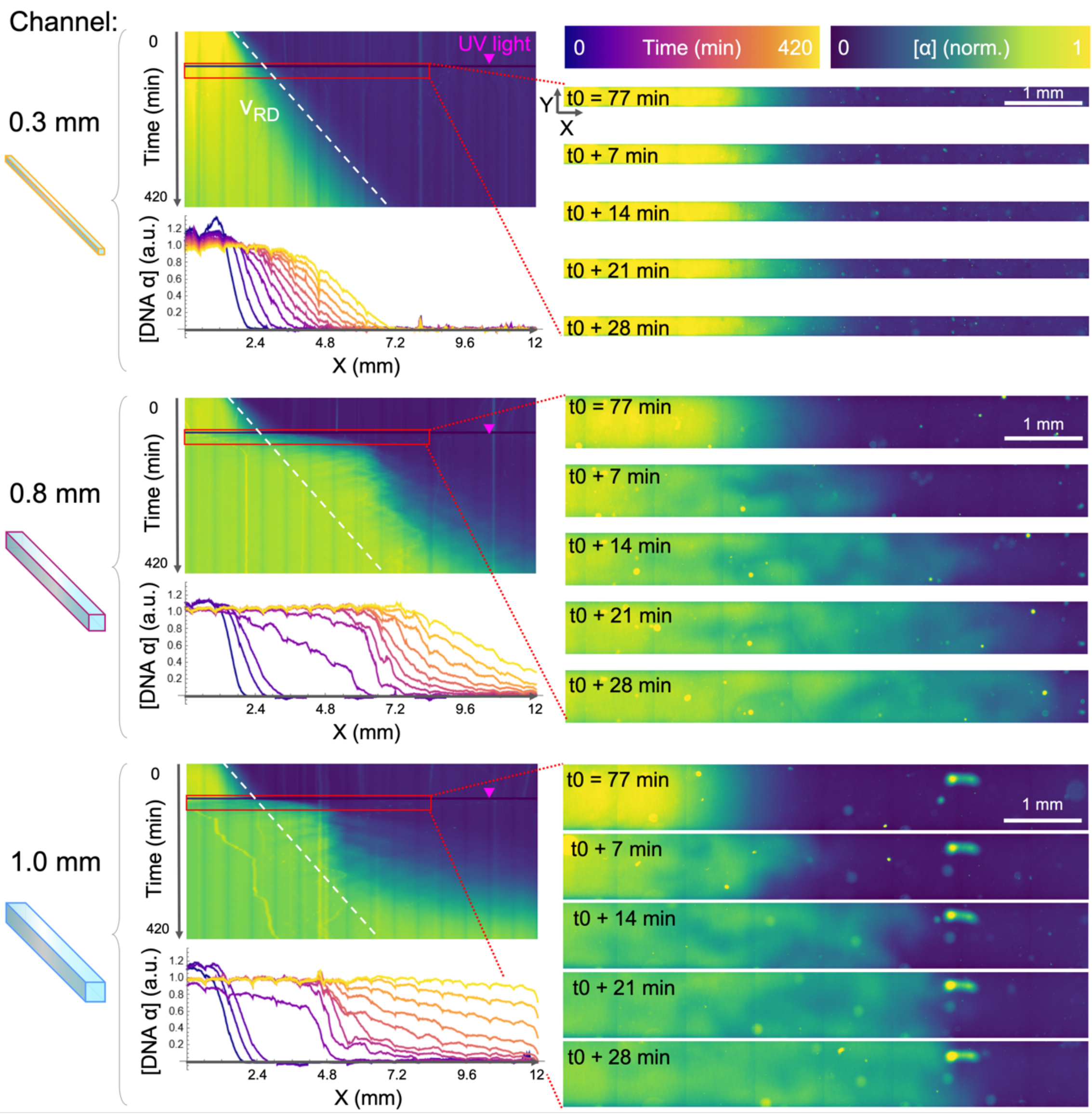

***Figure S13– Geometry-dependent Reaction-Diffusion-Advection.*** *Kymographs allow to compare the change in speed. Concentration profiles show the change in sharpness.*

# Active chaotic flows of kinesin-driven microtubules

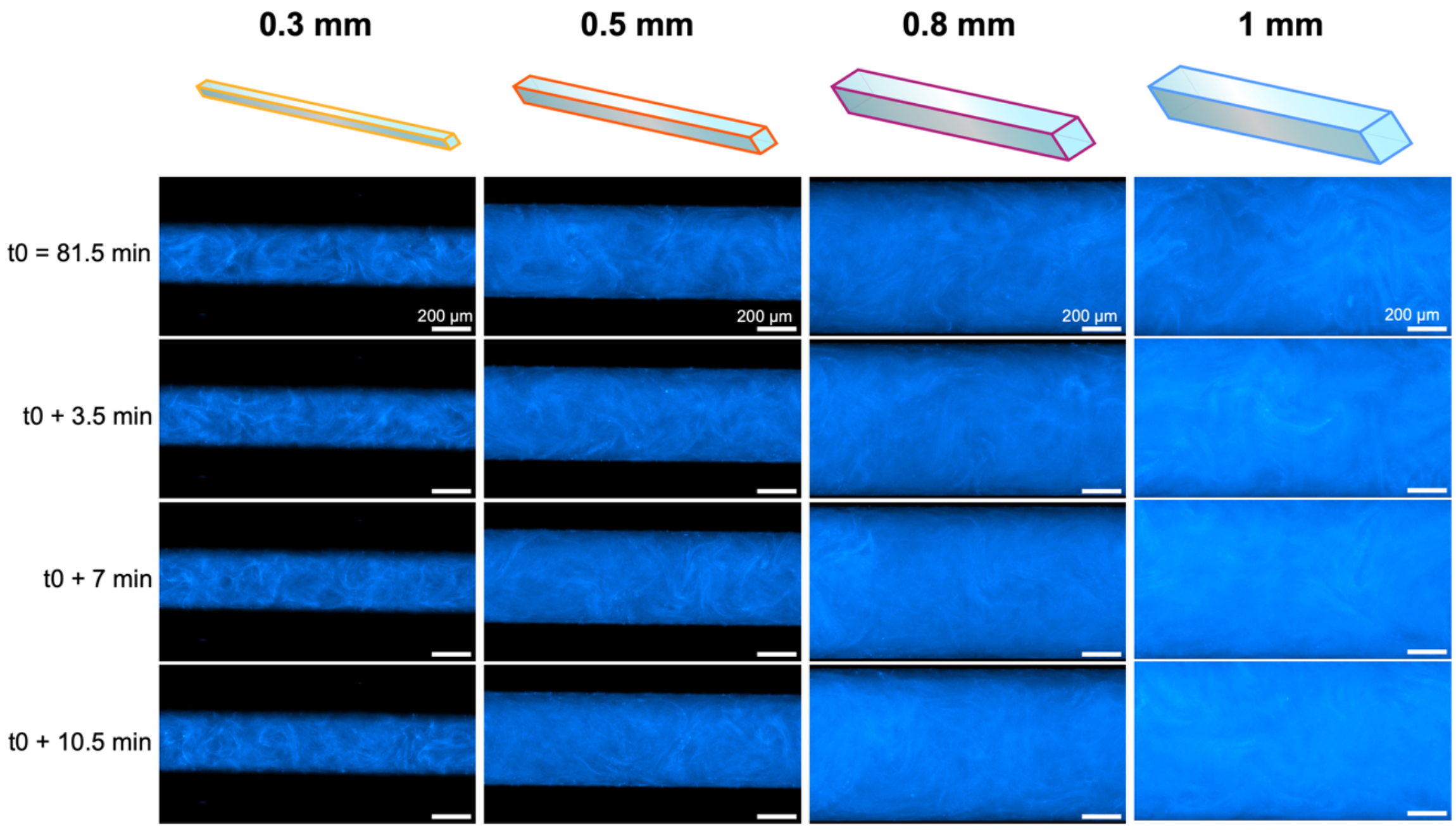

***Figure S14 – Time-lapse of motor-driven microtubules for varying channel widths.*** *Fluorescence is normalized by channel depth. Local configuration of microtubule bundles is very similar across channels. Microtubule bundles move in a chaotic manner.*

## Piecewise decomposition of front propagation over time

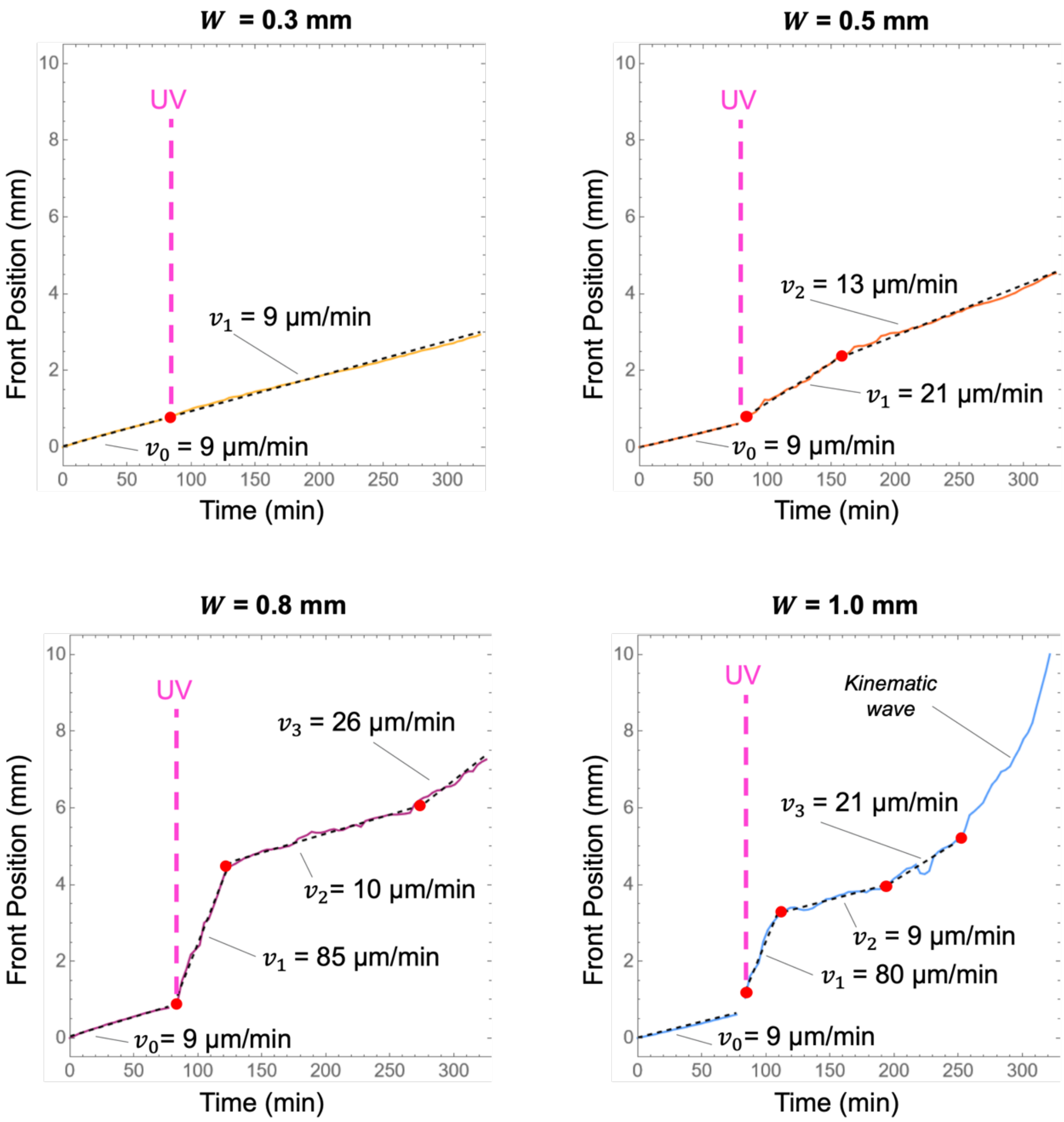

***Figure S15 – Piecewise linear fits of front propagation in channels of different widths***. *Evolution of front position for different channel widths (Threshold* $[\alpha]$ *= 0.5). The speeds of each different transport phase are marked for each channel. For each width, several front speeds are measured sequentially.* $v_0$ *is the speed before ATP was uncaged while* $v_{1-3}$ *correspond to the constant speeds measured at different transport phases.*

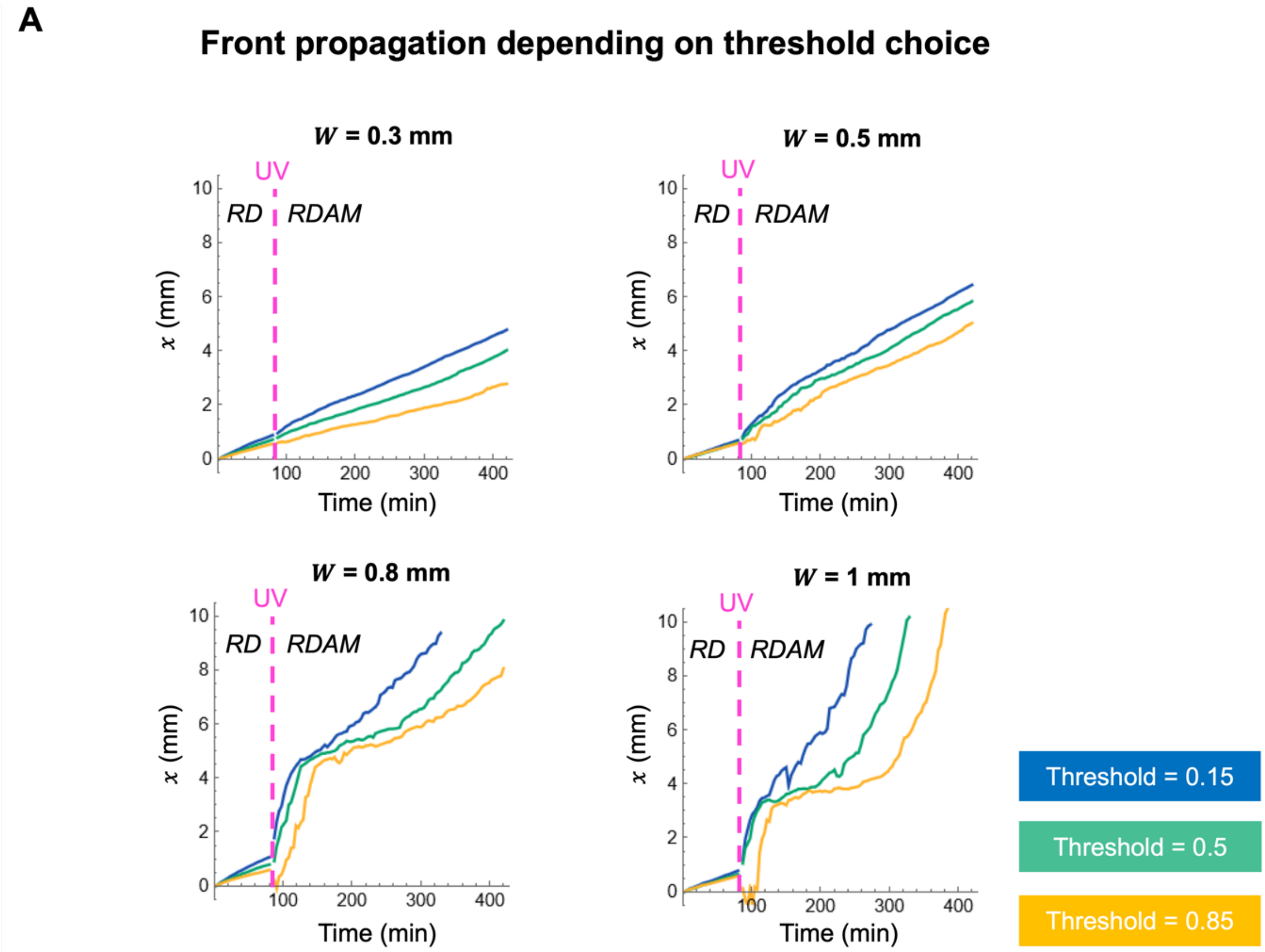

***Figure S16 – Front speed and width depending on definition choices. A:*** *Evolution of front position for different channel widths depending on concentration threshold.*

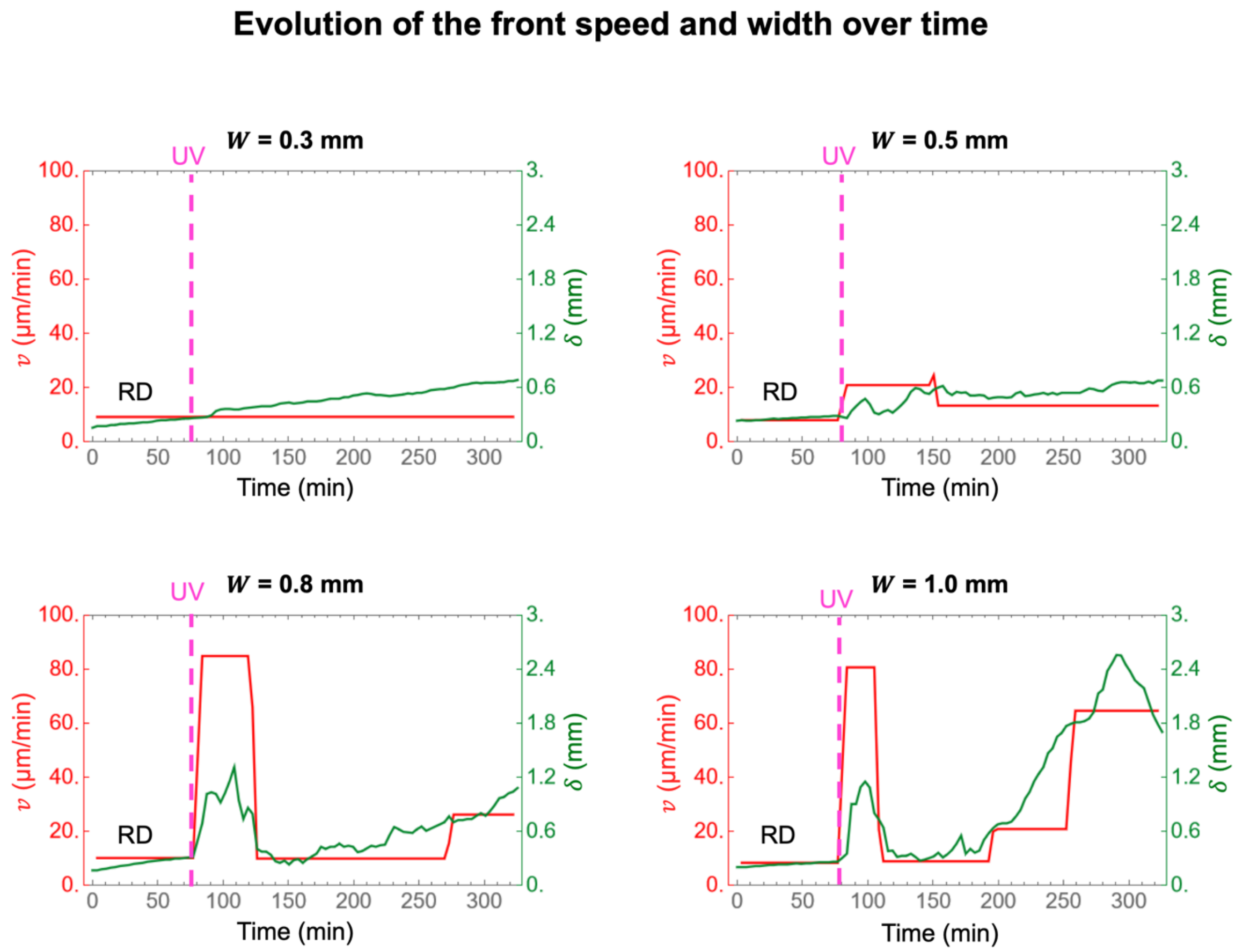

***Figure S17 – Effect of channel geometry on the evolution of front speed and width.*** *Evolution of front speed (left axis, red) and width (right axis, green) for the 0.3 mm, 0.5 mm, 0.8 mm and 1 mm channels. Pink dashed line indicates UV light exposure.*

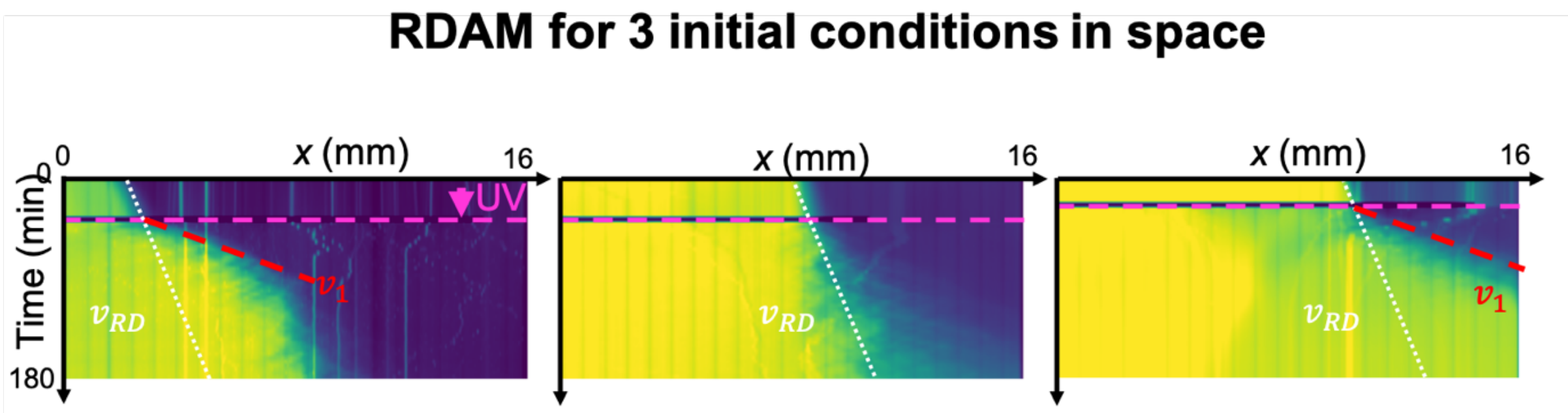

***Figure S18 – Propagation of DNA fronts starting at different positions in the channel*** *for $W$ = 0.8mm. The propagation of a front starting at the middle of the channel, and thus sitting at the border between two hydrodynamic cells, is not modified by active matter and its speed remains equal to $v_0 = v_{RD}$. Instead, a front starting on the right side of the channel, and thus sitting on the second hydrodynamic cell, propagates similarly to a front starting on the left side of the channel, and thus sitting on the first hydrodynamic cell, both travelling at $v_1$.*

## Kymographs of RD & RDAM in 17 additional channels

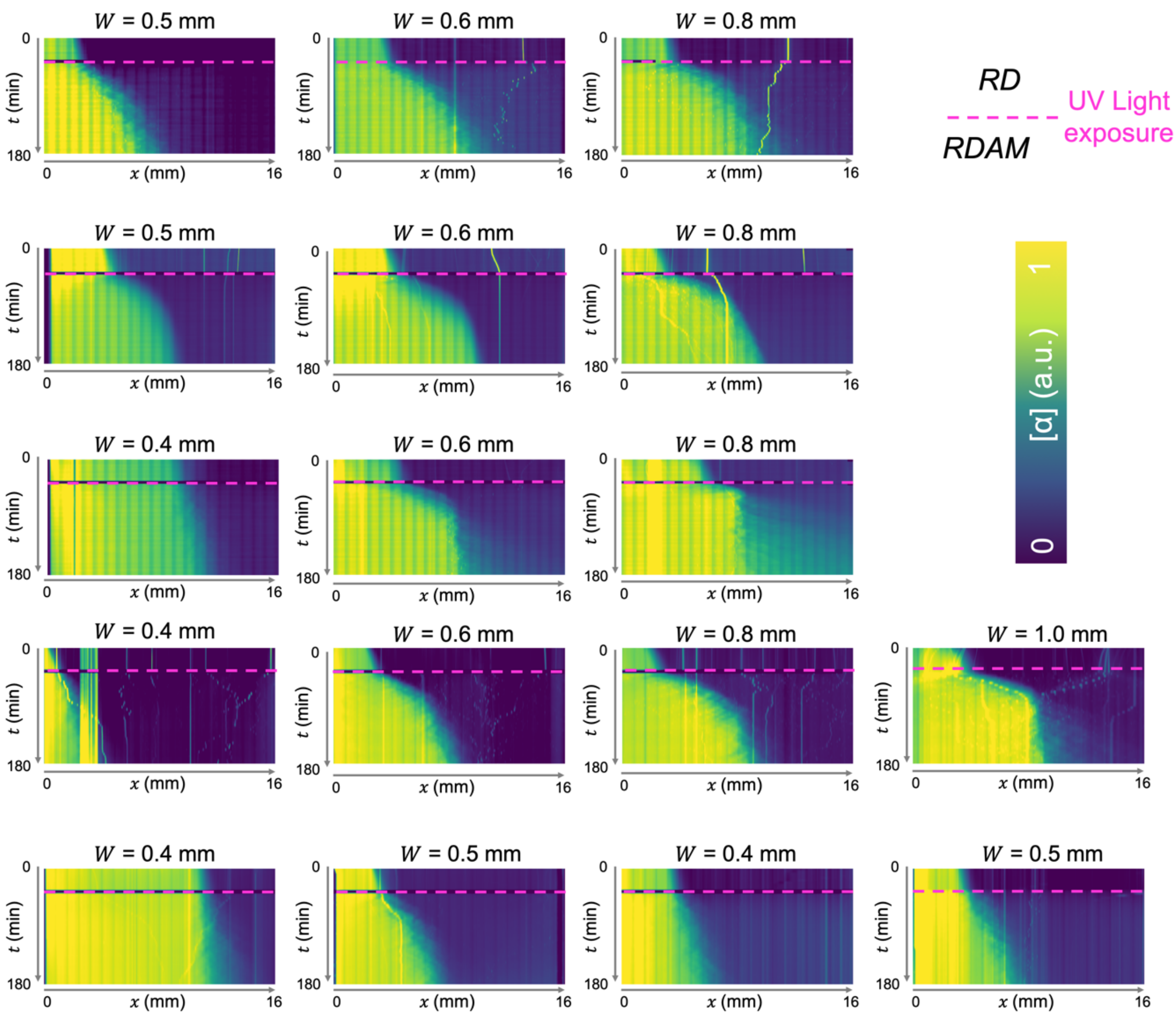

***Figure S19 – Additional experiments to measure speed and width when varying confinement.*** *The pink lines correspond to the UV exposure. Simultaneous experiments are presented in each line, with the exception of the last line corresponding to two experiments.*

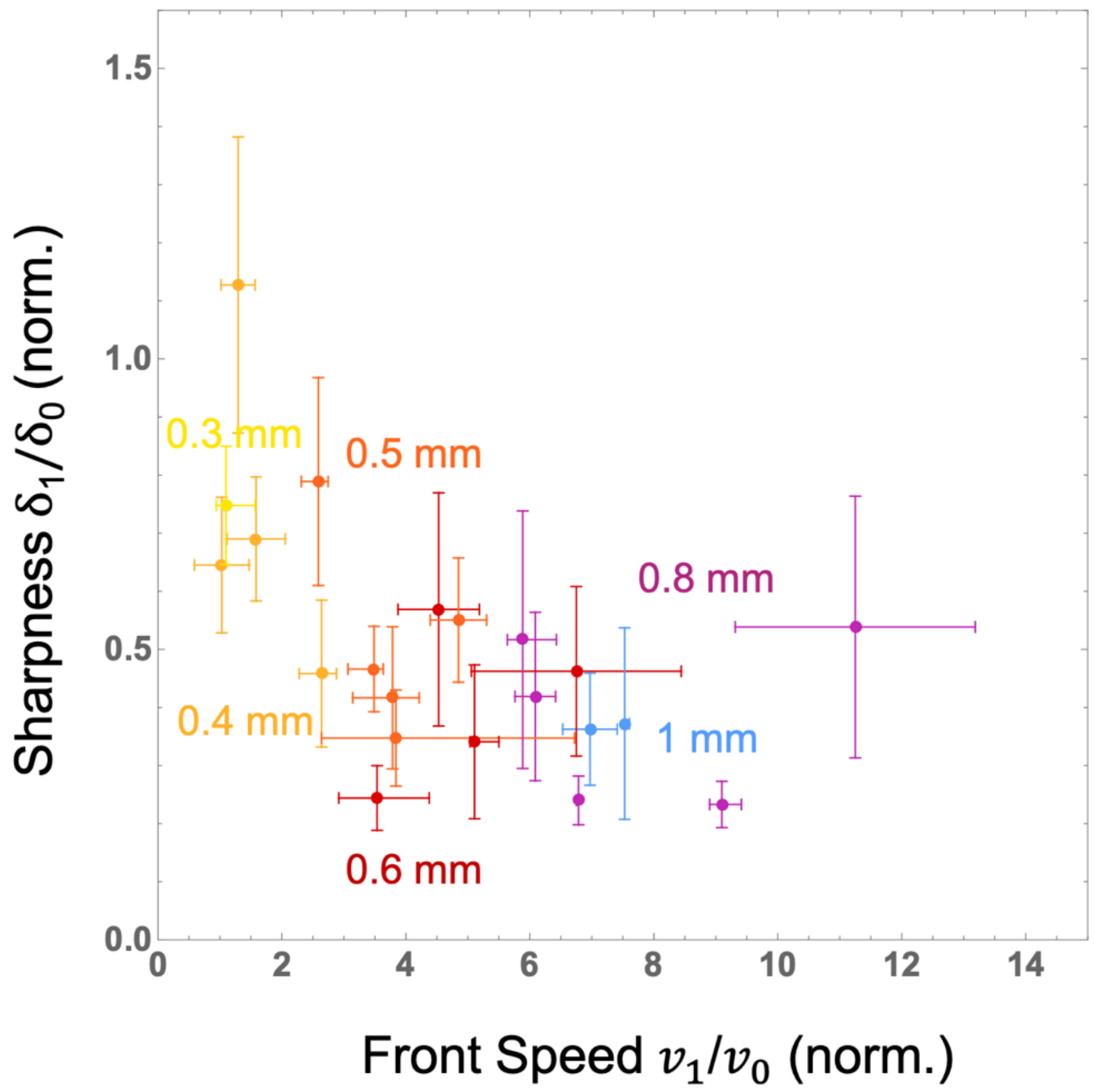

***Figure S20 - [DNA] front sharpness vs speed for varying channel widths*** *during the first phase for n = 21 experiments. The dots and the bars correspond respectively to the mean and the standard deviation, taken respectively over different thresholds for the speed or over time for the sharpness (see supplementary methods section).*

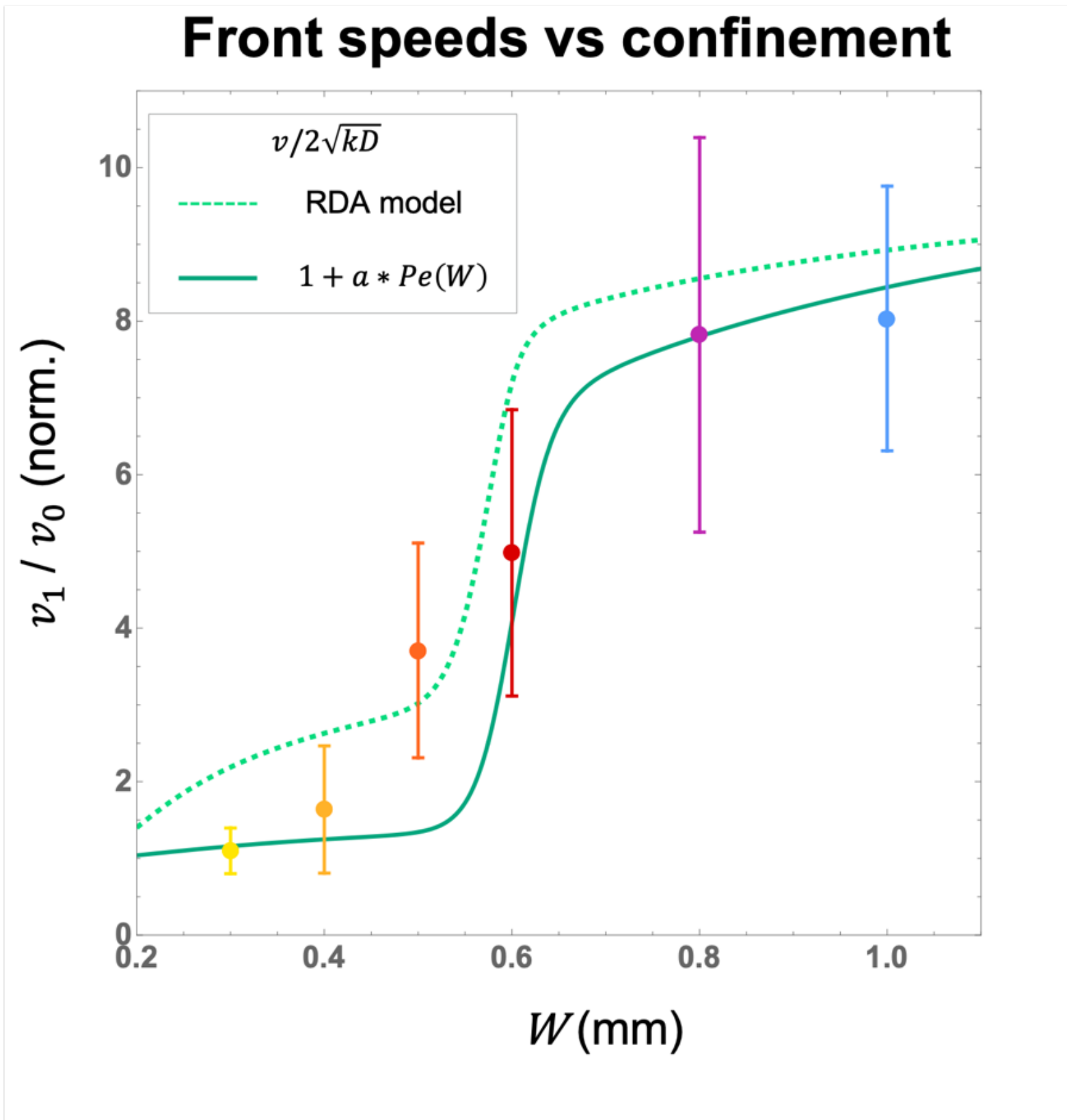

***Figure S21 - Geometry-dependent front propagation: experiments vs Péclet model**. Front speeds during the first phase (v1, hydrodynamic cell) vs. channel width for n = 21 experiments. The dots and the bars correspond respectively to the mean and the standard deviation. The green dashed line corresponds to taking the extrema values of $u_{max}$ and $\tau_{cor}$ as shown in Fig. S11. The green full line corresponds to a phenomenological model based linearly on the Péclet number, with $a = (v_1(W_0) - v_{RD})/Pe(W_0)$ a prefactor obtained empirically (here with the channel W = 0.8 mm). Experimental values are normalized by $v_0$, the RD speed. The models are normalized by $v_{FKPP}$*

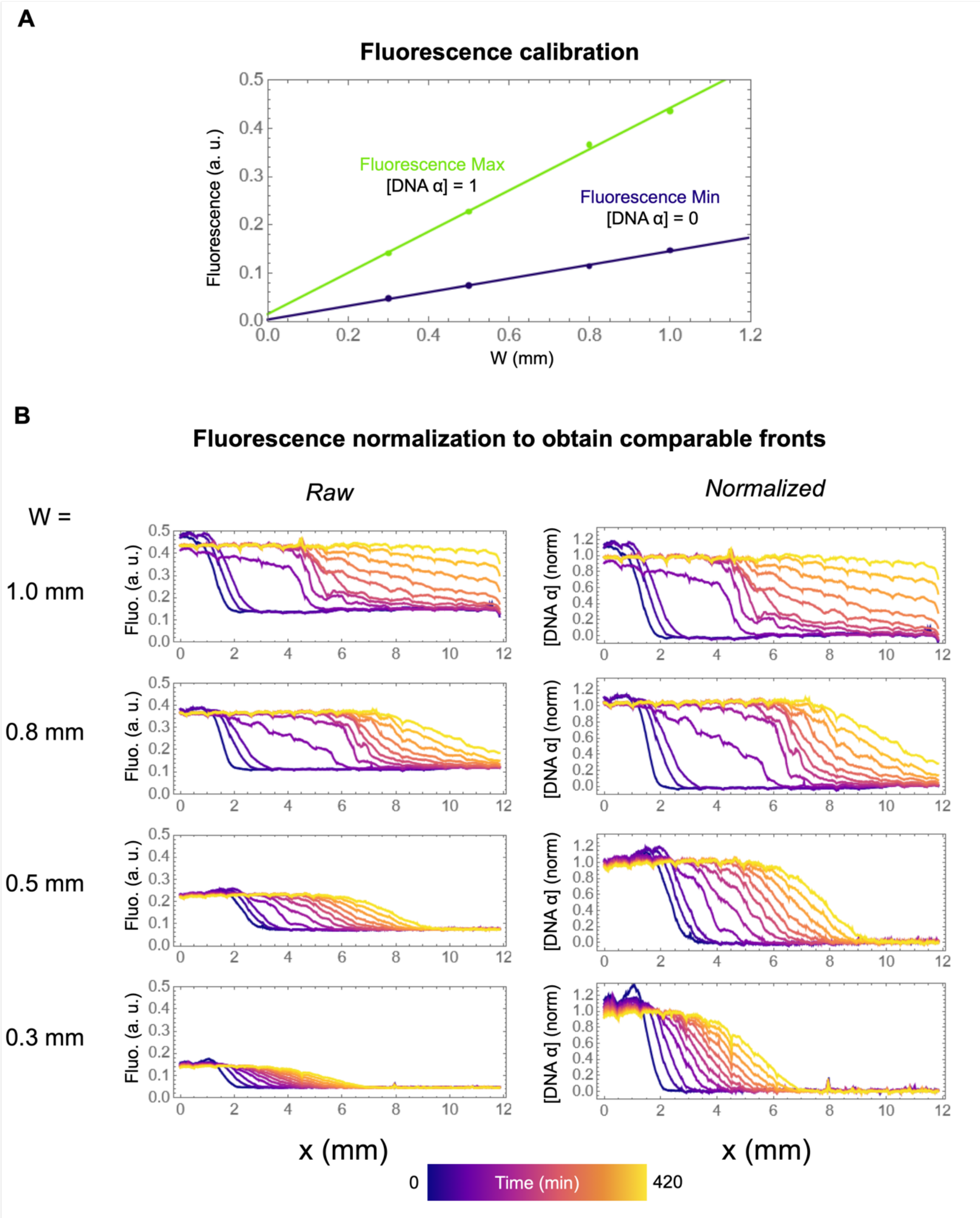

***Figure S22 – Fluorescence normalization taking into accounts the varying channel heights. A:*** *Calibration of the fluorescence in each channel using the high steady state and background fluorescence levels.* ***B:*** *Comparison of the raw and normalized fluorescence curves.*

## 2. Supplementary Tables

| Component | Concentration |
|---|---|
| PEM buffer (pH=6.9 KOH adjusted) | 0.6 x |
| PEM buffer (pH=7.2 NaOH adjusted) | 0.2 x |
| MgSO4 | 8 mM |
| Depletant (Pluronic) | 1.5% (w/v) |
| NPE-Caged ATP | 2 mM |
| Creatine Phosphate | 5 µg/mL |
| Creatine Kinase | 10 mM |
| DTT | 3 mM |
| Glucose oxidase | 150µg/mL |
| D-glucose | 20 mM |
| Catalase | 25 µg/mL |
| Trolox | 1 mM |
| BSA | 0.5 mg/mL |
| dATP | 10 µM |
| dGTP | 25 µM |
| dCTP | 500 µM |
| dTTP | 500 µM |
| SYBRGreenI | 0.4 x |
| $T_{\alpha\alpha}$ | 100 nM |
| α | 0 or 1 µM |
| Polymerase (BST Large Fragment) | 0.5 % (v/v) |
| Nickase (Nt. BstNBI) | 1.5 % (v/v) |
| Kinesin motors (BCCP kinesin) | 75 nM |
| Streptavidin | 75 nM |
| Microtubules | 1 mg/mL |

***Table S1: Standard Active mix composition.*** *This corresponds to Fig. 3-4. In Fig.1 α was injected at 10 µM for a better visualization and Polymerase or Kinesin were not added in some channels. In Fig. 2, $T_{\alpha\alpha}$ , α , enzymes were replaced with Beads for tracking.*

| Name | Sequence (5'-3') | Modif. 5' | Modif. 3' |
|---|---|---|---|
| α | CGAGTCTGTT | ∅ | ∅ |
| $T_{\alpha\alpha}$ | A*A*CAGACTCGAACAGACTCG | ∅ | Phosphate |

**phosphorothioate backbone modifications on the 5'*

***Table S2: Sequences of oligonucleotides***

## 3. Legends of Supplementary Movies

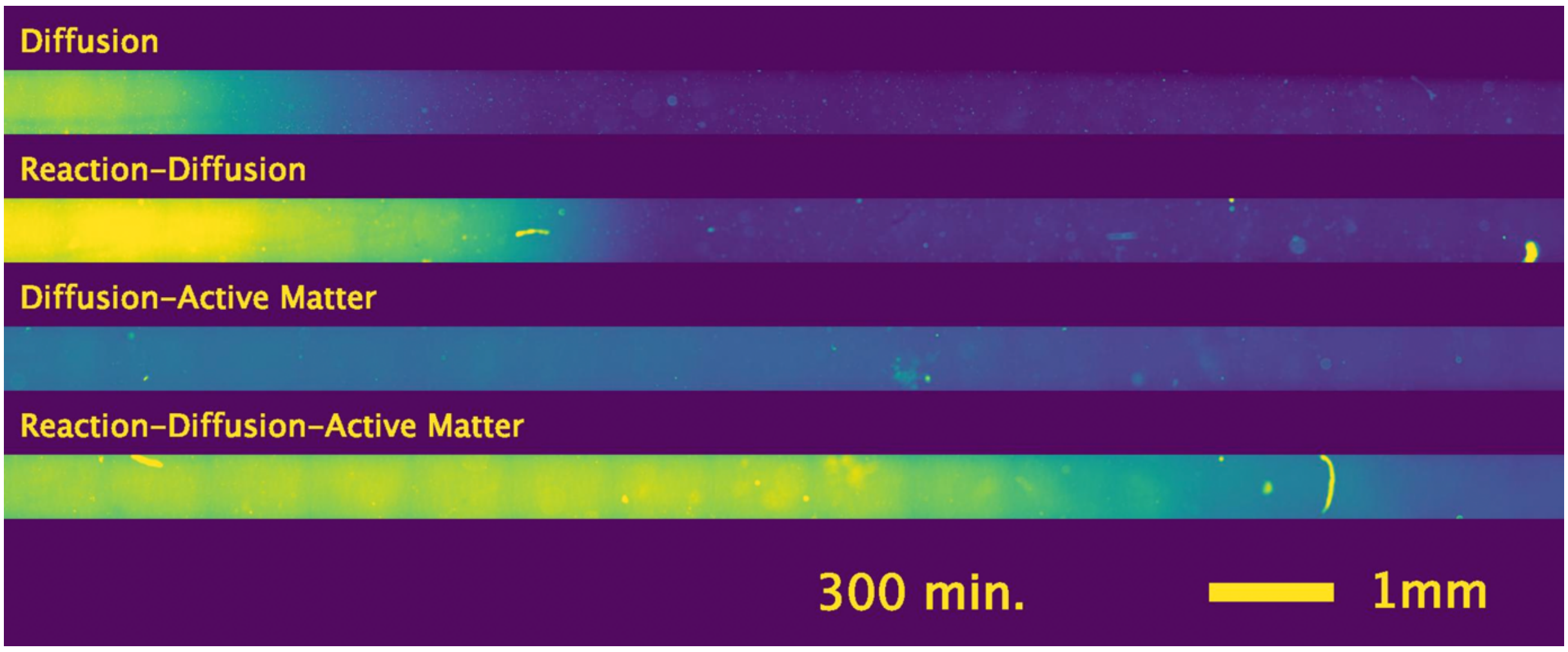

***Movie S1 - Comparison of different Reaction-Diffusion-Advection configurations.*** *Time-lapse images analyzed in Figure 1D,E. and Figures S3-6. Intensity corresponds to [DNA α] via fluorescence of Sybr Green I.*

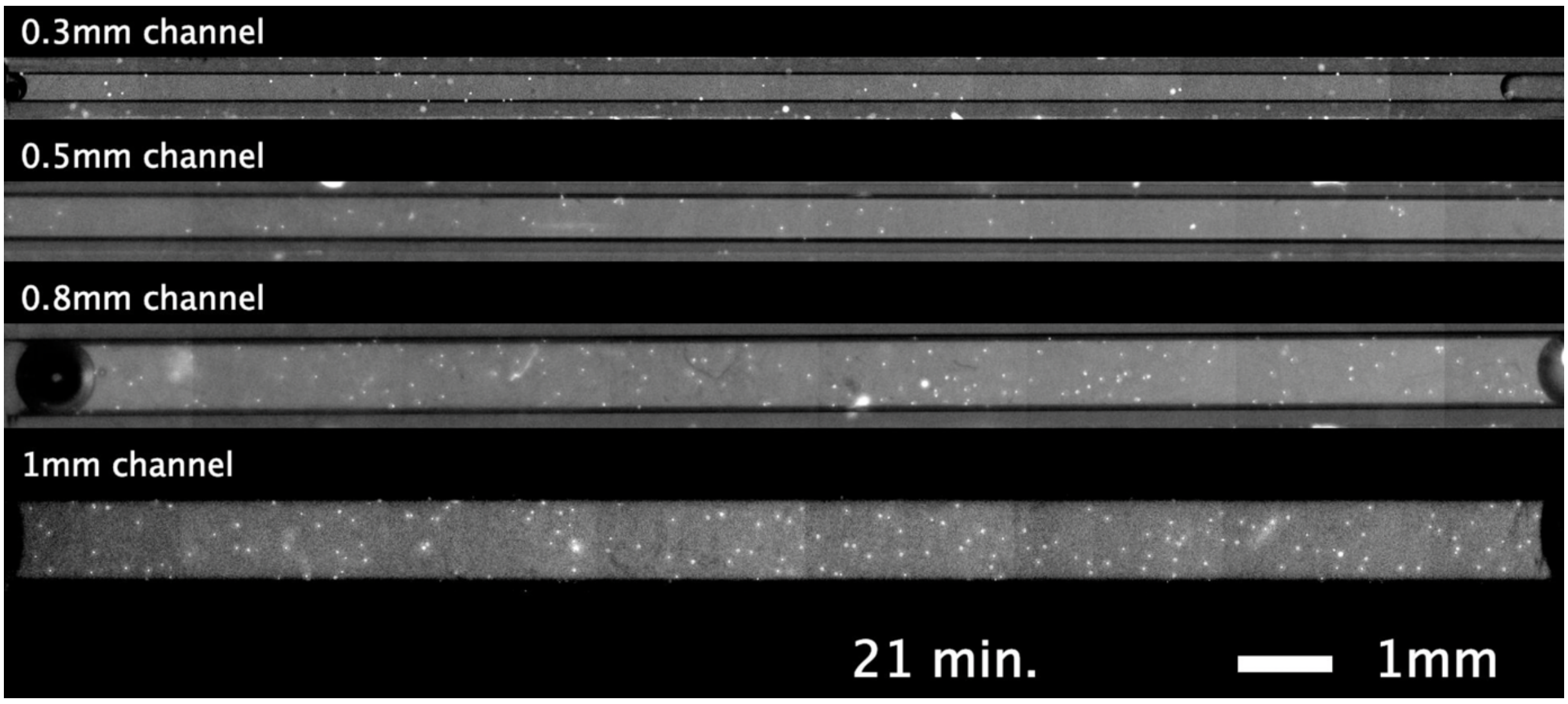

***Movie S2 –Trajectories of fiduciary beads in channels of different widths.*** *Time-lapse images analyzed in Figure 2. For W=0.8 mm and W=1mm channels, images taken at different heights are summed to avoid losing focus.*

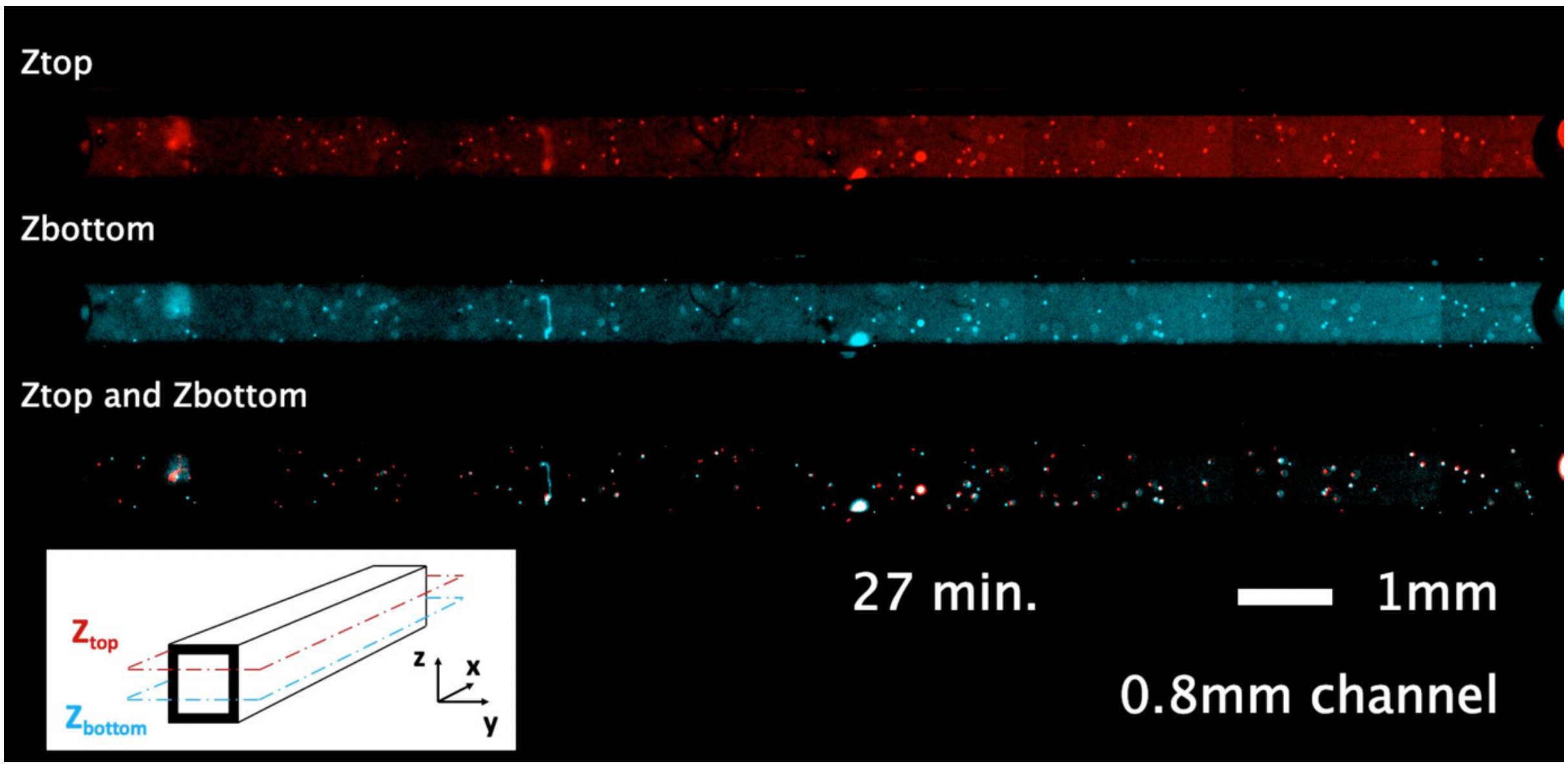

***Movie S3 - Trajectories of fiduciary beads in a 0.8 mm channel at two different heights.*** *Beads at different heights go in opposite directions. Time-lapse images analyzed in Figure 2.*

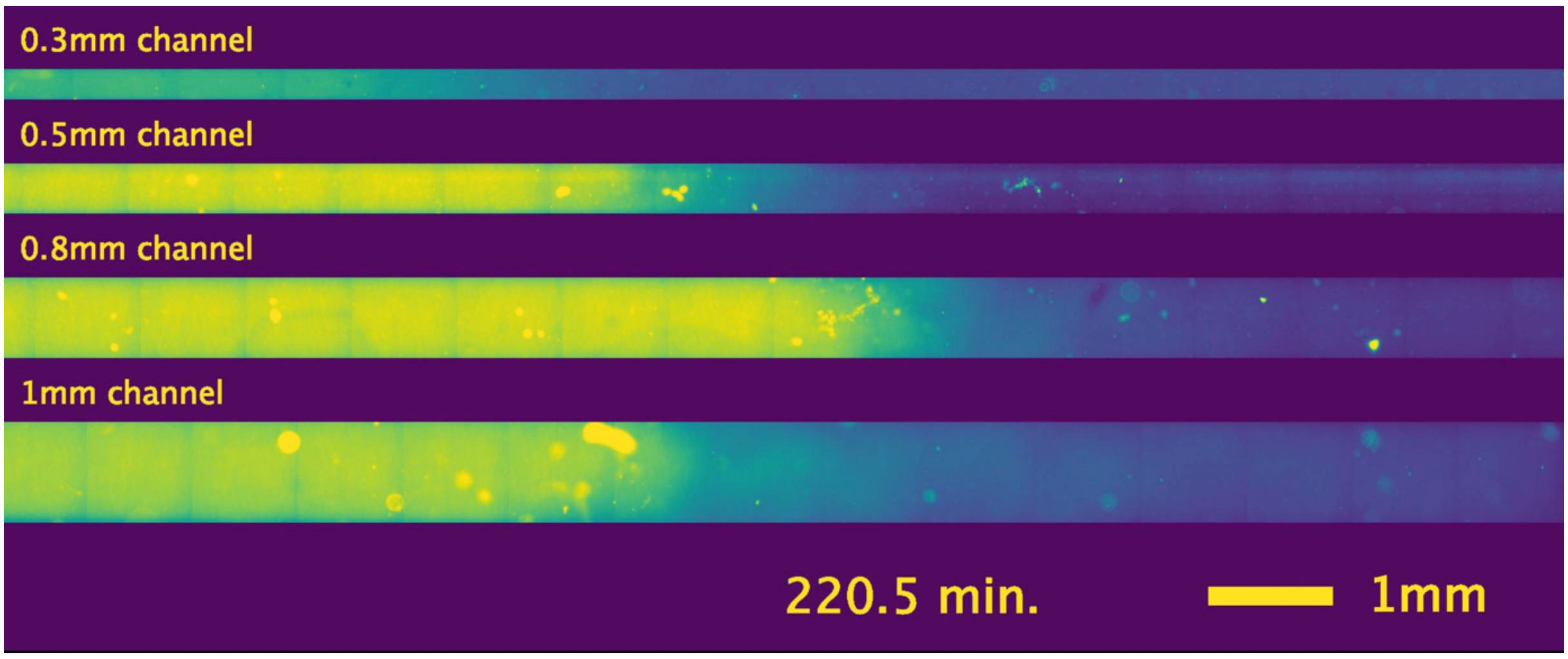

***Movie S4 - Geometry-dependent Reaction-Diffusion-Advection.*** *Time-lapse images analyzed in Figure 3-4. Intensity corresponds to [DNA α] via fluorescence of Sybr Green I. At* $t_0$ *= 77 min, the channel is exposed to UV light which triggers advection.*

# 4. Supplementary methods

**Extraction of the diffusivity *D***

We use the method of images allowing to transform our problem of diffusion from a single source next to a wall into a problem where the wall is replaced by an additional source that is the mirror image of the first one (see Figure S3C). At each time the shape of the front is the sum of two gaussians and given by:

$$\frac{A}{\sqrt{2\pi}}\left(\exp\left(\frac{-(x-x_0)^2}{2\sigma(t)^2}\right)+\exp\left(\frac{-(x+x_0)^2}{2\sigma(t)^2}\right)\right)$$

with $A$ the initial amplitude, $x_0$and $-x_0$ respectively the positions of the real and mirror sources, and $2\sigma(t)^2 = 4Dt$. Fitting the gaussians at each time and plotting $0.5\sigma(t)^2$ against $t$ allows to extract $D$ from a linear fit (see Figure S3D). The error on $D$ comes quasi exclusively from the linear fit as the error on the first gaussian fit is negligible compared to the variation of $\sigma$ over time. The range of values of *D* is assessed by taking the minimal and maximal slope that passes through the data.

**Extraction of the reaction rate *k***

Taking only at the reaction term in Eq (2) in the main text gives:

$$\frac{\mathrm{d}[\alpha](t)}{\mathrm{dt}} = \frac{V\cdot[\alpha](t)}{\mathrm{K_M}+[\alpha](t)}$$

which is integrated into

$$\frac{1}{V}\cdot\left([\alpha](t)-[\alpha](0)\right)+\frac{\mathrm{K_M}}{V}\cdot\ln\left(\frac{[\alpha](t)}{[\alpha](0)}\right)=t.$$

For any threshold value $[\alpha](t_s) = [\alpha]_s \gg [\alpha](0)$, we can derive the time over threshold, $t_s$ :

$$t_s = \left(\frac{1}{V}\cdot[\alpha]_s+\frac{\mathrm{K_M}}{V}\ln([\alpha]_s)\right)-\frac{\mathrm{K_M}}{V}\cdot\ln\left([\alpha](0)\right)$$

$$t_s = B - 1/k\cdot\ln(c_0)$$

with *B* a constant and $c_0$ the initial concentration of $\alpha$ and $k = V/K_M$.. Therefore, measuring $t_s$ as a function of $-\ln(c_0)$ allows to obtain $1/k$ from the slope of the linear variation (see Figure S4). Taking the extrema slopes that pass through our data allows to give a range of values for $1/k$.

**Kymographs (**Figs 1D, 3A, 3B, S19)

Kymographs were obtained using Fiji. Fluorescence is normalized to account for the fluorescence background and capillary thickness (Fig S22). and averaged along the y axis by resizing images to only 1 pixel in height ("adjust size"). The stack of images in time was then stitched along the y axis (using "make montage").

**Front propagation curves** (Figures 1E, S15 , S16)

Image analysis is performed using Mathematica. Images are resized to 1000 pixels in length. Fluorescence is averaged along the y axis, and normalized to account for the fluorescence background and capillary thickness (Fig S22). Normalized fluorescence corresponds to normalized concentration of $\alpha$ with $[\alpha]$ = 0, and $[\alpha]$ = 1 respectively corresponding to the high level and zero concentration levels. For a given time and concentration threshold 0<$[\alpha]_{th} < 1$, the position of the front is the position $x$ that minimizes $|[\alpha](x) - [\alpha]_{th}|$.

In Figure 1E, the threshold is chosen at $[\alpha] = 0.2$ in order to be measurable even in absence of reaction. Indeed in the DA case $[\alpha] < 0.5$, which prevents from choosing higher thresholds. In Figure 3C, the threshold is chosen at $[\alpha] = 0.5$.
Figure S. shows how propagation is affected by the threshold choice. $[\alpha] = 0.15$ and $[\alpha] = 0.85$ are the minimal and maximal thresholds imposed by our observation setup. Indeed choosing $[\alpha]_{th}<0.15$ or $[\alpha]_{th}>0.85$, corresponds to zones where the front is flatter and our measurement becomes very sensitive to the inhomogeneities in each field of view or due to dust.

**Speed of the front** (Figures 3E, 3F, 4, 6A, 6B, S15, S16, S20, S21)
For a given threshold $0 < [\alpha]_{th} < 1$, the speed of the front is taken as the distance traveled over the time of a particular studied phase.
In figures 3E, 3F, 4. By analogy with sigmoidal fronts, the position of the front is taken at half level $[\alpha] = 0.5$. In Figure 4 we also show the speeds for $[\alpha] = 0.15$ and $[\alpha] = 0.85$.

In Figures S21 we calculate the speed for 13 thresholds from $[\alpha] = 0.2$ to $[\alpha] = 0.8$ with a step of 0.05. We show the median speed (central dot) and standard deviation (bars).
Figure 6A and 6B corresponds to the aggregation of the results for each channel width. Error bars correspond to the standard deviation $\sigma_{tot}$, obtained from the standard deviation $\sigma(m_i)$ of all the means of the measurements $m_i$ and the mean of the standard deviations for each measurement $\sigma_i$ such that ${\sigma_{tot}}^2 = \sigma(m_i)^2 + \langle \sigma_i^2 \rangle$.

**Width of the front** (Figures 3E, 3F, 6B, S4, S17, S20, S21)
For a typical reaction diffusion, the width $\delta_{RD}$ corresponds to the characteristic length of a sigmoidal profile $1/(1 + e^{x/\delta_{RD}})$ and is measured using the difference in position of the thresholds $[\alpha] = 0.27$ and $[\alpha] = 0.73$ that are spaced by $2\delta_{RD}$ (Fig S4E).
As our reaction-diffusion-advection fronts still resemble a sigmoidal profile, we kept the same measurement method.
The variability in the measurement of the width $\delta$ comes from its variability in time more than a specific choice of thresholds. Therefore, in Figure S17, the central dot corresponds to the median of the width over time during the studied phase and the bars to the standard deviation. In Figure 6B, the standard deviation $\sigma_{tot}$ is obtained as previously for the speed.

**Fiduciary beads trajectories (Figures 2, S7-9)**
In order have all beads in focus at all times, Images were taken at several heights (up to 3 in 1mm channel) and summed. To remove dust and other immobile particles, the first images step was subtracted to all other images. Figures 2A, S were obtained using "Temporal color code" in Fiji.

Beads' trajectories coordinates used in Fig 2, were obtained using Trackmate plugin (Ref 59). Errors in tracking, mostly due to trajectory crossings, were corrected manually inside Trackmate to obtain at least 25 trajectories lasting more than 3h in each capillary. Beads trajectories were smoothed over 2 minutes to remove detection noise. Velocity is taken over 3 consecutive time steps.

**Extraction of maximal speed and correlation time**
$u_{max}$ and $\tau_{cor}$ are measured for each bead trajectory and the distribution of results for all beads are then presented with their median value as well as 25 % and 75% quantiles (Figures 2C,2D,2E).

For each bead $u_{max}$ is the maximal velocity along the x axis during the measured 3h.

$\tau_{cor}$ corresponds to the characteristic time between a direction switch along the x axis. To extract $\tau_{cor}$, each 180-minutes trajectory is duplicated into 30 trajectories of $\Delta t$ =150 minutes using a moving window of 30 minutes to cancel effect the effect of choosing a particular starting time $t_0$. For each $t_0$ we calculate the velocity correlator along the $x$ axis, $u_x(t_0) \cdot u_x(t_0 + t)$, and then average over the 30 different $t_0$ to obtain the velocity correlation function. These velocity correlation functions are shown in Figure 2B, S10.

For each of these velocity correlation functions, we obtain a list of times $\tau_i$ at which there is a direction switch. To obtain the characteristic time $\tau_{cor}$ related to a single bead, we extracted a weighted average of the times $\tau_i$, each time being weighted by itself so that $\tau_{cor} = \sum \tau_i^2 / \Delta t$. Doing so means that we consider that a correlation time is characteristic the larger portion of time a bead spends in the corresponding piece of trajectory. It allows to mitigate edge effects and the noise. Indeed, considering the following example $\{\tau_i\} = \{0.01, 0.49, 0.49, 0.01\}\Delta t$, we obtain $\tau_{cor}$=0.48$\Delta t$ which is within 2% of $0.49\Delta t$, the correlation time for 98% of our total measurement window. Taking a simple average would give $\langle \tau_i \rangle = 0.25\Delta t$, which would not be characteristic of any trajectory piece.

# 5. Theoretical supplements

**Derivation of the conveyor-belt RDA model**

The flow within the cell is modeled as a conveyor belt in the *xy* plane, with the upper stream having velocity $+u_0\vec{x}$ and the lower stream $-u_0\vec{x}$ . Species α can switch from one stream to the opposite stream through smaller scale vertices observed experimentally, with a characteristic time $\tau$. The corresponding equations are derived from Eq. (2) by considering a uni-dimensional problem where α switches between two populations, $\alpha_+$ and $\alpha_-$ through a fictional reaction of characteristic time $\tau$ (Fig. 5A). $\alpha_+$ is advected with velocity $+u_0$, while $\alpha_-$ is advected with velocity $-u_0$, both diffusing with coefficient $D$. This can be modeled as

$$\begin{cases} \dfrac{\partial[\alpha]_+}{\partial t} + u_0 \dfrac{\partial[\alpha]_+}{\partial x} = k[\alpha]_+ - \dfrac{1}{\tau}([\alpha]_+ - [\alpha]_-) + D\Delta[\alpha]_+ & (1) \\ \dfrac{\partial[\alpha]_-}{\partial t} - u_0 \dfrac{\partial[\alpha]_-}{\partial x} = k[\alpha]_- + \dfrac{1}{\tau}([\alpha]_+ - [\alpha]_-) + D\Delta[\alpha]_- & (2) \end{cases}$$

Defining $[\alpha] \coloneqq [\alpha]_+ + [\alpha]_-$ and $\widetilde{[\alpha]} \coloneqq [\alpha]_+ - [\alpha]_-$ , (1) and (2) become

$$\begin{cases} \dfrac{\partial[\alpha]}{\partial t} + u_0 \dfrac{\partial\widetilde{[\alpha]}}{\partial x} = k[\alpha] + D\Delta[\alpha] & (3) \\ \dfrac{\partial\widetilde{[\alpha]}}{\partial t} + u_0 \dfrac{\partial[\alpha]}{\partial x} = k\widetilde{[\alpha]} - \dfrac{2}{\tau}\widetilde{[\alpha]} + D\Delta\widetilde{[\alpha]} & (4) \end{cases}$$

**First asymptote: $k\tau \ll 1$**

for $k\tau \ll 1$, $\widetilde{[\alpha]}$ relaxes fast and we consider it reaches a stationary state that we will inject in the equation for $[\alpha]$.

$$\left[\left(k-\frac{2}{\tau}\right)+D\frac{\partial}{\partial x^2}-\frac{\partial}{\partial t}\right]\widetilde{[\alpha]}=u_0\frac{\partial[\alpha]}{\partial x}$$

We define

$$\gamma \coloneqq k-\frac{2}{\tau}$$

$$\widetilde{[\alpha]}=\frac{u_0}{\gamma}\left[1+\frac{D}{\gamma}\frac{\partial}{\partial x^2}-\frac{1}{\gamma}\frac{\partial}{\partial t}\right]^{-1}\frac{\partial[\alpha]}{\partial x}$$

For $k\tau \ll 1$, $|\gamma| \sim \left|-\frac{2k}{k\tau}\right| \gg 1$ and thus $\left|\frac{1}{\gamma}\right| \ll 1$. This allows us to take the Taylor expansion of $(1+x)^{-1}=1-x+o(x^2)$ which gives:

$$\widetilde{[\alpha]}=\frac{u_0}{\gamma}\left[1-\frac{D}{\gamma}\frac{\partial}{\partial x^2}+\frac{1}{\gamma}\frac{\partial}{\partial t}\right]\frac{\partial[\alpha]}{\partial x}$$

Keeping only terms up to 2nd order derivatives we obtain

$$\widetilde{[\alpha]}=\frac{u_0}{\gamma}\frac{\partial[\alpha]}{\partial x}+\frac{u_0}{\gamma^2}\frac{\partial^2[\alpha]}{\partial t\partial x}$$

We inject this value of $\widetilde{[\alpha]}$ in equation (3)

$$\frac{\partial[\alpha]}{\partial t}+\frac{{u_0}^2}{\gamma}\frac{\partial^2[\alpha]}{\partial x^2}+\frac{{u_0}^2}{\gamma^2}\frac{\partial^2[\alpha]}{\partial x^2\partial t}=k[\alpha]+D\frac{\partial^2[\alpha]}{\partial x^2}$$

$$\frac{\partial[\alpha]}{\partial t}=-\frac{{u_0}^2}{\gamma}\frac{\partial^2[\alpha]}{\partial x^2}-\frac{{u_0}^2}{\gamma^2}\frac{\partial^2[\alpha]}{\partial x^2\partial t}+k[\alpha]+D\frac{\partial^2[\alpha]}{\partial x^2}$$

As we restrict to 2nd order derivatives, when replacing $\frac{\partial[\alpha]}{\partial t}$ in $\frac{\partial^2[\alpha]}{\partial x^2\partial t}$ we will only consider the $k[\alpha]$ term such that $\frac{\partial}{\partial x^2}\frac{\partial[\alpha]}{\partial t}=\frac{\partial}{\partial x^2}\,k[\alpha]$, which gives

$$\frac{\partial[\alpha]}{\partial t}=k[\alpha]+\left(D-\frac{{u_0}^2}{\gamma}-\frac{{u_0}^2k}{\gamma^2}\right)\frac{\partial^2[\alpha]}{\partial x^2}$$

Defining $D_{eff} \coloneqq D-\frac{{u_0}^2}{\gamma}-\frac{{u_0}^2k}{\gamma^2}$, we recover the FKPP equation which gives us the speed of the front as:

$$v=2\sqrt{kD_{eff}}$$

**Second asymptote: $k\tau \gg 1$**

For $k\tau \gg 1$, we neglect transfer between + and – streamlines. Without transverse transfer, in the vicinity of the front the concentration $[\alpha]_-$ can only come from low concentration side and we can therefore neglect it. We now consider $[\alpha] \sim [\alpha]_+$. Equation (1) becomes

$$\frac{\partial[\alpha]}{\partial t} + u_0 \frac{\partial[\alpha]}{\partial x} = \left(k - \frac{1}{\tau}\right)[\alpha] + D\frac{\partial^2[\alpha]}{\partial x^2}$$

We place ourselves in the referential of the front moving at speed $v$ and define $\eta := x - vt$

$$-(v - u_0)\frac{\partial[\alpha]}{\partial \eta} = \left(k - \frac{1}{\tau}\right)[\alpha] + D\frac{\partial^2[\alpha]}{\partial \eta^2}$$

This corresponds exactly to FKPP equation for $V := v - u_0$, and $k_{eff} := k - \frac{1}{\tau} \gg 1$. Applying FKPP solution gives us

$$V = 2\sqrt{k_{eff}D}$$

The speed of the front is then

$$v = u_0 + 2\sqrt{k_{eff}D}$$

**General case**

For the general case we also place ourselves in the in the reference frame co-moving with the front at speed $v$ our equations for $[\alpha]$ and $\widetilde{[\alpha]}$ become:

$$\begin{cases} -v\dfrac{\partial[\alpha]}{\partial\eta} + u_0\dfrac{\partial\widetilde{[\alpha]}}{\partial\eta} = k[\alpha] + D\dfrac{\partial^2[\alpha]}{\partial\eta^2} \\ -v\dfrac{\partial\widetilde{[\alpha]}}{\partial\eta} + u_0\dfrac{\partial[\alpha]}{\partial\eta} = \left(k - \dfrac{2}{\tau}\right)\widetilde{[\alpha]} + D\dfrac{\partial^2\widetilde{[\alpha]}}{\partial\eta^2} \end{cases}$$

We now assume the front has an exponential decay shape and we look for solutions in the form of $A\,\mathrm{e}^{-\mu\eta}$ travelling at speed $v_\mu$. Doing so taking the derivative $\frac{\partial}{\partial\eta}$ means multiplying by $-\mu$, and we get the following matrix equation.

$$\begin{pmatrix} k + D\mu^2 - v_\mu\mu & u_0\mu \\ u_0\mu & \left(k - \dfrac{2}{\tau}\right) + D\mu^2 - v_\mu\mu \end{pmatrix}\begin{pmatrix} [\alpha] \\ \widetilde{[\alpha]} \end{pmatrix} = 0$$

Finding non-trivial solutions requires that

$$\det\begin{pmatrix} k + D\mu^2 - v_\mu\mu & u_0\mu \\ u_0\mu & \left(k - \frac{2}{\tau}\right) + D\mu^2 - v_\mu\mu \end{pmatrix} = 0$$

This gives us two families of solutions for the speed of the front.

$$v_\mu(\mu) = \frac{k}{\mu} + D\mu - \frac{1}{\tau\mu}\left(1 \pm \sqrt{1 + (\tau u_0\mu)^2}\right)$$

If we now remove the terms related to advection we recover FKPP

$$\det\begin{pmatrix} k + D\mu^2 - v_\mu\mu & 0 \\ 0 & k + D\mu^2 - v_\mu\mu \end{pmatrix} = 0 \;\Rightarrow\; v_\mu(\mu) = \frac{k}{\mu} + D\mu$$

$v_{FKPP}(\mu) \coloneqq \frac{k}{\mu} + D\mu$ is the exact expression obtained in the conventional FKPP derivation and $v_{FKPP} = \min_{\mu} v_{FKPP}(\mu)$

Our families of solutions can be seen as extensions of the FKPP model

$$v_\mu(\mu) = v_{FKPP}(\mu) - \frac{1}{\tau\mu}\left(1 \pm \sqrt{1 + (\tau u_0\mu)^2}\right)$$

The speed of the front is the maximum of these solutions; it can also be shown that this solution yields the KPP one in the limit of no flow:
However only one family gives $\lim_{u_0 \to 0^+} v_\mu(\mu) = v_{FKPP}(\mu)$

$$v_\mu(\mu) \underset{u_0 \to 0^+}{\sim} v_{FKPP}(\mu) - \frac{1}{\tau\mu}\left(1 \pm \left(1 + \frac{1}{2}(\tau u_0\mu)^2\right)\right)$$

$$\begin{cases} v_{\mu_+}(\mu) \underset{u_0 \to 0^+}{\sim} v_{FKPP}(\mu) - \dfrac{2}{\tau\mu} < v_{FKPP}(\mu) \\[2ex] v_{\mu_-}(\mu) \underset{u_0 \to 0^+}{\sim} v_{FKPP}(\mu) + \dfrac{1}{2}\tau\mu\, u_0{}^2 \underset{u_0 \to 0^+}{\longrightarrow} v_{FKPP}(\mu) \end{cases}$$

We therefore take

$$v_\mu(\mu) = \frac{k}{\mu} + D\mu \; - \frac{1}{\tau\mu}\left(1 - \sqrt{1 + (\tau u_0\mu)^2}\right)$$

and by analogy with FKPP* we define the speed of the front $v$ as

$$v = \min_{\mu}\left[\frac{k}{\mu} + D\mu \; - \frac{1}{\tau\mu}\left(1 - \sqrt{1 + (\tau u_0\mu)^2}\right)\right]$$

*Van Saarloos, Wim. "Front propagation into unstable states." *Physics reports* 386.2-6 (2003): 29-222.

## Theoretical supplement: Linear stability analysis of aligned active fluids in three-dimensional channels

In this supplementary section, we describe the flows and instabilities of a three-dimensional channel filled with microtubule filaments and kinesin motors. The microtubules are initially aligned by flow while filling the channel with the gel and, in the absence of *active* motors, remain aligned. We model the microtubule gel as a three-dimensional nematic fluid oriented along $\hat{x}$ and consider the effect of activity on this state [1–4]. This is somewhat distinct in detail (though, ultimately not in the end result) from [5–7] who assume that the orientational order decays with time. We believe that this distinction arises from the differing microscopic details in the experiments such as the bundle thickness and the density. Indeed, we and others [4, 8–10], showed in earlier works that changing the composition of the gel—such as the amount of ATP or motor concentration—dramatically changes its physical properties.

We characterise the ordered state by the nematic order parameter $\mathbf{Q} = \mathbf{nn} - (\mathbf{I}/3)$, where $\mathbf{I}$ is the identity tensor and $\mathbf{n}$ is the director which is a unit vector that points along the microtubules. The fluctuations about the nematic state is parametrised by the director as $\mathbf{n} \approx (1, \delta n_y, \delta n_z) \equiv (1, \delta\mathbf{n}_\perp)$, with the perfectly ordered state being $\mathbf{n}_0 = \hat{x}$.

The evolution of the fluctuations of the director field $\delta\mathbf{n}_\perp$ about its ordered state is described by the standard dynamical equation [11, 12]

$$\partial_t \delta\mathbf{n}_\perp = \frac{1-\lambda}{2}\partial_x \mathbf{v}_\perp - \frac{1+\lambda}{2}\nabla_\perp v_x + \Gamma K \nabla^2 \delta\mathbf{n}_\perp, \tag{1}$$

where $\mathbf{v}$ is the velocity field of the microtubule gel and $K\nabla^2\mathbf{n}_\perp = -\delta F/\delta\mathbf{n}_\perp$ with the free energy $F = (K/2)\int(\nabla\delta\mathbf{n}_\perp)^2$ in the single Frank constant approximation [11, 13]. The coupling of the velocity field to the director fluctuations has two parts [11, 12]: i. the director field is rotated by the vorticity field $(1/2)[\partial_x\mathbf{v}_\perp - \nabla_\perp v_x]$ and ii. the director field also experiences a torque in response to a strain-rate [11, 12] $(1/2)[\partial_x\mathbf{v}_\perp + \nabla_\perp v_x]$. The coefficient $\lambda$ for this latter coupling is phenomenological (and not fixed by rotation invariance unlike the coupling to vorticity); if $|\lambda| < 1$, the system is flow-tumbling and if $|\lambda| > 1$ it is flow-aligning.

For slow flows, relevant for all microbiological materials, inertia is irrelevant and we balance viscous forces with other active and passive force densities due to the motor-microtubule gel to construct the linearised, constitutive equation for the velocity field:

$$\eta\nabla^2\mathbf{v} = \nabla\Pi - \zeta(\nabla_\perp\cdot\delta\mathbf{n}_\perp\hat{x} + \partial_x\delta\mathbf{n}_\perp) + \zeta_2\nabla^2[(1/3)\nabla_\perp\cdot\delta\mathbf{n}_\perp\hat{x} - (2/3)\partial_x\delta\mathbf{n}_\perp] + \frac{1+\lambda}{2}\nabla_\perp\cdot\frac{\delta F}{\delta\mathbf{n}_\perp}\hat{x} - \frac{1-\lambda}{2}\partial_x\frac{\delta F}{\delta\mathbf{n}_\perp}, \tag{2}$$

where $\Pi$ is the pressure that enforces the incompressibility constraint $\nabla\cdot\mathbf{v} = 0$. There are two active force densities in this equation. The first $\propto \zeta$ is the standard active dipolar force density, with $\zeta < 0$ being extensile (which corresponds to motor-microtubule gels) and $\zeta > 0$ being contractile, and the second $\propto \zeta_2$ is a higher order in gradients active force density. It is irrelevant in a bulk fluid where the dynamics is controlled by $\zeta$, but *not in a confined geometry.* There are also other active force densities at this order in gradients, such as $\nabla^2(\nabla_\perp\cdot\delta\mathbf{n}_\perp\hat{x} + \partial_x\delta\mathbf{n}_\perp)$. We choose to retain just this because i. while we do want to highlight the importance of retaining higher order in gradient active stresses for understanding the dynamics of confined active fluids, we don't aim at a quantitative calculation of the active flow in such a geometry and ii. it was shown [4] that this active force density qualitatively modifies the character of the active instability—changing its threshold and even eliminating it completely—and active flows in confined geometries. Note that both $\zeta$ and $\zeta_2$ are likely to depend on the same parameters microscopically. Dimensionally, $\zeta_2$ contains two extra powers of lengths which are likely to be proportional to a lengthscale associated with the microtubule filaments. The last two terms in (2) are passive force densities arising from director distortions and is required for relaxation to a steady state distribution $\propto e^{-F}$ in the absence activity [11, 12].

We now Fourier transform (1) and (2) in space and time, with $\omega$ being the frequency and $\mathbf{q} \equiv (q_x, q_y, q_z) \equiv (q_x, \mathbf{q}_\perp)$ being the wavevector of perturbation. Solving for the velocity field in terms of the director perturbation by eliminating the pressure using the transverse projector $\mathcal{P}_{ij} = \delta_{ij} - q_i q_j/q^2$, we obtain two coupled equations for $\delta n_y$ and $\delta n_z$ fluctuations. From this, we obtain the two eigenfrequencies which correspond to twist-bend and splay-bend modes (i.e., twist modes and splay modes with an admixture of twist):

$$\omega_{tb} = -i\left[\frac{\zeta}{2\eta}(1-\lambda)\frac{q_x^2}{q^2} - \frac{\zeta_2}{3\eta}(\lambda-1)q_x^2 + K\Gamma q^2 + \frac{K}{4\eta}(1-\lambda)^2 q_x^2\right] \tag{3}$$

$$\omega_{sb} = -i\left[\frac{\zeta}{2\eta}\frac{q_x^2 - q_\perp^2}{q^2}\left(1 - \lambda\frac{q_x^2 - q_\perp^2}{q^2}\right) + \frac{\zeta_2}{6\eta}\frac{2q_x^2 + q_\perp^2}{q^2}\{(\lambda-1)q_x^2 - (\lambda+1)q_\perp^2\} + K\Gamma q^2 + \frac{K}{4q^2}\{(\lambda-1)q_x^2 - (\lambda+1)q_\perp^2\}^2\right]. \tag{4}$$

The *negativity* of $-i\omega_{tb}$ and $-i\omega_{sb}$ ensures the stability of the orientationally ordered gel. That is, if *either* of the quantities in square brackets in (3) and (4) turn negative for some set of parameter values, the ordered state of the gel is unstable for that set of parameter values. Random fluctuations excite modes at all wavenumbers. An inspection of (4) reveals that to lowest order in wavenumber (i.e., for $q \to 0$), the eigenfrequency $\omega_{sb}$ is always destabilising for *some* wavevector directions (i.e., either $q_x \lesssim |q_\perp|$ or $q_x \gtrsim |q_\perp|$) irrespective of parameter values [2]. However, for *extensile* $\zeta < 0$ active systems when $\lambda < 1$ [5, 6], $\omega_{tb}$ in (3) is *also* destabilising; in this case, the growth rate of the twist-bend mode dominates over the splay-bend one [5, 6] because it is unstable for all wavevectors with $q_x \neq 0$ while $\omega_{sb}$ interpolates to splay and is thus stabilised [14].

The value of $\lambda$ need to be measured experimentally. There is no a priori reason to assign a specific value or even sign to $\lambda$ for a particular system. It has, at times, been argued that $\lambda$ must be small (i.e., $|\lambda| < 1$) in the ordered state. This is not necessarily justified. However, in equilibrium systems, in general, molecular nematics are flow-aligning (i.e.$|\lambda| > 1$) and larger objects are flow-tumbling (i.e., $|\lambda| < 1$). Further, rod-like objects usually have $\lambda < 0$ and disc-like objects have $\lambda > 0$ [15]. Since microtubule bundles are rod-like and fairly large, and since multiple experiments [5, 6] and simulations in three-dimensional active fluids suggest that $\lambda < 1$ in motor-microtubule gels, we use this value in the discussion that follows.

In the experiment, the extensile active gel is in a channel with a square cross-section of size $W$. To examine the growth rate of the gel with this constraint, we replace $q_\perp^2$ by the smallest value it can assume: $q_\perp^2 = 2\pi^2/H^2$. Using this and noting that the channel is at least one order of magnitude longer than $H$, we obtain the wavelength of the fastest growing mode as

$$\lambda_f^2 = \frac{2W^2}{(W/l_a) - 1} \tag{5}$$

where we have defined the active lengthscale

$$l_a = \pi\sqrt{\frac{3K[4\Gamma\eta + (\lambda - 1)^2] + 4\zeta_2(\lambda - 1)}{3\zeta(\lambda - 1)}}. \tag{6}$$

Eq. (5) has the same form as in [5]; the main difference is in $l_a$ which now contains a second activity coefficient $\zeta_2$ and fully accounts for passive stresses. From (5), we see that $W$ must be larger than $l_a$ for $\lambda_f$ to be real i.e., the ordered state of the gel to be unstable. $\lambda_f$ decreases with increasing $W$ till $W = 2\ell_a$ and then increases with $W$, scaling as $\lambda_f \sim \sqrt{Wl_a}$ for $W \gg l_a$. Note that $\zeta_2$ is important for obtaining both the threshold beyond which the gel is unstable and the crossover scale that determines whether $\lambda_f$ decreases with $W$ or increases with $W$. In particular, $\zeta$ and $\zeta_2$ are likely to depend similarly on motor density. Therefore, simply increasing the motor density may not lead to a decrease in $l_a$ and, in a confined system, an active gel may not get destabilised for any motor concentration, as discussed in [4]. Determining the values of $\zeta$, $\zeta_2$ etc. individually presents a difficult experimental challenge. Instead, $l_a$ may, in principle, be determined from measurements of $\lambda_f$. Also recall that this calculation is only valid when the dominant instability is due to the twist-bend mode; if the dominant instability is via the splay-bend mode (i.e., is $\lambda > 1$), the wavelength of the fastest growing mode has a different, more complicated dependence on $W$.

Note further that for $W \gtrsim l_a$, $\lambda_f$ is very large. Since the channel in our experiment has a finite length $L$, $\lambda_f < L$ for this mode to exist. For channels smaller than that, the ordered state of an active gel is unstable if and only if the mode $q_x^2 = \pi^2/L^2$ is unstable.

In our experiments, the flows get smoother as $W$ is increased. This is clearly seen from the measurement of the temporal correlation of the direction of motion of Lagrangian tracers; the persistence time of the tracers goes from being around a minute to around 20min as $W$ increases from 0.3 or 0.5mm to 0.8 or 1 mm. This immediately implies that the lengthscale of flow structures also significantly increase (by at least an order of magnitude) as $W$ is increased. Since one may reasonably expect the size of vortices to scale as $\lambda_f$ (though this is not necessary: the scale of structures formed at long times need not be given by the fastest growing mode of an instability that destroyed the original state; indeed, it is not in our system), this increase in the scale of flow structures should be an indication of an increasing $\lambda_f$ with $W$ which means $W \gg 2\ell_a$ in our experiments even for the smallest $W$.

While the qualitative increase in the size of flow structures with increasing channel dimensions is accounted for by the simple linear stability analysis displayed here, understanding detailed features of the flow require a numerical calculation of the active gel model in the channel geometry, which will be presented elsewhere. Here, we qualitatively discuss some features of the flow. Experimental measurements of the active flow demonstrate that the persistence time of the Lagrangian tracers and, therefore, the scale of vortices are roughly the same for channels with $W = 0.3$mm and $W = 0.5$mm. The vortices are much larger for channels with $W = 0.8$mm and $W = 1$mm, but again, roughly of the same size for these two values. This may be rationalised by assuming that the flow is organised in a vortex array aligned along the $x$ axis. Vortex arrays have been regularly obtained in simulations of active fluids [16, 18–20]. The number of clockwise and anticlockwise rotating vortices must be the same to maintain zero net vorticity [16]. Further, number of vortices should be a function of the ratio $L/W$ on general grounds and should increase with $L/W$. Ref. [16]) obtains a linear relationship for a polar fluid in a two-dimensional rectangular channel with strong anchoring condition. Whatever the functional dependence of the vortex number on $L/W$, the vortex number only increases in the increment of 2. Therefore, when $W$ doesn't change by a lot, the vortex number may not change and the flow structure may remain qualitatively unchanged. The fact that there seems to be two well-defined "cells" in 0.8mm and 1mm channels in which the velocity field is roughly along (or opposite to) $\hat{x}$ separated by a region in which the velocity field is mostly in the directions perpendicular to $\hat{x}$ supports the existence of two oppositely rotating vortices in these channels. At smaller confinements, there seems to be more vortices. A more quantitative and detailed theoretical and experimental account of the active flow will be presented elsewhere.

To summarise, the linear stability analysis of an ordered active fluid in a channel geometry demonstrates that for $W \gg \ell_a$, the wavelength of the fastest growing mode grows with $W$. Beyond the instability, we believe that the experimental observations point to the existence of an array of vortices along $\hat{x}$. Given that the wavelength of the fastest growing mode increases with $W$, it is natural for there to be fewer vortices at larger $W$; in fact, we seem to observe just a pair of vortices in the 0.8 and 1mm channels. We believe that a channel with a cross section of 1mm would support a steady flowing state [17] were it not closed at the end. Since the vortex number cannot increase continuously but has to increase by 2, it is also natural that channels with similar $W$ have the same number of vortices. However, it is important to note that the stability analysis fails to predict both the structure of the flow and the scaling of vortex size with $W$. The vortex size increases 10-fold upon a 3-fold increase of $W$, signifying it scales as $W^2$, in contrast to the stability analysis which would predict a $\sqrt{W}$ scaling.

[1] M. C. Marchetti et al., Rev. Mod. Phys **85**, 1143 (2013)
[2] R. A. Simha, S. Ramaswamy, Phys. Rev. Lett. **89**, 058101 (2002)
[3] R. Voituriez, J.-F. Joanny, J. Prost, Europhys. Lett. **70**, 404 (2005)
[4] A. Maitra et al., Proc. Natl. Acad. Sci 115, 6934 (2018)
[5] P. Chandrakar et al., Phys. Rev. Lett. **125**, 257801 (2020)
[6] M. Varghese et al., Phys. Rev. Lett. **125**, 268003 (2020)
[7] L. M. Lemma et al., PNAS Nexus **2**, 5 (2023)
[8] G. Sarfati et al. Soft Matter 18, 3793 (2022)
[9] B. Najma et al., Nat. Commun. **13**, 6465 (2022)
[10] D. A. Gagnon et al., Phys. Rev. Lett. **125**, 178003 (2020)
[11] P. G. de Gennes, J. Prost, The Physics of Liquid Crystals, Clarendon Press (1995)
[12] H. Stark, T. C. Lubensky, Phys. Rev. E **67**, 061709 (2003)
[13] P. M. Chaikin, T. C. Lubensky, Principles of Condensed Matter Physics, Cambridge University Press (2000)
[14] R. Chatterjee et al., Phys. Rev. X **11**, 031063 (2021)
[15] S. A. Edwards, J. M. Yeomans, Europhys. Lett. **85**, 18008 (2009)
[16] P. Gulati, S. Shankar, M. C. Marchetti, Front. Phys. **10**, 948415 (2022)
[17] K.-T. Wu et al., Science **355**, 1284 (2017)
[18] A. Doostmohammadi et al., Nat. Commun. **8**, 15326 (2017)
[19] A. Doostmohammadi et al., Nat. Commun. **7**, 10557 (2016)
[20] S. Chandragiri et al., Phys. Rev. Lett., **125**, 148002 (2020)